\documentclass{aastex61}

\usepackage{graphicx}
\usepackage{epstopdf}

\received{January 12, 2018}
\revised{May 13, 2018}
\accepted{May 16, 2018}
\submitjournal{ApJ}

\shortauthors{Medina et al.}
\begin{document}

\title{An Automated tool to detect variable sources in the Vista Variables in the V\'{\i}a L\'actea Survey. The VVV Variables (V$^{4}$) catalog of tiles d001 and d002.}

\correspondingauthor{Nicol\'as Medina}
\email{nicolas.medina@postgrado.uv.cl, nicomedinap@gmail.com}

\author[0000-0002-0786-7307]{N. Medina}
\affil{Millennium Institute of Astrophysics (MAS), Santiago, Chile.}
\affil{Instituto de F\'isica y Astronom\'ia, Universidad de Valpara\'iso, Av. Gran Breta\~na 1111, Playa Ancha, Casilla 5030, Chile.}

\author{J. Borissova}
\affil{Instituto de F\'isica y Astronom\'ia, Universidad de Valpara\'iso, Av. Gran Breta\~na 1111, Playa Ancha, Casilla 5030, Chile.}
\affil{Millennium Institute of Astrophysics (MAS), Santiago, Chile.}

\author{A. Bayo}
\affil{Instituto de F\'isica y Astronom\'ia, Universidad de Valpara\'iso, Av. Gran Breta\~na 1111, Playa Ancha, Casilla 5030, Chile.}
\affil{N\'ucleo Milenio Formaci\'on Planetaria - NPF, Universidad de Valpara\'iso, Av. Gran Breta\~na 1111, Valpara\'iso, Chile.}

\author{R. Kurtev}
\affil{Instituto de F\'isica y Astronom\'ia, Universidad de Valpara\'iso, Av. Gran Breta\~na 1111, Playa Ancha, Casilla 5030, Chile.}
\affil{Millennium Institute of Astrophysics (MAS), Santiago, Chile.}

\author{C. Navarro-Molina}
\affil{Millennium Institute of Astrophysics (MAS), Santiago, Chile.}
\affil{Instituto de F\'isica y Astronom\'ia, Universidad de Valpara\'iso, Av. Gran Breta\~na 1111, Playa Ancha, Casilla 5030, Chile.}

\author{M. Kuhn}
\affil{Instituto de F\'isica y Astronom\'ia, Universidad de Valpara\'iso, Av. Gran Breta\~na 1111, Playa Ancha, Casilla 5030, Chile.}
\affil{Millennium Institute of Astrophysics (MAS), Santiago, Chile.}

\author{N. Kumar}
\affil{Centre for Astrophysics, University of Hertfordshire, College Lane, Hatffeld, AL10 9AB, UK.}

\author{P. W. Lucas}
\affil{Centre for Astrophysics, University of Hertfordshire, College Lane, Hatffeld, AL10 9AB, UK.}

\author{M. Catelan}
\affil{Instituto de F\'isica, Facultad de F\'isica, Pontificia Universidad Cat\'olica de Chile, Casilla 306, Santiago 22, Chile}
\affil{Millennium Institute of Astrophysics (MAS), Santiago, Chile.}

\author{D. Minniti}
\affil{Departamento de Ciencias F\'isicas, Facultad de Ciencias Exactas, Universidad Andr\'es Bello, Av. Fernandez Concha 700, Las Condes, Santiago, Chile.}
\affil{Millennium Institute of Astrophysics (MAS), Santiago, Chile.}
\affil{Vatican Observatory, V00120 Vatican City State, Italy.}

\author{L. C. Smith}
\affil{Institute of Astronomy, University of Cambridge, Madingley Road, Cambridge, CB3 0HA, UK}
\affil{Centre for Astrophysics Research, University of Hertfordshire, College Lane, Hatfield AL10 9AB, UK.}

\date{\today}

\begin{abstract}
Time-varying phenomena are one of the most substantial sources of astrophysical information and their study has led to many fundamental discoveries in modern astronomy. We have developed an automated tool to search and analyze variable sources in the near infrared $\rm K_{s}$-band, using the data from the Vista Variables in the V\'ia L\'actea (VVV) ESO Public Large Survey. This process relies on the characterization of variable sources using different variability indices, calculated from time series generated with Point Spread Function photometry of sources under analysis. In particular, we used two main indices: the total amplitude $\rm \Delta K_s$ and the eta index, $\eta$, to identify variable sources. Once variable objects are identified, periods are determined with Generalized Lomb-Scargle periodograms, and the Information Potential Metric. Variability classes are assigned according to a compromise between comparisons with VVV Templates and the period of the variability.
The automated tool is applied on VVV tiles d001 and d002 and led to discovery of 200 variable sources. We detected 70 irregular variable sources and 130 periodic ones. 
In addition nine open cluster candidates projected in the region are analyzed, the infrared variable candidates found around these clusters are further scrutinized by cross-matching their locations against emission star candidates from VPHAS+ survey $\rm H_{\alpha}$ color cuts.
\end{abstract}

\section{Introduction}\label{introduction}

Time-varying phenomena are arguably one of the most powerful sources of astrophysical information. In the last decades, the development of astronomical instrumentation and automation has enabled many time-domain surveys, such as for example the wide-field optical imaging surveys: the Catalina Real-time Transient Survey~\citep{Drake2009}; Pan-STARRS~\citep{Kaiser2002}; and GAIA~\citep{Perryman2005}. In the near future, even more ambitious programs, such as the Large Synoptic Survey Telescope (LSST,~\citealt{Krabbendam2010}) are planned to start monitoring the optical sky. While optical surveys are getting wider and deeper, the extension and the systematic exploration of the variable sky toward the infrared, is also under development, in order to better cope with the problem of the interstellar extinction. The VISTA Variables in the V\'ia L\'actea survey (VVV;~\citealt{Minniti2010,Saito2012}) is one of these infrared surveys and is comparable to the optical ones both in areal and time-domain coverage (e.g.,~\citealt{Arnaboldi2007,Arnaboldi2012}). It has been designed to catalog $\sim 10^{9}$ sources, where a great part of those are expected to be variable stars. All these sources will be used to map the structure of the optically obscured Galactic disk and bulge by using some main distance indicators such as the red-clump giants and pulsating variable stars (RR Lyrae stars, classical Cepheids, anomalous Cepheids and Miras, and semi-regular variables), as well as to provide a census of Young Stellar Objects (YSOs) across the southern Galactic plane. Some focused studies have been carried out in the southern disc region in search of variable stars using these data, for example:~\cite{Contreras_Pena2017,Contreras_Pena_b2017}, cataloging high amplitude variable stars, with emphasis on YSOs; ~\cite{Borissova2016} searching for YSOs around young stellar clusters;~\cite{Dekany2015b} searching Classical Cepheids in the bulge;~\cite{Elorrieta2016,Gran2016,Minniti2017} focusing on the RR Lyrae stars. However, the systematic and uniform searches of the variability phenomena in the VVV disc area are still missing. On the other hand, the VVV multi-epoch observations produced a huge amount of information, a dataset of challenging size. It is highly necessary to develop tools, processes and techniques ables to perform sophisticated analysis in an automated way in order to efficiently exploit this unique dataset.
In this paper we present an automated tool designed to search, classify and analyze variable sources in the near infrared $\rm K_{s}$-band. Our tool is fed VVV tile images to extract time series and identify different types of variable sources. The main goal is to understand the behavior of $\rm K_{s}$-band variability in large regions of the sky, with the ultimate goal of processing the 1.8 squared degrees images of the VVV observations. The identified variables will be used to derive properties in active star forming regions, to determine distances using the periodic stars with available period-luminosity relation, as RR Lyrae and Cepheids, and to identify parameters of different variables stars. The information gathered from these sources will be collected in the "VVV Variables ($\rm V^{4}$)" catalog.

The structure of the paper is as follows:
In section~\ref{Data} we present the photometry and calibration process on the VVV tiles and characteristics of the extracted time series. Then in section~\ref{Methodology} the methodology is explained, where we focus on identifying irregular and periodic variables, mainly using different variability indices and periodograms. Next in section~\ref{results}, the preliminary classification of the selected sources is presented. This classification is based on principal properties of the sources, as the shape of time series and light curves, and the period in the case of variable sources. We determined general properties of selected variables sources, as characteristic features of the light curves and locations on the color-magnitude and color-color diagrams. Also, we described the environment of variable sources in the Young stellar clusters candidates projected on VVV tiles d001 and d002.
Finally, in section~\ref{Summary} we present the catalog of variable sources in these tiles. The individual characterization of the variable objects is beyond the scope of this paper, and they will be analyzed in an up-coming work, once the follow-up spectroscopic analysis is completed.

\section{The VVV data}\label{Data}
The VVV survey is an ESO Infrared Large Public survey~\citep{Minniti2010,Saito2012DR1} which uses the 4-meter VISTA telescope located at Cerro Paranal Observatory, Chile. The survey was designed for mapping 562 deg$\rm ^{2}$ in the Galactic bulge and the southern disk in five near-infrared broad-band filters: $\rm Z\ (\lambda_{eff}=0.87\mu m)$, $\rm Y\ (\lambda_{eff}=1.02\mu m)$, $\rm H\ (\lambda_{eff}=1.25\mu m)$, $\rm J\ (\lambda_{eff}=1.64\mu m)$, $\rm K_{s}\ (\lambda_{eff}=2.14\mu m)$, with a time coverage spanning over five years between 2010 and 2015 in the $\rm K_{s}$-band. The telescope has a near-infrared camera, VIRCAM~\citep{Dalton2006}, consisting of an array of 16 detectors with $\rm 2048\times2048$ pixels. A set of single exposures (a paw-print) are combined into a tile, covering $\rm 1.5\times1.1$ degrees in the sky. To cover the VVV area, the disk field was divided into 152 tiles and the bulge into  196 tiles (see~\citealt{Saito2012} for more details). 

\begin{figure}[htbp]
      \centering
      \noindent\makebox[\textwidth]{\includegraphics[height=6.2cm,width=18cm]{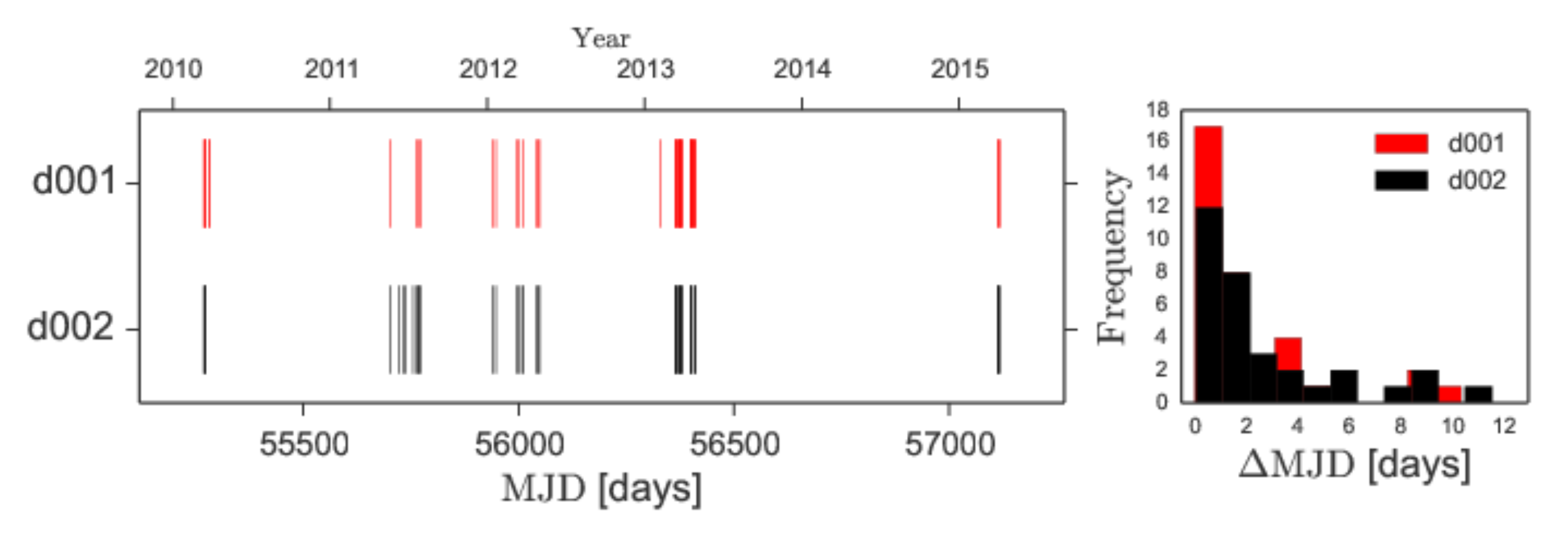}}
      \caption{Left side: The log of observations of VVV tiles d001 and d002 between 2010 and 2015, based on the log representation of~\cite{Rebull2014}. Each photometric measurement is marked by a '$|$' symbol. The bars are thicker in places with high cadence. Right side: Histogram of differences between consecutive observations $\rm \Delta MJD$. Is possible to see that the typical time interval between observations is between 0.3 and 2 days.}
	  \label{cadencia}
\end{figure}

To test our method we choose the first two VVV disk tiles, namely d001 and d002, due to their low crowding and interstellar reddening when compared to the rest of the VVV disk area. The preliminary reduced images were retrieved from the VISTA Science Archive\footnote{http://horus.roe.ac.uk/vsa} (VSA) database~\citep{Cross2012}, keeping the quality flags. In total, we analyzed up to 55 and 41 $\rm K_{s}$ images for the tiles d001 and d002, respectively. Figure~\ref{cadencia} represents the log of observations of the VVV tiles.

\subsection{Photometry and calibrations}\label{photometry}
The Point Spread Function (PSF) photometry was obtained using the $\mathtt{Dophot}$ software~\citep{Schechter1993, Alonso2012} in all available tile images in the field of view (FoV). We based this procedure in the method explained in~\cite{Navarro2016}. We assessed the reliability of the photometry, using the $\rm \mathtt{Dophot}$ parameter $chi$, which quantifies the PSF quality. The sources with $chi>3$ were rejected, due to the large associated uncertainty. The calibration process to the VISTA system was done using the aperture photometry catalogs produced by the Cambridge Astronomical Survey Unit\footnote{http://casu.ast.cam.ac.uk} (CASU). We selected sources with Stellar (``-1'') or border-line stellar (``-2'') morphological classification to perform the cross-match using $\rm \mathtt{STILTS}$~\citep{Taylor2006}, using the catalog of the first epoch as reference with a 0.''34 tolerance (VIRCAM pixel size). The conversion factors and uncertainties were estimated via a 2-sigma clipping linear fit to the $\rm \mathtt{Dophot}$ PSF photometry vs. the CASU isolated selected sources. By following this procedure, we have found 624,983 sources in d001 and 683,643 sources in d002 in common, in $\rm K_{s}$-band. The photometry in J and H band was performed using $\rm \mathtt{Dophot}$ in a similar manner, i.e., using the CASU catalogs to calibrate the PSF photometry.

\subsection{Cadence of the observations}\label{Cadence}
As it has been pointed out, the tiles d001 and d002 accumulate up to 55 and 41 epochs observed between 2010 and 2015, respectively. Left side of Figure~\ref{cadencia} shows the gaps, the baseline, and the maximum size of the time-step between epochs. The right side of Figure~\ref{cadencia} shows the distribution of the difference of consecutive observations $\rm \Delta MJD$, zoomed up to $\rm \Delta MJD>12$ days. The minimum time interval between the observation is $\sim$0.35 days, with distribution maximum between 0.35 and 2 days. Thus, from the time cadence of the observations, we can expect to detect variability related to timescale accretion variations, star spots, episodic accretion events, rotational modulation and variable extinction in the YSOs~\citep{Contreras_Pena2017,Rebull2014}.
On the other hand, VVV produces unevenly spaced light-curves, which provides its challenges, but still we expect to identify different types of periodic variability in a wide range of time-scales (see for example~\citealt{Elorrieta2016,Gran2016,Minniti2017}). 

\subsection{The $\rm K_{s}$-band time series}\label{Ks_lightcurve}
\begin{figure}[!tbp]
  \centering
  \noindent\makebox[\textwidth]{
  \includegraphics[height=11cm, width=12cm]{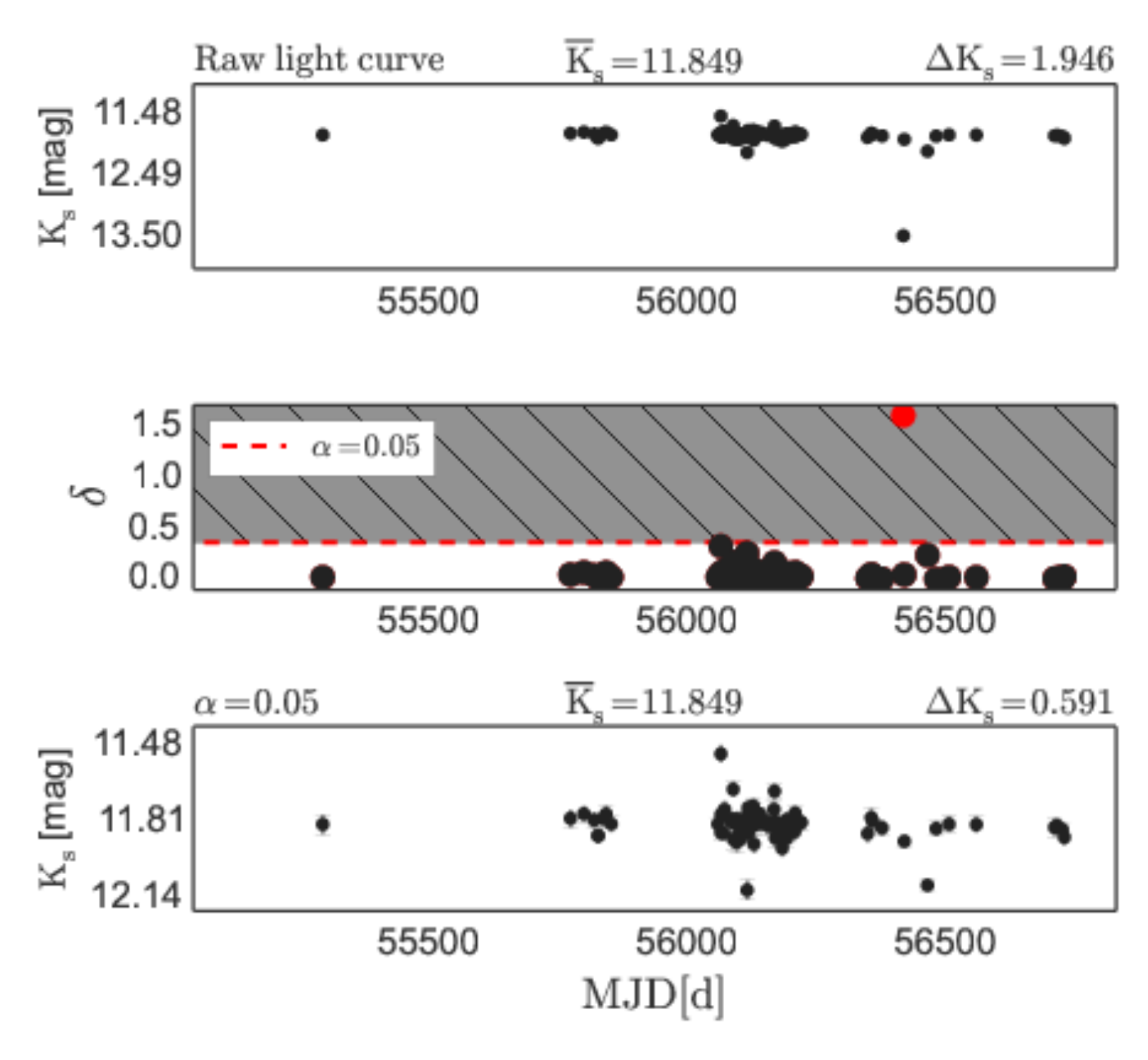}}
  \caption{Performance of Thompson $\tau$ technique acting on a time series. Top plot: Raw light curve extracted from the photometric process. Middle plot: Absolute deviation $\delta$ of each measurements in function of $\rm MJD$. The dashed red line indicates the rejection region (gray region) using $\alpha=0.05$. The red point that fall in this region is removed from the sample. Lower plot: Modified time series, which will be used in the analysis.}
  \label{Thompson_tau_technique}
\end{figure}

The $\rm K_{s}$-band time series of the sources were constructed by cross-correlation of all the catalogs for all available epochs. We filtered our initial sets of time series with some ad-hoc quality and robustness criteria: 1) A minimum of 25 photometric measurements. 2) A total amplitude $\rm \Delta K_{s}>0.2 $ mag, where $ \rm \Delta K_{s}=(K_{s}^{max}-K_{s}^{min})$. 3) An upper limit in flux of $\rm K_{s}>11$ mag (to avoid objects that may suffer from saturation in some VVV epoch). The first restriction represents the minimum number of epochs which allows to search for reliable periods. The second is motivated by a conservative estimation of the errors of photometry and transformation to the standard system. These initial filters reduced the source numbers obtained from photometry by approximately 30\% (for example from 669825 to 433102 for d001). Moreover, the photometric measurements are prone to be affected by systematic errors that are hard to clarify and quantify, given atmospheric or instrumental problems. For example ~\cite{Alonso2015} reported a problem related with highly variable PSFs in the tile images, due to different geometric distortions in the combination process of the paw-print images. Thus, to remove outliers in the time series, we implement the modified Thompson $\tau$ technique, which is based on the definition shown in~\cite{Thompson1985}. The modified Thompson $\tau$ statistic is defined as:

\begin{equation}
\rm \tau = \displaystyle \frac{t_{\alpha/2}(n-1)}{\sqrt{n}\sqrt{n-2+t^{2}_{\alpha/2}}},
\end{equation}

\noindent where $\rm n$ is the number of data points, and $t_{\alpha/2}$ is the critical value of Student's t-distribution given a confidence parameter $\alpha$. For each photometric data point $\rm K_s$ in a time series, the standard deviation of the time series $\rm \sigma_{ts}$ and the absolute deviation $\rm \delta_{i}=|Ks_{i}-\overline{K}s|$ is calculated, where $\rm \overline{K}$s is the mean $\rm Ks$ magnitude. Individual photometric measurements were removed from time series when $\rm \delta_{i} > \tau \sigma_{ts}$, using a confidence level of 95\% (i.e. $\alpha=0.05$). One of the consequences of this approach is that we will remove poorly sampled transients event from our time series. Figure~\ref{Thompson_tau_technique} shows the performance of this method acting on a time series.

\section{Methodology}\label{Methodology}
In time series analysis, it is frequent to use different sets of statistics, commonly called ``variability indices'', to quantify changes in luminosity with time. Depending on the definition of these indices, a population of variables that have  similar behavior can be identified and one can try to separate stochastic variability from ``further organized'' flux variations. Examples of these indexes are the Welch-Stetson $\rm I_{ws}$~\citep{Welch1993} and Stetson $\rm J_{Stet}$ and $\rm K_{Stet}$ indexes~\citep{Stetson1996}. This quantities have been used to identify sources that exhibit large photometric variations along the time. Defined in this manner, sources with larger $\rm J_{stet}$ values are the most probable variable sources. Different authors in the literature define particular limit values of $\rm J_{stet}$ for this task (such as $\rm J_{stet}\geq 0.55$~\citealt{Carpenter2001}, $\rm J_{stet}\geq 0.9$~\citealt{Rebull2014}). In the literature, it is possible to find many more different indices (or features) tailored to identify different types of variable sources. Thus, deciding which index is useful to detect a specific type of variability does not only depend on the definition of the index itself, but also on the properties of the available data. Here, we briefly summarize some variability indices applicable when only one photometric band is available; and the type of variability that they can detect. 

\begin{table}[!tbp]
\centering
\caption{Variability indices computed in this analysis}
\begin{tabular}{cc}
\hline
\multicolumn{2}{c}{Set of variability indices}                    \\ \hline 
Index                                & Reference                  \\ \hline \hline
$\eta$ index                         & ~\cite{Neumann1941}        \\
Stetson J                            & ~\cite{Stetson1996}        \\
Stetson K                            & ~\cite{Stetson1996}        \\
$\rm \sigma_{ts} / \mu$ ratio        & ~\cite{Shin2009}           \\
Classical $\chi^{2}$                 & ~\cite{Rebull2014}         \\ 
Total amplitude $\rm \Delta K_{s}$   & ~\cite{Contreras_Pena2017} \\ \hline
\label{Table_indices}
\end{tabular}
\end{table}

\begin{figure}[htbp]
\centering
\noindent\makebox[\textwidth]{
\includegraphics[height=5.5cm,width=15.5cm]{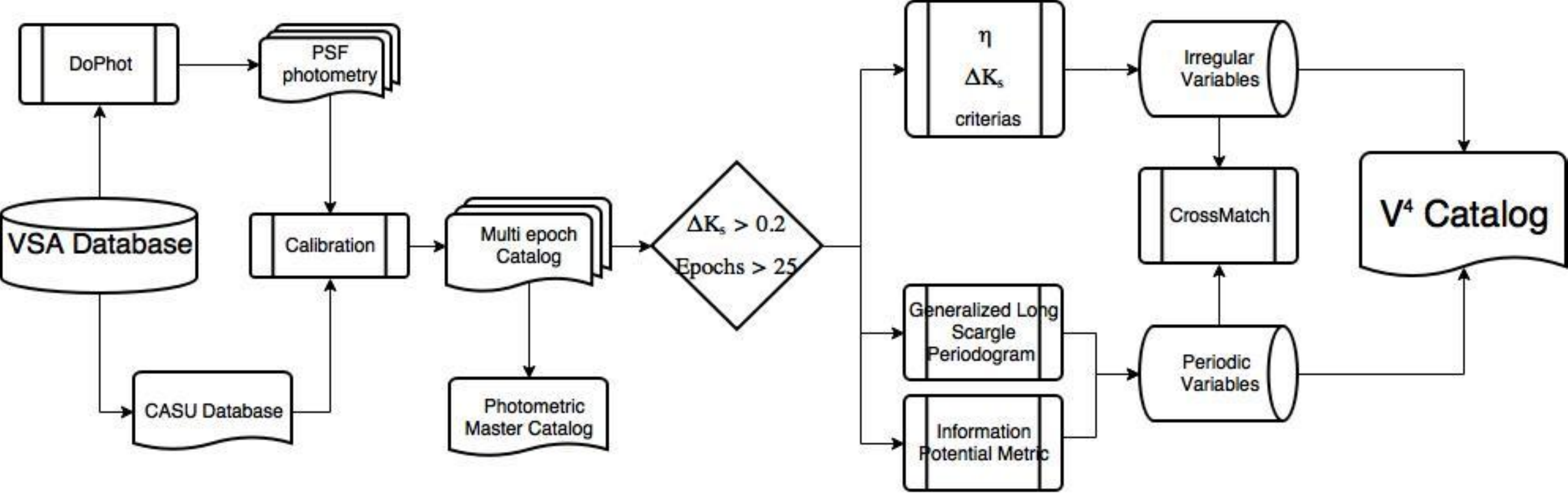} }
\caption{Schematic view of the automated process developed and used in this study to categorize the sources in the $\rm V^{4}$ catalog.}
\label{scheme}
\end{figure}

In this work, we considered a set of six variability indices (see the references in Table~\ref{Table_indices}) in order to characterize the behavior of the variable sources along time. Mainly, we used the amplitude $\rm \Delta K_s$ and the $\eta$ index to identify irregular variables. All these parameters are estimated directly from their $\rm K_{s}$-band time series. All the procedures, and the automated process that we developed were summarize in the flux diagram of Figure~\ref{scheme}. 

\subsection{Identifying irregular variable sources}\label{Irregular_methods}

Eruptive Pre-main Sequence (PMS) stars are traditionally classified as FU Orionis Types (FUors, \citealt{Herbig1966}) and EX Lupi (EXors, \citealt{Herbig89}) types. Their variations occasionally have high amplitudes (up to 2-6 magnitudes in optical bands) in a short time scales. Different physical processes have been proposed to explain the variations of these objects: accretion or variable extinction induced by their circumstellar disks, among others. These high amplitude variables are potential tracers of new generation of stars, so we expect to detect them within and close to the star forming regions (SFR) of the Galaxy.

As an example, in~\cite{Contreras_Pena2017}, the total amplitude $\rm \Delta K_{s}$ was used as a discriminant to identify variable sources with $\rm \Delta K_{s}>1 $ magnitude. This method was very useful to help identify likely YSOs amongst irregular and periodic variables stars projected against SFRs. The ``Amplitude index'' also performs satisfactorily in identifying sources with a large amplitude $\rm \Delta K_{s}$, like some Eclipsing Binary systems and likely pulsating asymptotic giant branch (AGB) stars as Miras and semi-regulars sources, which has periods longer than $ \rm P \geq 100$ days, and frequently are grouped under the name ``long period variables'' (LPV).

\begin{figure}[!tbp]
  \centering
  \noindent\makebox[\textwidth]{
  \includegraphics[height=11.5cm, width=6.5cm]{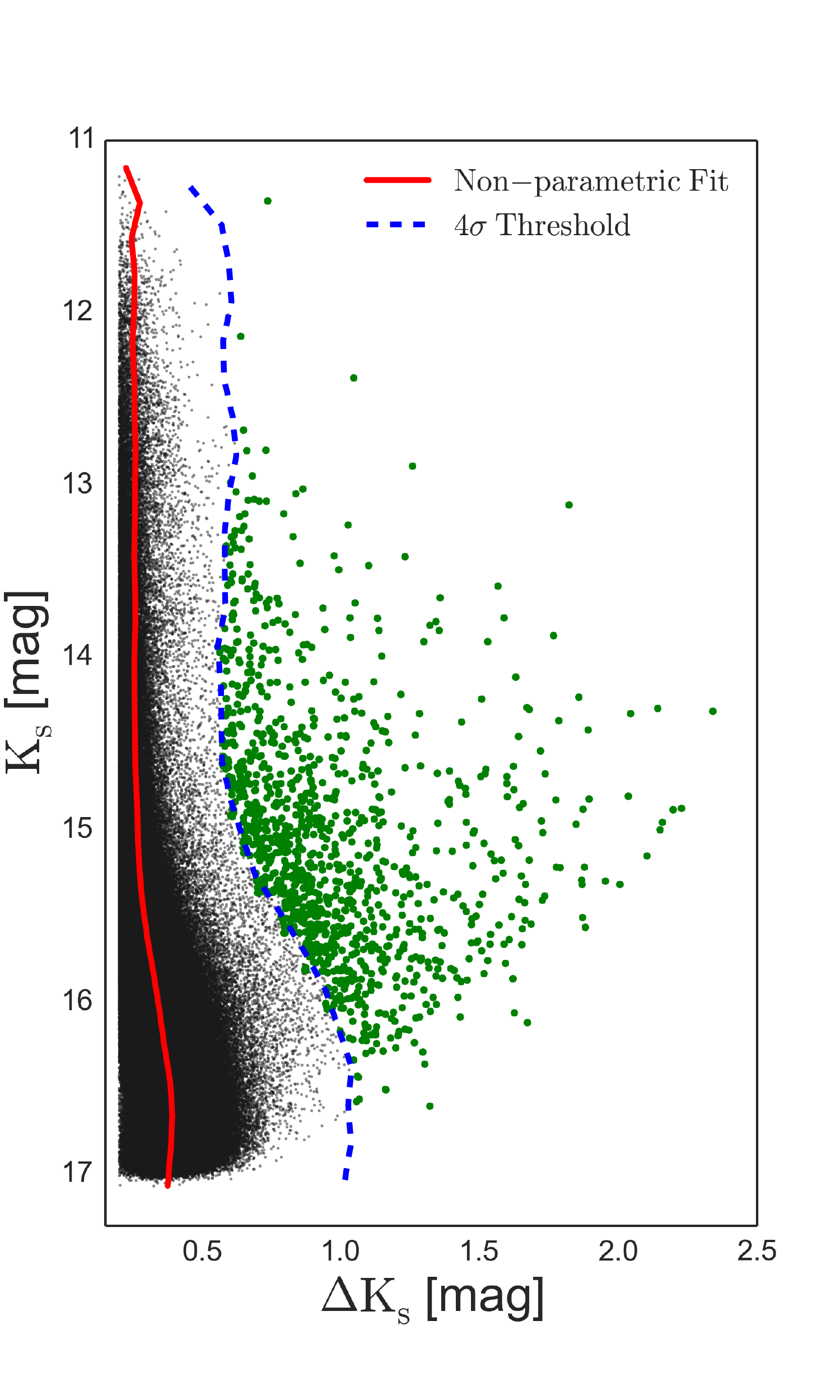}
  \includegraphics[height=10.5cm, width=9.5cm]{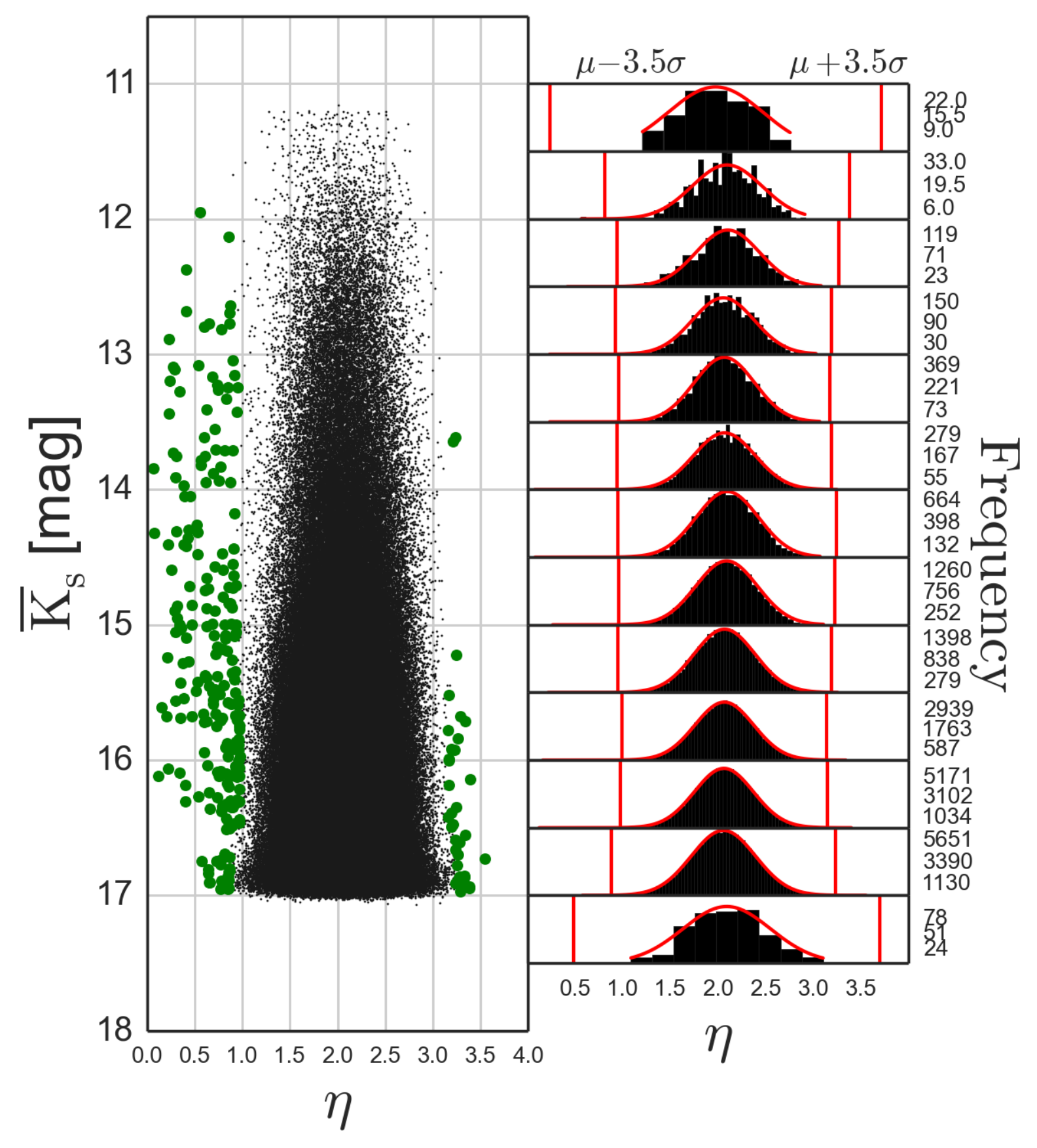}}
  \caption{Representation of the two selection criteria aiming at identifying irregular variables. The dark points are all stars in the final photometric catalog, and the green filled circles highlight the selected sources by each method. Left panel: Total amplitude $\rm \Delta K_{s}$ selection. The red solid line represent the non-parametric fit, while the dashed blue line displays for the 4$\rm \sigma$ threshold.  
Right panels: $\eta$ index selection. The red lines in the bins of 0.5 mag represent the $3.5\sigma$ threshold of the Gaussian fit.}
  \label{Criterios_seleccion}
\end{figure}

Following the same idea, we made a non-parametric fit on our photometric catalogs of d001 and d002. In order to quantify the behavior of $\rm \Delta K_{s}$ as a function of mean magnitude $\rm \overline{K}_{s}$, we measured the dispersion \textbf{in bins}, and selected those above 4$\rm \sigma$. This allowed us to define dynamical thresholds, taking into account that the estimated $\sigma$ depends on the stellar population projected on the different tiles (for example a tile containing a projected star forming region will have a different threshold assigned, than a tile with Population II stars). The left side of Figure~\ref{Criterios_seleccion} shows the amplitude $\rm \Delta K_{s}$ selection for the sources found in tile d001.

To complement the previously described criterion, we performed an additional selection using the $\eta$ index~\citep{Neumann1941,Shin2009,Sokolovsky2017}. This statistic is defined as the squared addition of successive differences between adjacent observations in a time series:

\begin{equation}
 \rm \eta = \displaystyle{\frac{1}{(N-1)\sigma^{2}_{ts}} \sum_{i=1}^{N-1} (m_{i+1}-m_{i})^{2}},
\end{equation}

\noindent where $\rm m_{i}$ are magnitude measurements, $\rm N$ the available number of epochs. The properties of the $\eta$ index are well known for a stationary Gaussian distribution, but not for astronomical time series, because usually, they have an unequal sampling. For time series with uncorrelated photometric measurements (i.e. time series with an uncorrelated normally distributed measurements), the $\eta$ index would have a value $\sim 2$, and extreme values for series with long time variability trends. Given the aforementioned properties and the volume of the data, we expect that index $\eta$ has a Gaussian distribution centered in $\eta \approx 2$. We separated $\rm \overline{K}_{s}$ into 0.5 magnitudes bins and considered $3.5\sigma$ confidence intervals on every bin to identify variable sources. The right two panels of Figure~\ref{Criterios_seleccion} show the $\eta$ selection, and the fit of Gaussian functions in each histogram generated by separating the distribution in bins of 0.5 mag. 

Each irregular variable should satisfy both $\rm \Delta K_{s}, \eta$ criteria in order to be included in the paper as an irregular variable. All selected candidates have visual confirmation on the corresponding images, the candidates with close (less than 0.4 arcsec) companions are removed.

\subsection{Identifying periodic variables stars}
One of the main goals of the VVV Survey is to obtain a complete census of pulsating stars, such as RR Lyrae, Cepheids, Semi-Regular and Mira variables across the Milky Way. These sources provide useful information in their quality as standard candles (using the developed Period-Luminosity relations in the near-IR), to determine the structure of our Galaxy. These stars are also useful to map the extinction affecting the projected areas. Another potential set of targets to be identified are the Eclipsing Binary (EBs) systems, which are known to provide the most robust/model free estimates of the fundamental stellar parameters. 

In this study, we have implemented two methods to identify periodic sources within VVV data:

\begin{itemize}
\item The Generalized Lomb-Scargle Periodogram (GLS, ~\citealt{Zechmeister2009}): A least-squares spectral analysis method based on the classical Lomb-Scargle Periodogram~\citep{Lomb1976,Scargle1982}. In particular, we used its implementation in the $\mathtt{astroML}$\footnote{ http://www.astroml.org/} python library~\citep{astroML}

\item The Informatic Potential Metric $\rm Q_{m}$ (IP metric, \citealt{Huijse2011}): A discriminant designed to identify the fundamental period of a time series using information theory. Within this framework, different set of time series $\rm \{x_{n}\}$ are assumed to be realizations of a continuous random variable $\rm X$. The Informatic Potential is defined as follows:

\begin{equation}
\rm IP_{X}(\{x_{n}\}) = \displaystyle\frac{1}{N^{2}}\frac{1}{\sqrt{2\pi}\sigma}\displaystyle\sum_{i=1}^{N} \sum_{j=1}^{N} \exp\left(\displaystyle-\frac{\|x_{i} - x_{j} \|^{2}}{2\sigma^{2}} \right).
\end{equation}

We used a grid of trial periods $\rm P_{t}$ to fold the time series into phase space. The folded light curves are then segmented in $\rm H$ bins and IP is computed for every bin ($\rm h$). The IP metric $\rm Q_{m}$ is computed as the squared differences between the information potential of each bin and the global IP:

\begin{equation}
\rm Q_{m}(P_{t}) = \frac{1}{H}\sum_{h=1}^{H} \left[IP_{X}(\{x_{n}\})-IP_{X}(\{x_{n}\}_{n \in h})\right]^{2}.
\end{equation}
\end{itemize}

To estimate the reliability of the periods found with both approaches, we calculated the statistical significance for the spectral power peaks in GLS and IP metric. Only objects with peak-significance greater that 99.9\%, were considered for further analysis and characterization. We note that this formal peak-significance assumes the uncertainties are described by uncorrelated Gaussian noise.

\subsection{Classification of periodic stars}
To determine the variability type of the identified periodic stars, we consider the shape of the light curve, and the period $\rm P$ determined by the methods previously described. A common tool to quantify the shape of a light curve is using templates, where the light curves of periodic sources are compared with templates to assign a well define variability class, using one or a set of statistics to relying this classification. These templates could be collected from public archives, literature and other databases, in order to create ``training sets'' that points to an automated classification. In this context, a significant (and still increasing) number of infrared light curves templates have been assembled in the VVV Templates Project~\citep{Angeloni2014}, where the main goal is to develop and test machine-learning algorithms for the automatic classification of VVV light curves.
Nevertheless, we need to consider that in the NIR, RR Lyrae stars are a special case, given that the amplitudes of light curves decreases from optical to infrared wavelengths, this leads to being more difficult to differentiate between RR Lyraes in fundamental-mode (RRab) and first-overtone (RRc) subtypes using only Templates. In this context, we used two criterion to classify these periodic sources:

\subsubsection{Using the VVV template project}
We expect low accuracies in the period estimation due to the relatively small number of epochs, as we pointed out in section~\ref{Cadence}. With this in mind, we have used templates of those variables that we expected to find in this galactic longitude, such as Classic Cepheids and Eclipsing variables. The $\delta$ Scuti stars were not considered in this analysis given that just two light curve templates are available in $\rm K_s$-band. Nevertheless, their periods are less than 0.2 days, and may have similar characteristics to EBs~\citep{Dong2017}. Several $\delta$ Scuti sources were identified around open clusters in $\rm K_s$-band~\citep{Palma2016}, so we need more information for a well characterization of this type of variables.

\begin{figure}[!tbp]
  \centering
  \noindent\makebox[\textwidth]{
  \includegraphics[height=13cm,width=3.5cm]{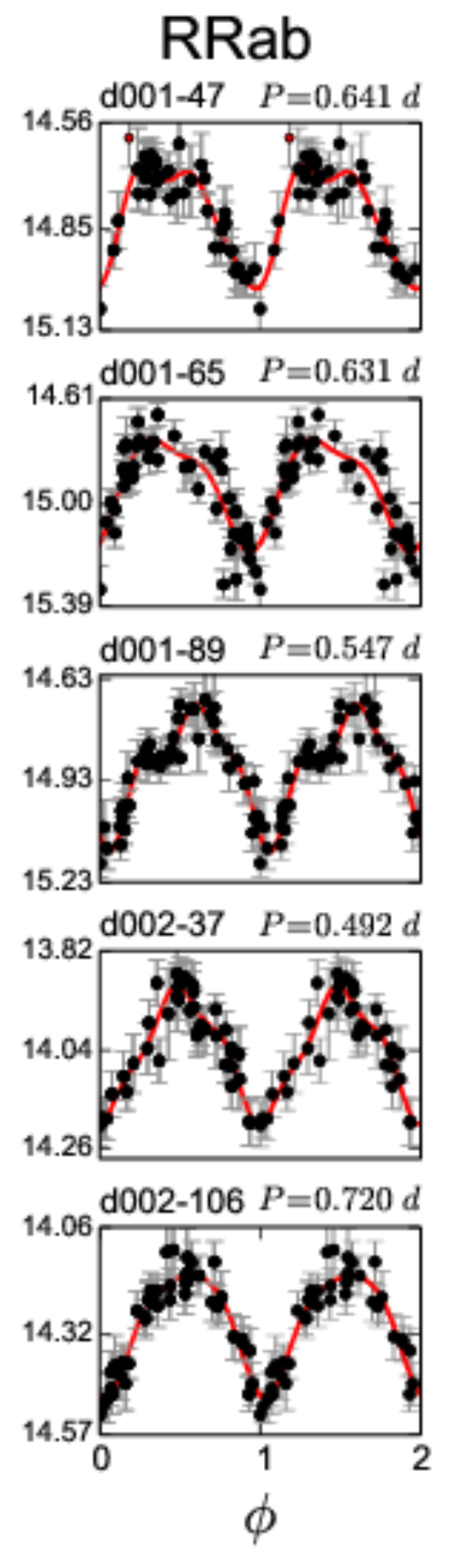}
  \includegraphics[height=13cm,width=3.5cm]{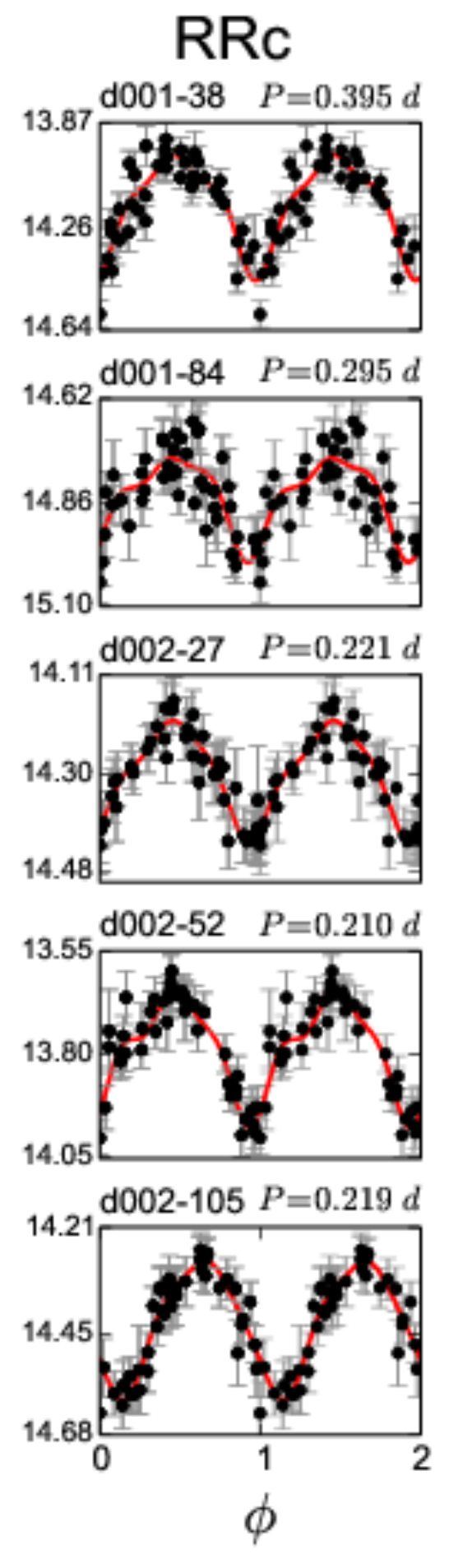}
  \includegraphics[height=13cm,width=3.5cm]{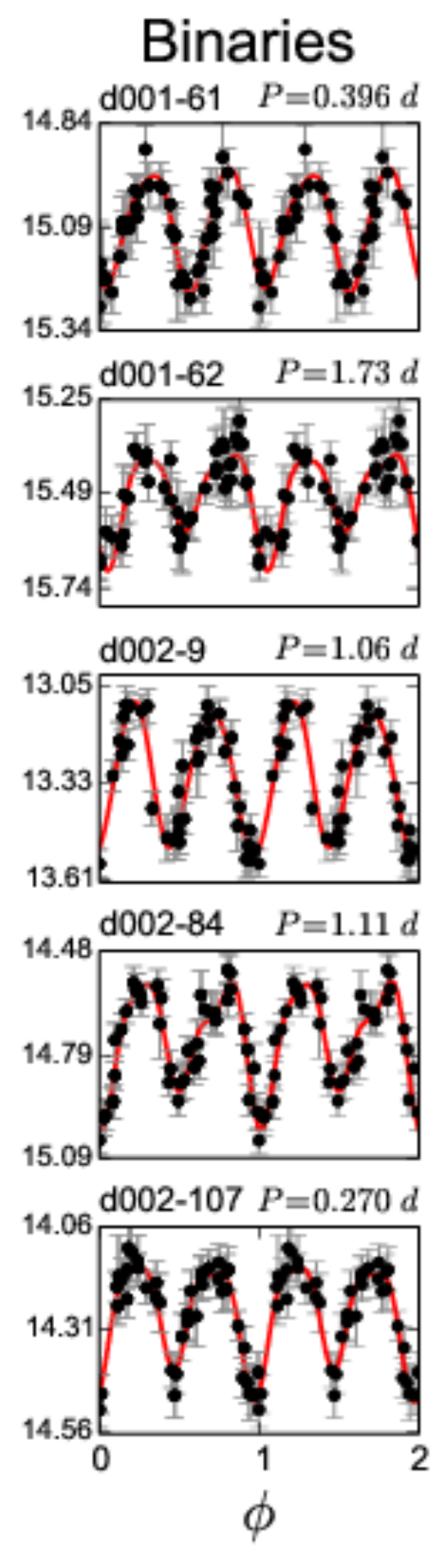}
  \includegraphics[height=13cm,width=3.5cm]{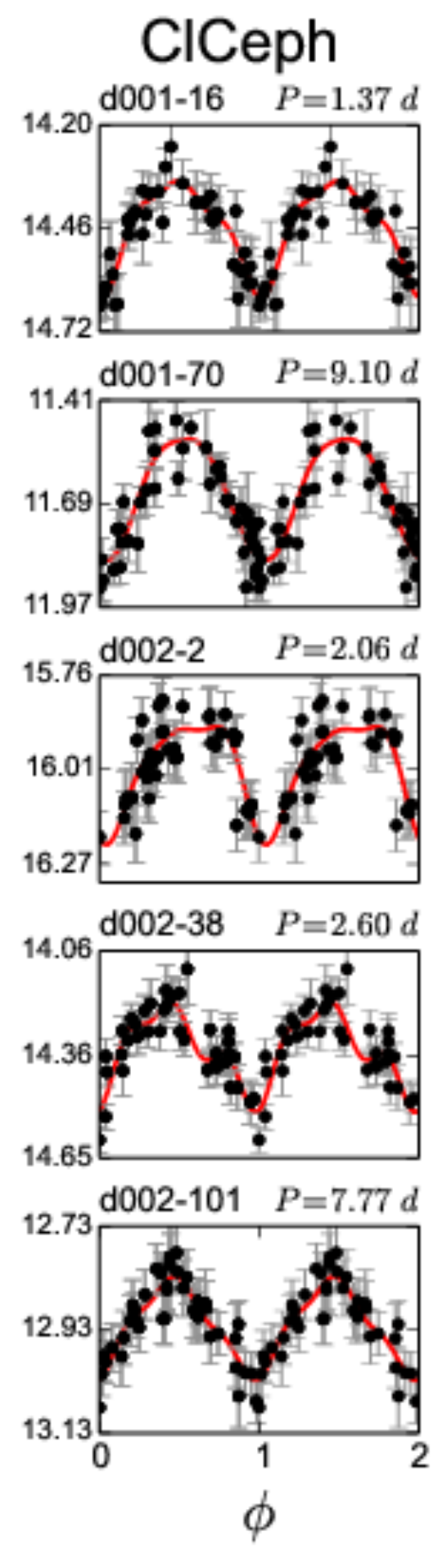}
  \includegraphics[height=13cm,width=3.5cm]{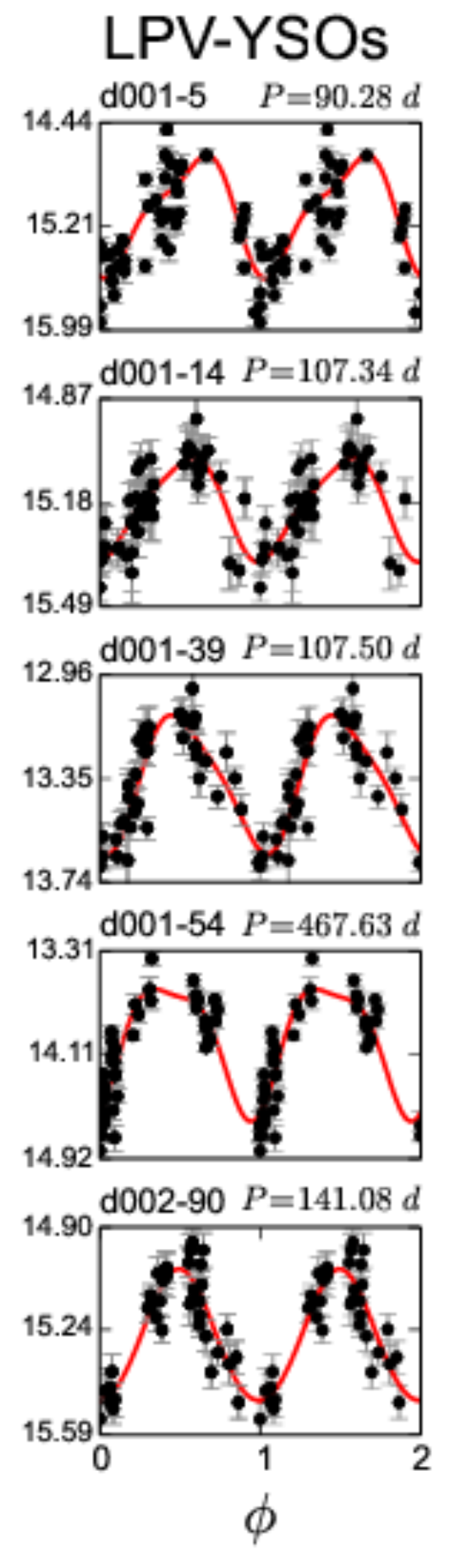}
  }
  \caption{Examples of light curves of the periodic stars of RRab, RRc, Cepheid, Binary and LPV types. Red points in the phase diagram are outliers of the Fourier fit. In the top of each plot the identification ID and obtained period $\rm P$ are shown.}
  \label{Periodics_phase}
\end{figure}

We used templates of Classic Cepheid (103 templates in $\rm K_s$, with $\rm 0.97<P<133.90$) and Eclipsing Binaries (76 templates in $\rm K_s$, with $\rm 0.305<P<16.092$) to compare with light curves of our objects when its period $\rm P$ is contained in the indicated period range of the templates. To quantify these comparisons, we performed a Fourier fit of $\rm N$ harmonics into the phase space for each template and variable, given its period $\rm P$ and average magnitude $\rm \langle m \rangle$ (see Figure~\ref{Periodics_phase}). The amplitude $\rm A_{k}$ and the phase $\rm \phi_{k}$ for each $k$ harmonic were determined. The Fourier series $\rm f(t)$ at time $\rm t$ is given by:

\begin{equation} 
 \rm f(t)= \langle m \rangle + \displaystyle{\sum_{k=1}^{N} A_{k}\sin \left( \frac{2\pi kt}{P} + \phi_{k}\right)}.
\end{equation}

Each ``synthesized'' template was then normalized subtracting the integral of the obtained Fourier series. The harmonic number $\rm N$ used to fit the model to periodic sources is $\rm N=4$. The periodic stars are classified using the template that had the best goodness of the fit, using the reduced $\rm \chi_{red}^{2}$ statistics as the criterion. If a source has $\rm \chi_{red}^{2} > 1$, will remain as not classified.

\subsubsection{Classifying RR Lyrae stars}
RR Lyrae stars can be sub-classified by their locations in the Bailey diagram~\citep{Bailey1902}, given that RRc sources have shorter periods than RRab type. Figure 5 of~\cite{Gavrilchenko2014} shows that RRab and RRc categories are located in different places of Bailey diagram. They also discuss the arbitrary limit $\rm P=0.4$ days to discriminate between these categories. Using this, we define the RRc region ($\rm 0.2<P<0.4$ days) and the RRab region ($\rm 0.4<P<1$ days), and they are shown in Figure~\ref{bailey}.

\section{Application of the automated process on d001 and d002 VVV tiles: The V$^{4}$ catalog.}\label{results}
\subsection{Irregular variable sources}
We had identified 72 variable sources that fulfill both the $\rm \Delta K_s$ and $\eta$ criteria. If a source presents periodicity, it is removed from the sample and add to the periodic sample. This was the case of two sources (d001-79, d002-103) which present a large amplitude and period, typical signatures of dust-enshrouded AGB stars, invisible in the optical range, due to its thick circumstellar envelope, a product of its high mass-loss rate.

\begin{figure}[htbp]
\begin{center}
      \noindent\makebox[\textwidth]{\includegraphics[height=6.5cm,width=20cm]{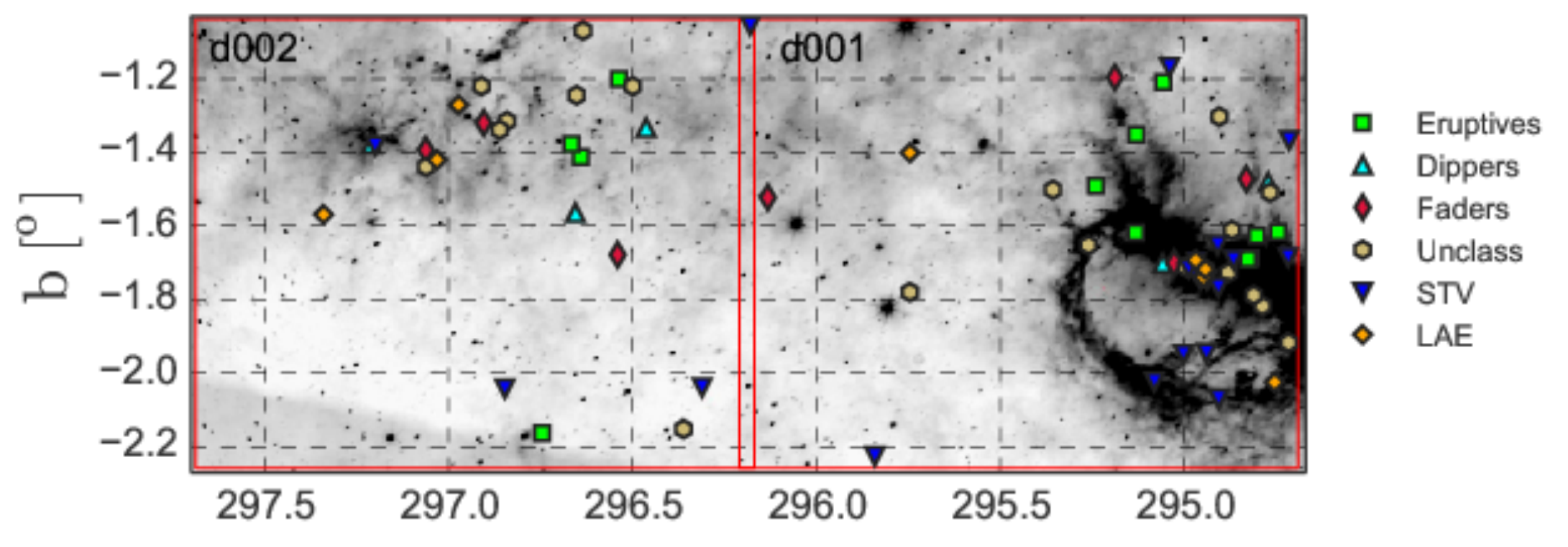}}      
      \noindent\makebox[\textwidth]{\includegraphics[height=7cm,width=20cm]{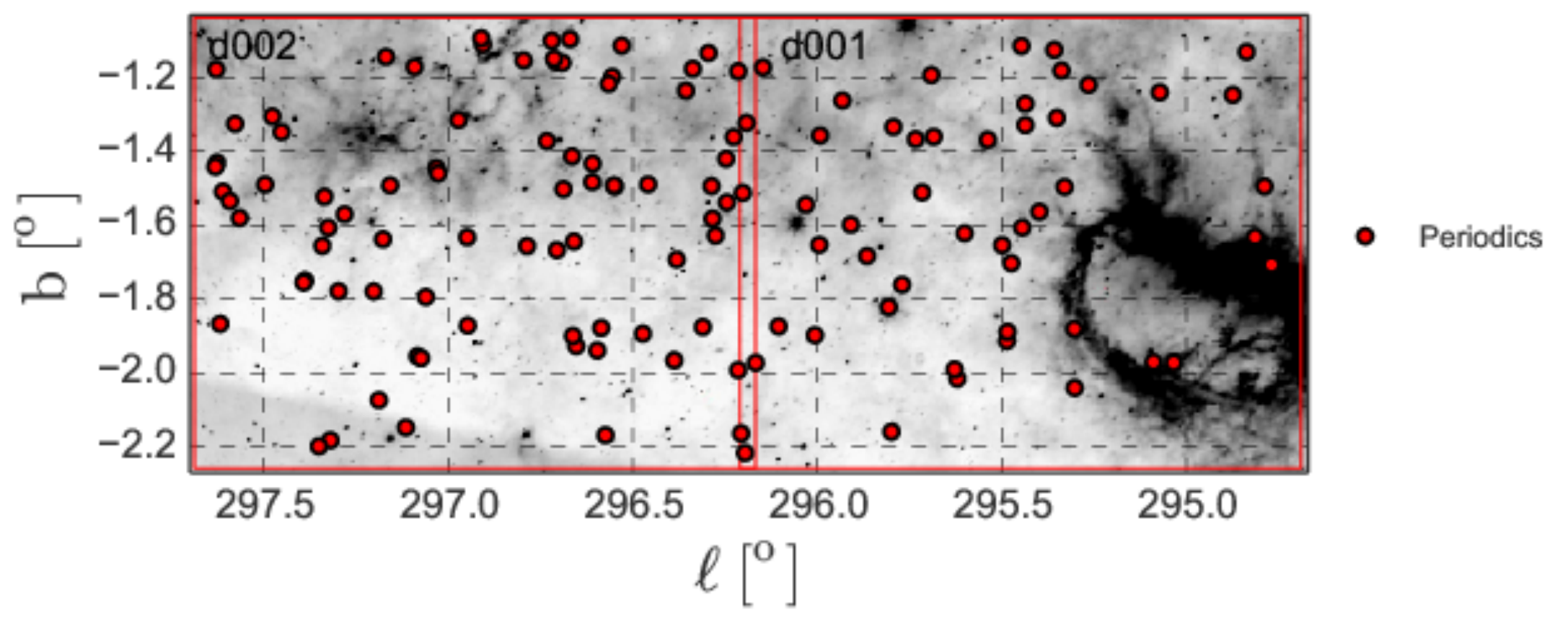}}
      \caption{Covered region by d001 and d002 tiles from VVV survey, showing the spatial distribution of identified sources. Galactic north is up, galactic east is to the left. The symbols represent different types of variable stars found in the FoV, and are explained in Section~\ref{results}. In background, W3 ($\rm \lambda_{eff}=12\mu m$) band WISE image is shown to illustrate the cold gas/dust distribution in the FoV.}
      \label{FoV}
\end{center}
\end{figure}

\begin{figure}[!tbp]
  \centering
  \noindent\makebox[\textwidth]{
  \includegraphics[width=9cm,height=15cm]{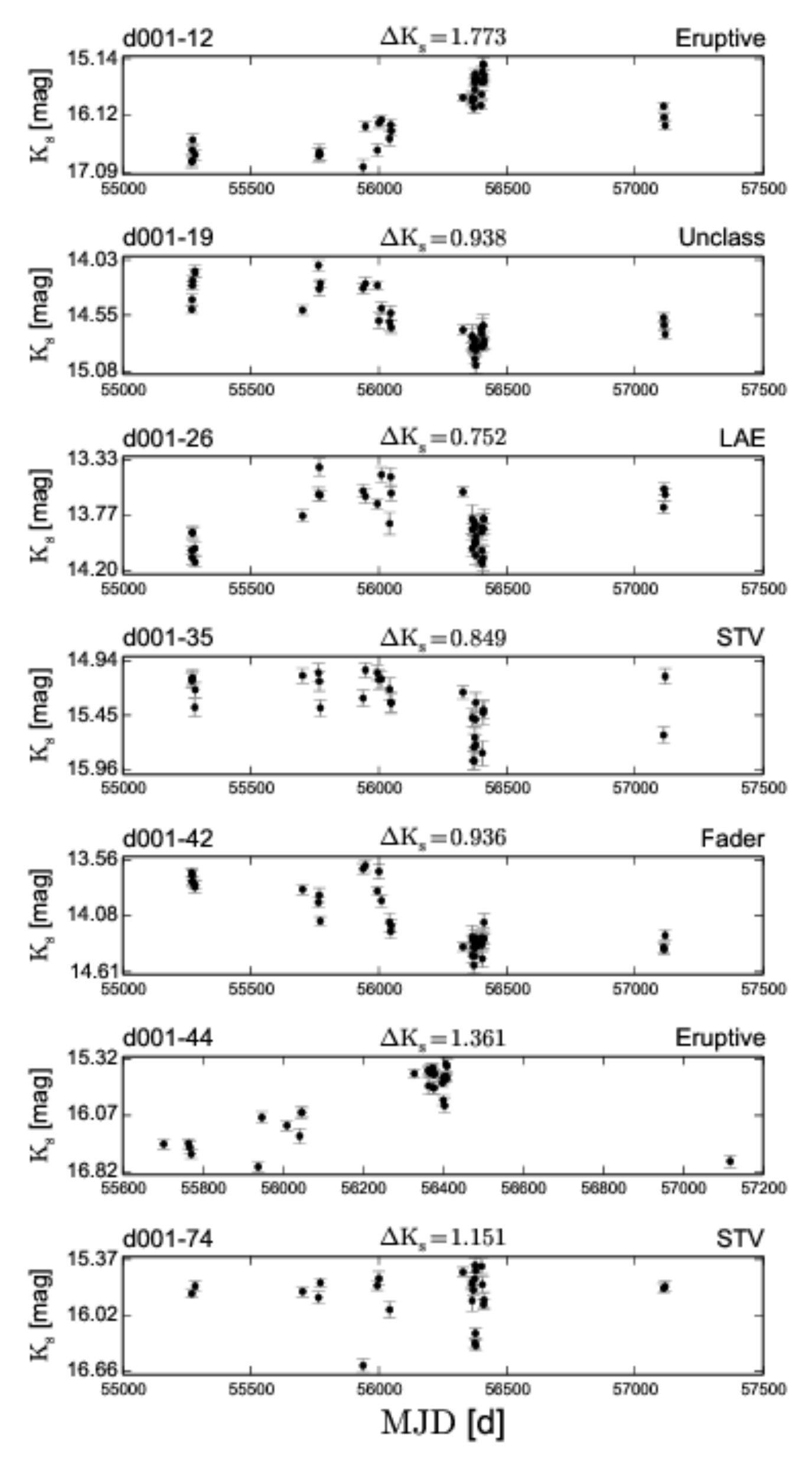}
  \includegraphics[width=9cm,height=15cm]{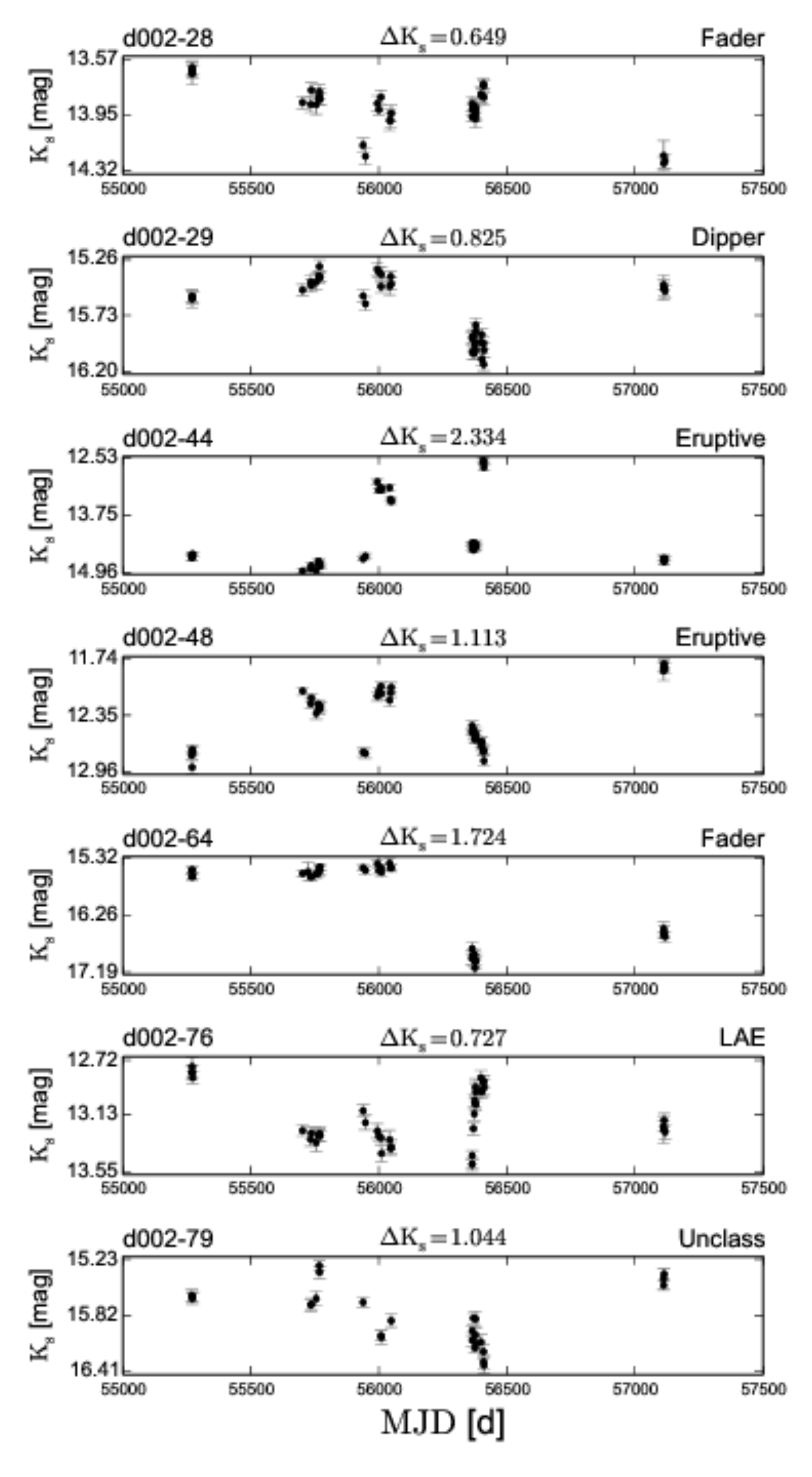}}
  \caption{Examples of $\rm K_{s}$-band time series of irregular variables. In the top of each plot the identification ID, mean magnitude $\rm \overline{K}_{s}$, amplitude $\rm \Delta K_{s}$ and variable type are shown.}
  \label{Plot_irregulares}
\end{figure}

Finally, we identified 70 irregular variable sources, 45 of them belonging to d001 and 25 to d002. Almost two thirds of them (64.28\%) are projected in d001 tile and follow the cold gas/dust distribution as traced by the W3 ($\rm \lambda_{eff}=12\mu m$) band WISE image (the background of Figure~\ref{FoV}). The highest over-density is observed at the borders of the star forming region, where the nebulosity is overwhelming, thus suggesting active star formation. All objects in the sample are, to the best of our knowledge, reported here for the first time. Some of these sources are shown in the Figure~\ref{Plot_irregulares}. 

The sample can be separated in two groups:
  
\begin{itemize}
\item Short-term irregular sources:
As discussed in~\ref{Cadence}, the cadence of VVV can reveal objects with large changes in their magnitudes in short periods of time, often associated with YSOs or PMS stars. Examples of such time series are shown in Figure~\ref{Plot_irregulares}. As noted previously, intrinsic changes in the sources can be explained by variable accretion (see e.g.~\citealt{Meyer1997,Rebull2014,Cody2014}) referred to as bursts if there is a brief, well defined event and then a return to quiescence. Variable extinction is also possible. The individual characterization of the objects is beyond the scope of this paper, and they will be analyzed in an up-coming work, once the follow-up spectroscopic analysis is completed.  

\item Long-term irregular variables:
This kind of sources have a slow change of their magnitudes over the time series, reaching large amplitudes $\rm \Delta K_{s}$ in longer time intervals. In general, the sources do not exhibit large amplitude changes in short time intervals, but increasing/decreasing their luminosities monotonically, and then in certain cases returning to their mean magnitude. Possible mechanisms here are the eruptive episodes or long-term extinction events, followed by quiescent periods. 
These time series can also reveal stellar sources as supernovae, microlensing events~\citep{Minniti2015}, LPVs, aperiodic long-term variability objects, and even extragalactic variable sources such quasars. Examples of long-term irregular time series can be seen on the right panels of figure~\ref{Plot_irregulares}.
\end{itemize}

\begin{table}
\centering
\caption{Characterization of the irregular variables in the categories proposed by ~\cite{Contreras_Pena2017}}
\label{caracterizacion}
\begin{tabular}{cc}
\hline
Class     & Description   \\ \hline \hline
Dippers   &   Shows fading events, to then return to their normal magnitude. \\ \hline
Eruptives & \begin{tabular}[c]{@{}c@{}}Shows sources with outbursts with amplitude $>1$ mag \\ and  duration longer than a few days and typically at least a year.\end{tabular}  \\ \hline

\begin{tabular}[c]{@{}c@{}}Low Amplitude \\ Eruptive \end{tabular} & \begin{tabular}[c]{@{}c@{}}Sources that present outbursts with amplitude lower than 1 mag \\ and duration typically longer than a year\end{tabular}.\\ \hline

LPV-YSOs  & \begin{tabular}[c]{@{}c@{}}Sources with a measured period, but with short-timescale \\ scatter in the time series.\end{tabular}  \\ \hline

\begin{tabular}[c]{@{}c@{}}Short Timescale \\ Variables \end{tabular} & \begin{tabular}[c]{@{}c@{}}Sources with fast and constant scatter in their time series. They also \\ can show brief rises in the magnitude in time scales of weeks.\end{tabular}     \\ \hline

Faders    & \begin{tabular}[c]{@{}c@{}}Shows a continuous decrease in brightness (t \textgreater1 yrs), or a big decrease \\ in its brightness in a source with relatively constant luminosity.\end{tabular} \\ \hline
\end{tabular}
\end{table}

In order to analyze the morphological behavior of VVV Irregular variable sources,~\cite{Contreras_Pena2017}, influenced by previous works such as ~\cite{Findeisen2013}, proposed the following classifications for their sample of high amplitude variables ($\rm \Delta K_s>1$ mag): Faders, Dippers, Short Time-scale Variables (STV) and Eruptives. However, half of our irregular variables sample has an amplitude $\rm \Delta K_{s}< 1$ mag (right side of figure~\ref{Amplitude_distribution}). Similar variable sources with low amplitude have been reported in literature (see for example ~\cite{Wolk2013} and ~\cite{Carpenter2001}) with light curves similar to those reported in this work. Therefore, given that classification for irregular variables is defined mainly by the shape of the time series, we extended the classification proposed in~\cite{Contreras_Pena2017} for low amplitude sources ($\rm \Delta K_{s}< 1$ mag), making an exception with the Eruptive classification, which describes sources with eruptions on timescales of hours to days. Sources that present eruptive long-timescales variability ($t > 1$ year) and $\rm \Delta K_{s}< 1$ mag, will be classified as 'Low Amplitude Eruptive' (LAE). We used this classification to characterize our sources, when applicable. In Table~\ref{caracterizacion}, the main characteristics of the different proposed classes are explained.  
\subsection{Periodic variables stars}\label{periodicity}
Our automated tool detected 22 periodic variable sources by the generalised Lomb-Scargle periodogram (GLS) analysis, both in d001 and d002 tiles, with range of periods between $4.03<P<1400$ days. Given the cadence of our data, and the limitations of the GLS, we can not obtain reliable periods shorter than 2 days. The IP metric, on the other hand, shows a great performance identifying periodic sources over the entire range of periods. Several stars have been detected by both methods, showing practically identical periods within the uncertainties. Thus, taking into account that the IP metric is computationally expensive, we used this method to search for shorter periods creating a grid of periods between 0.05 and 5 days. The IP metric identified 108 sources with $0.2<P<3.1$ days. 

\begin{figure}[!tbp]
  \centering
  \noindent\makebox[\textwidth]{
  \includegraphics[height=8cm,width=13cm]{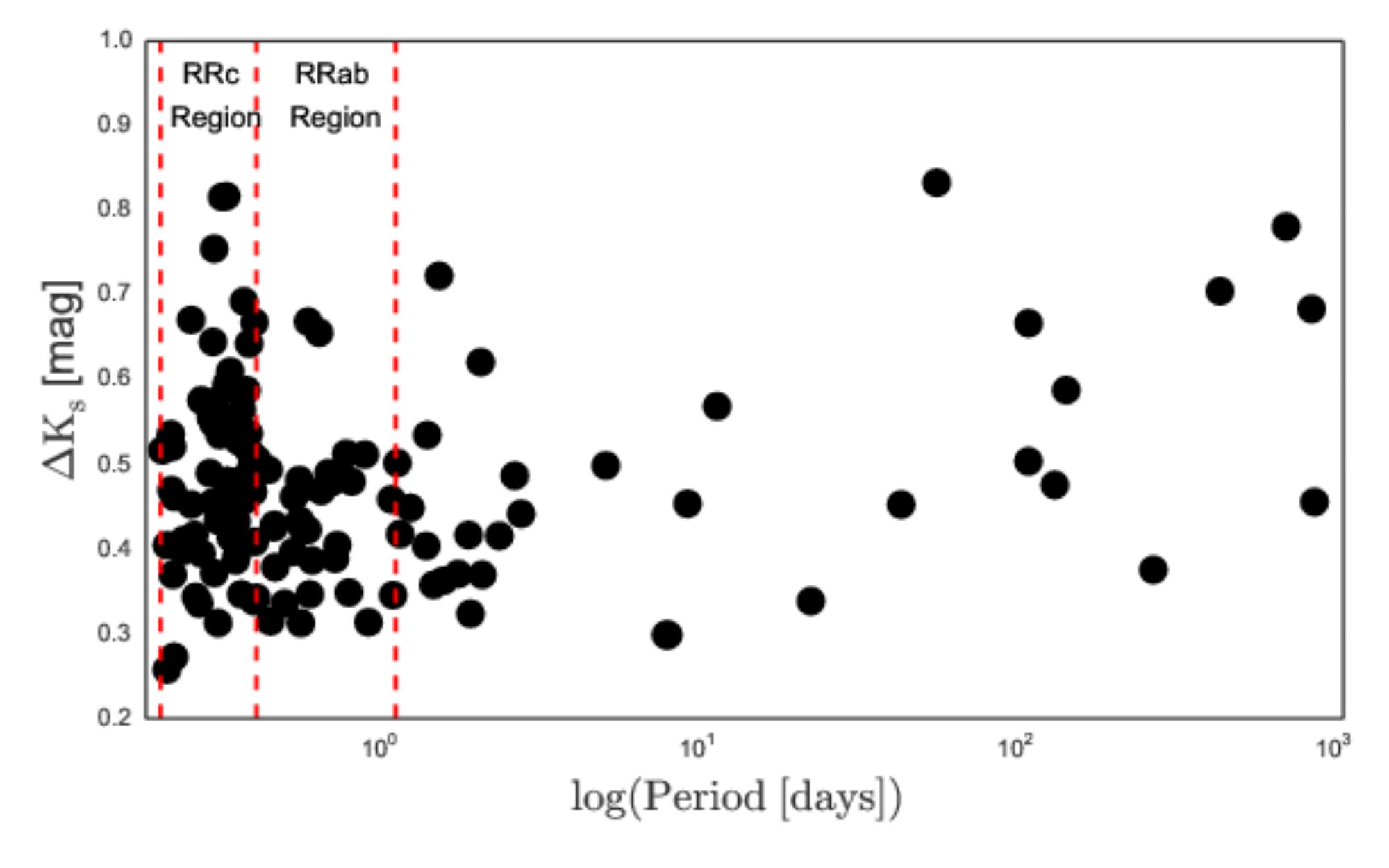}}
  \caption{Period - Amplitude diagram of periodic variables in our catalog.}
  \label{bailey}
\end{figure}

Figure~\ref{Periodics_phase} shows examples of the periodic variables, while in Figure~\ref{bailey} we show the Period-Amplitude diagram. As can be seen, most of the stars have short periods, typical for RR Lyrae stars.  

In some cases it was not possible to distinguish between different variability types. For example LPVs and YSOs (Figure~\ref{Periodics_phase}), thus we put mixed classification LPV-YSO, working on additional (spectroscopic) data to clarify. 

\subsection{General properties of selected sources}

\begin{figure}[htbp]
\begin{center}
      \noindent\makebox[\textwidth]{\includegraphics[height=6cm,width=16cm]{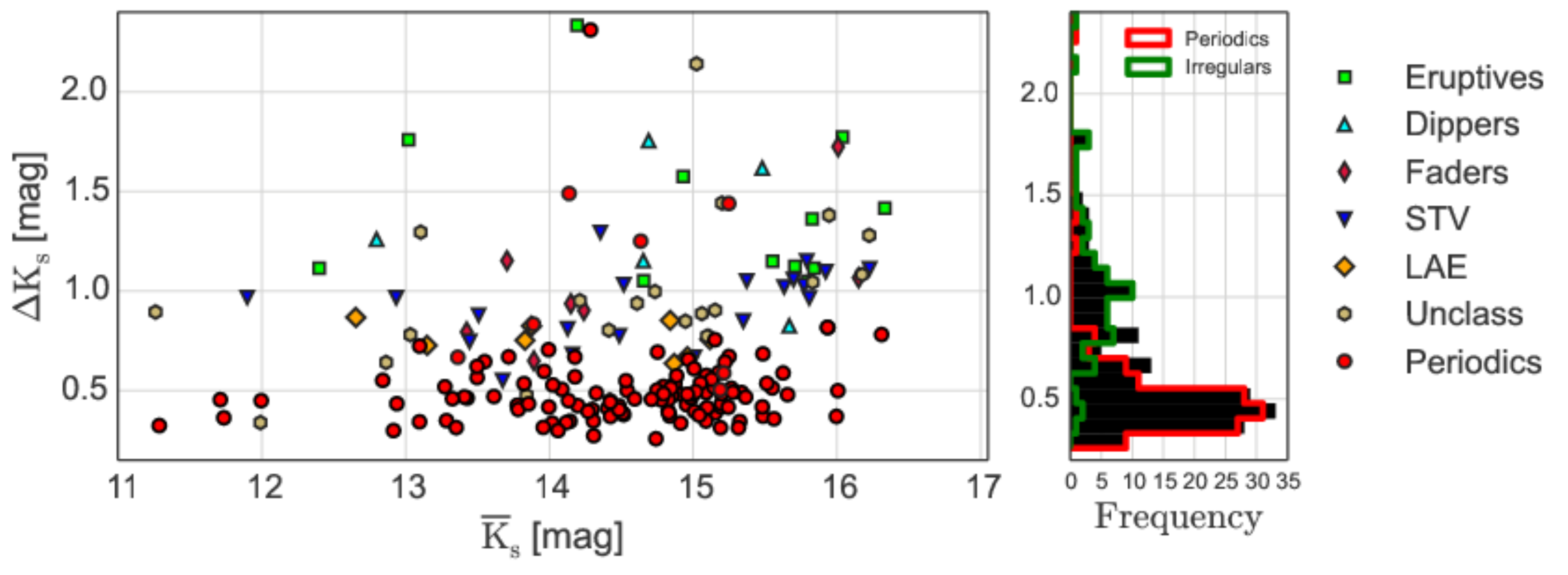}}
      \caption{Left: The $\rm \Delta K_{s}$ distribution of the selected sources vs. $\rm \overline{K}_{s}$. Right: The histogram of $\rm \Delta K_{s}$ distribution. In this figure, $\rm \Delta K_{s}$ is truncated at 2.3 magnitudes, given that the source d001-79 has much larger amplitude than other sources ($\rm \Delta K_{s}=3.2$ mag). The symbols are the same as in the Figure~\ref{FoV}.}
      \label{Amplitude_distribution}
\end{center}
\end{figure}

Figure~\ref{Amplitude_distribution} shows the $\rm \Delta K_{s}$ distribution of the selected sources as a function of the mean $\rm \overline{K}_{s}$ magnitude. The amplitude interval is between $\rm 0.5<\Delta K_{s}<3.2$ mag an average value of $\rm \left<\Delta K_{s}\right>=1.1$ mag. The histogram is influenced mainly by the periodic sources. 

\begin{figure}[htbp]
  \centering
  \noindent\makebox[\textwidth]{
  \includegraphics[height=8.5cm,width=8.5cm]{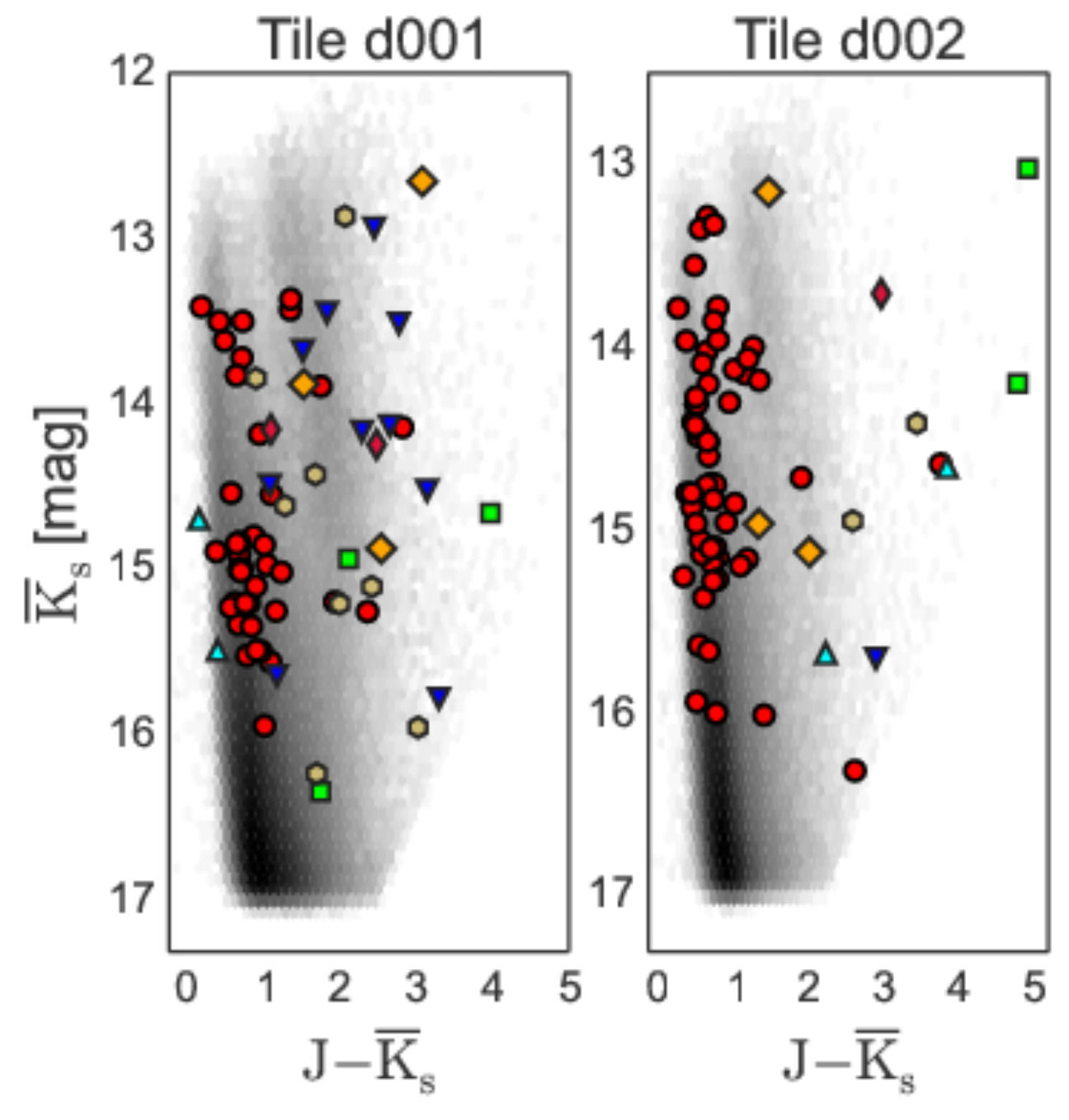}
  \includegraphics[height=8.5cm,width=10cm]{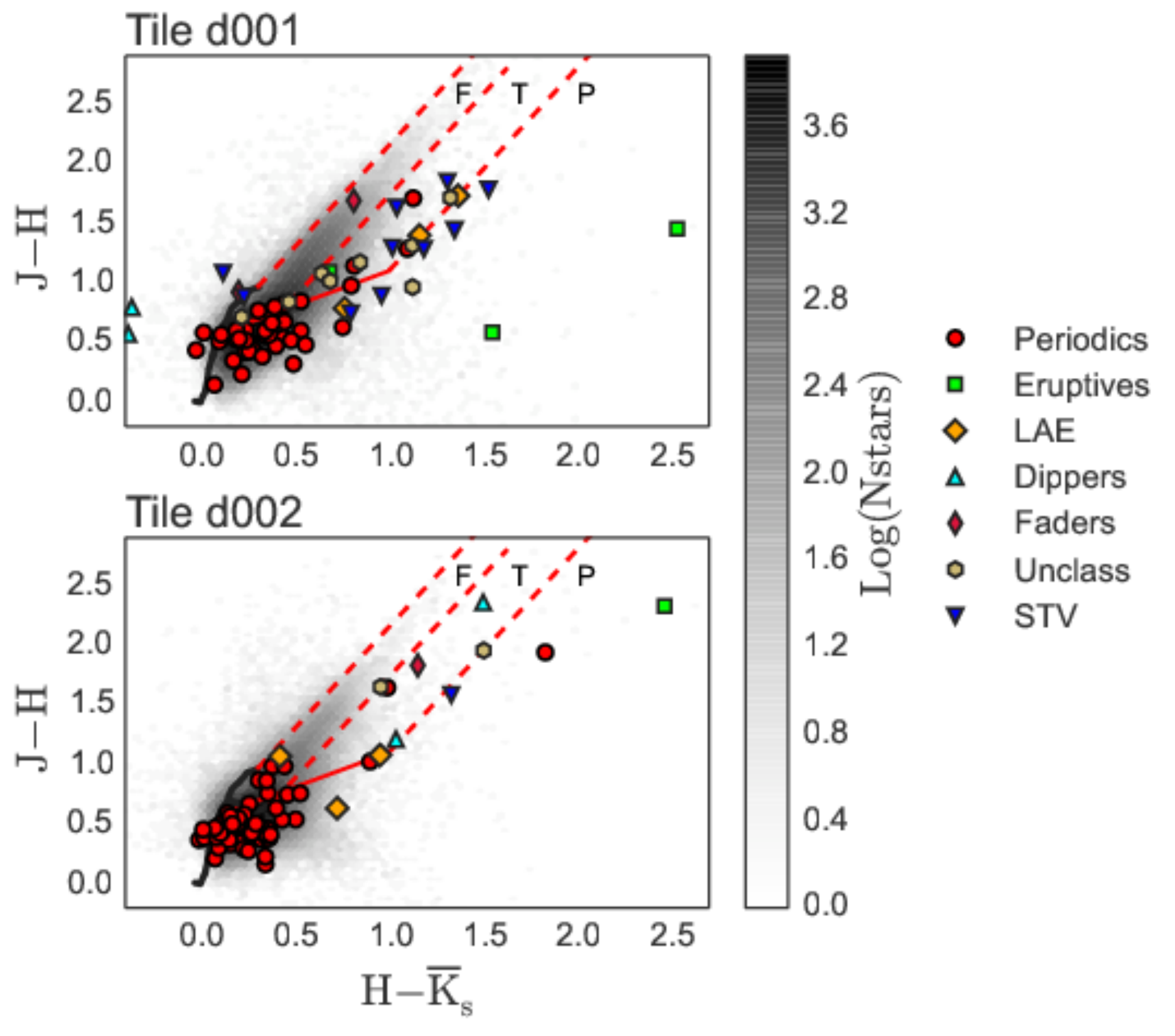}
  }
  \caption{The color-magnitude and color-color diagrams of all variable stars in d001 and d002. The solid black lines are the intrinsic colors of dwarf and giants stars from~\cite{Bessell1988}, the solid red line is the locus of un-reddened CTTs ~\citep{Meyer1997}, and the dashed red lines are the reddening vectors of the early spectral type dwarfs, giants stars and un-reddend CTTs, assuming a visual extinction $\rm A_{V}=15$ magnitudes. The symbols are the same as in the Figure~\ref{FoV}.}
	\label{ccd}
\end{figure}

Figure~\ref{ccd} shows the position of the variable stars in the color-magnitude and color-color diagrams. To plot the non-variable stars we used the first epoch of $\rm J$ and $\rm H$ band tile images taken in 2010. To analyze this, we have followed the method, described in~\cite{Ojha2004}. Three regions are defined: The 'F' region, which is located between the reddening vectors of giant and dwarf stars. The 'T' region is between the reddening vector of giant stars and CTTs locus, where  the Class II YSOs objects and Herbig Ae/Be stars~\citep{Hillenbrand1992} can be identified. In the so-called 'P' region, located below the reddening vector of CTTS the likely proto-stellar objects are situated. Thus, the corresponding variability types are assigned. Column 32 of Table~\ref{v4_catalog} contains the information of region in color-color diagram for each source.

\subsection{The completeness and accuracy of the catalog}
\begin{figure}[!tbp]
  \centering
  \noindent\makebox[\textwidth]{
  \includegraphics[height=8cm,width=8cm]{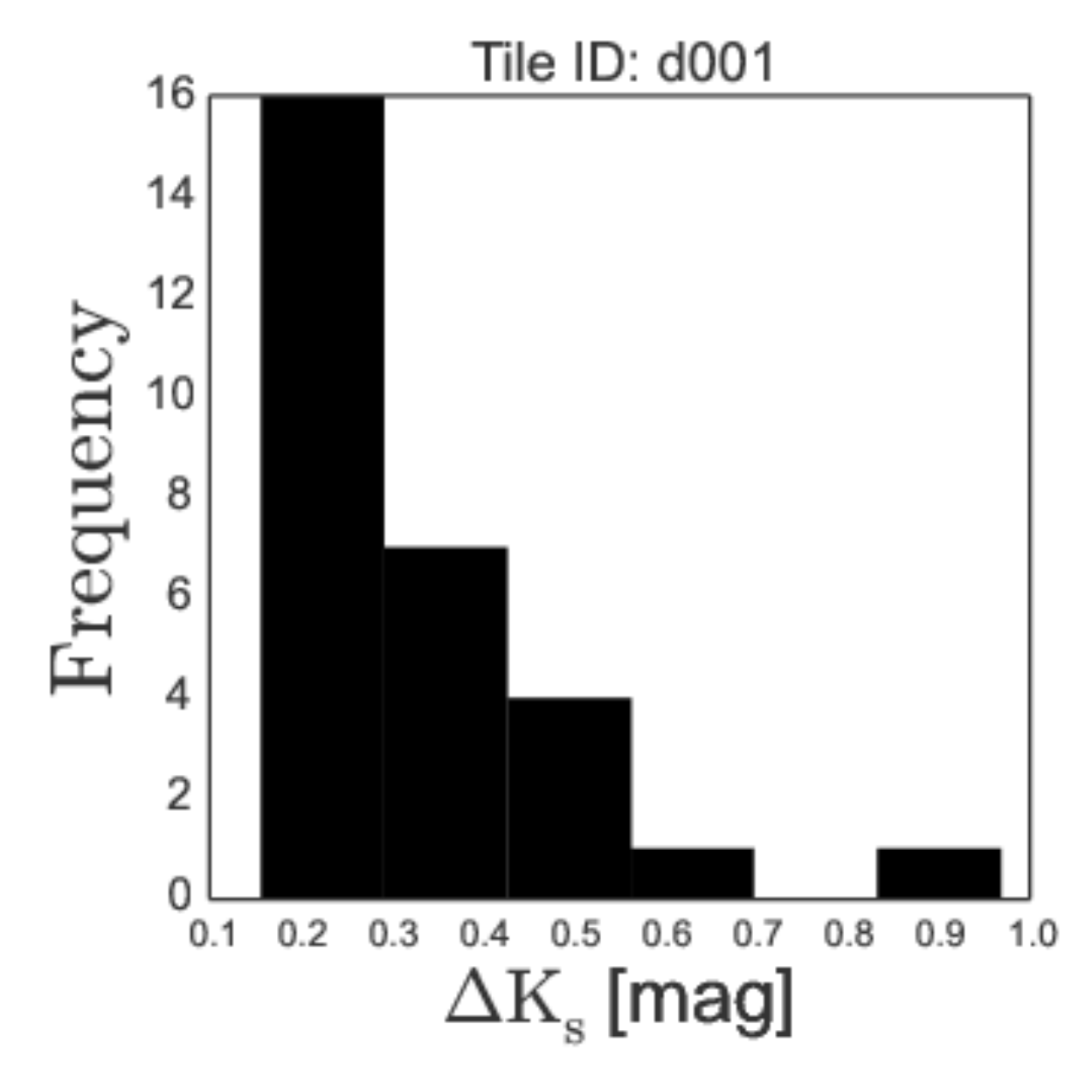}
  \includegraphics[height=8cm,width=8cm]{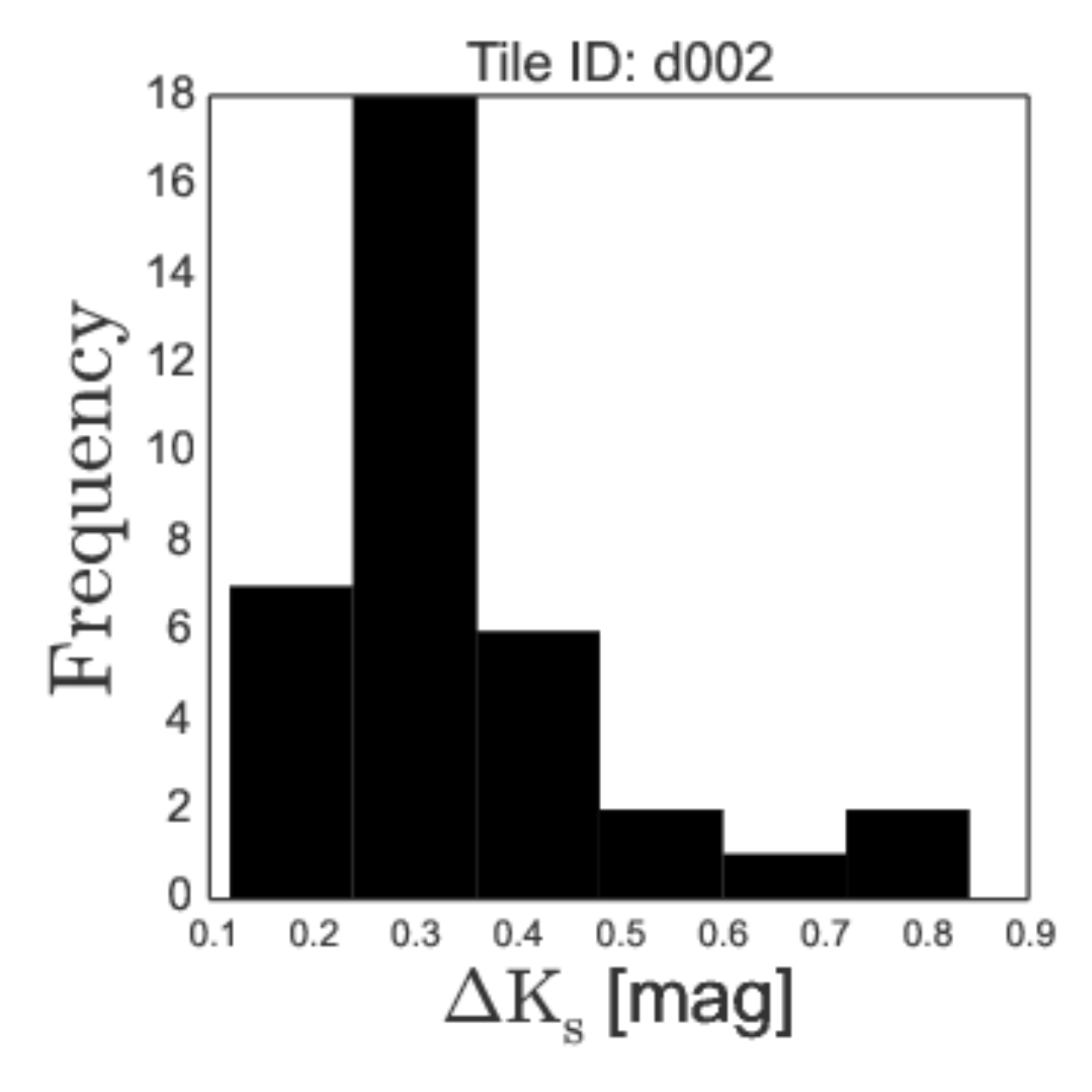}}
  \caption{\textbf{Histograms of the amplitudes} of know variable stars from  ``International Variable Star Index Catalog'' for d001 and d002. }
    \label{vsx_amp}
\end{figure}

The catalog is limited to sources brighter than $\rm K_s = 11$ mag, given the saturation of limit of VVV. We are only sensitive to the stars fainter than this magnitude limit. In the International Variable Star Index Catalog\footnote{https://www.aavso.org/vsx/index. php?view=search.top} (VSX) are listed 57 variables in d001 and 46 in d002 fainter than $\rm K_s=11$ mag. From these, 30 sources in d001 and 36 in d002 are identified in the photometry catalog, but all of them have a typical $\rm \Delta K_{s}$ around 0.2-0.3 mag, which is close to our conservative amplitude limit described in~\ref{Ks_lightcurve}. This range of amplitudes is relatively low in comparison of the average amplitude $\rm \left<\Delta K_{s}\right>=1.1$ mag of the selected variable sources. (see figure~\ref{vsx_amp}). Taking into account that these variables are detected in the optical wavelengths, the amplitudes are too low to be detected in  $\rm K_s$-band with our searching method. Four sources from the VSX are recovered in our catalog, namely d001-20, d001-81, d002-8 and d002-60, the rest of the stars with amplitudes greater than 0.3 have least than 25 measurements. Thus, in the magnitude interval $\rm 11 <K_{s}< 15.5$ mag we recovered only 6\% of the known optical variable stars, and more than 90\% of our discoveries are new. 
In the context of previous $\rm K_{s}$-band studies,~\cite{Minniti2017} and Eyheramendy et al. (private communication) reported 1 and 13 RRab stars in d001 and d002, respectively. The ~\cite{Minniti2017} RRLyrae star 'd002-0143595' has been re-discovered in our catalog as source d002-20, with practically the same period $P=0.456794$ (with a discrepancy lower than 0.01\%). Twelve of 13 RR Lyrae stars from Eyheramendy et al. list are re-discovered in our catalog. The only source missed by our procedure has $\rm \Delta K_{s}=0.195$ mag and is rejected from our initial conditions. Thus, the completeness of the catalog $\rm  11 <K_{s} < 15.5$ mag is very close to 90\%. We are expecting drops of the completeness for the fainter than $\rm K_{s} > 15.5$ mag objects, but it is hard to estimate due to lack of literature data. 

\begin{figure}[!tbp]
  \centering
  \noindent\makebox[\textwidth]{
  \includegraphics[height=6cm,width=18cm]{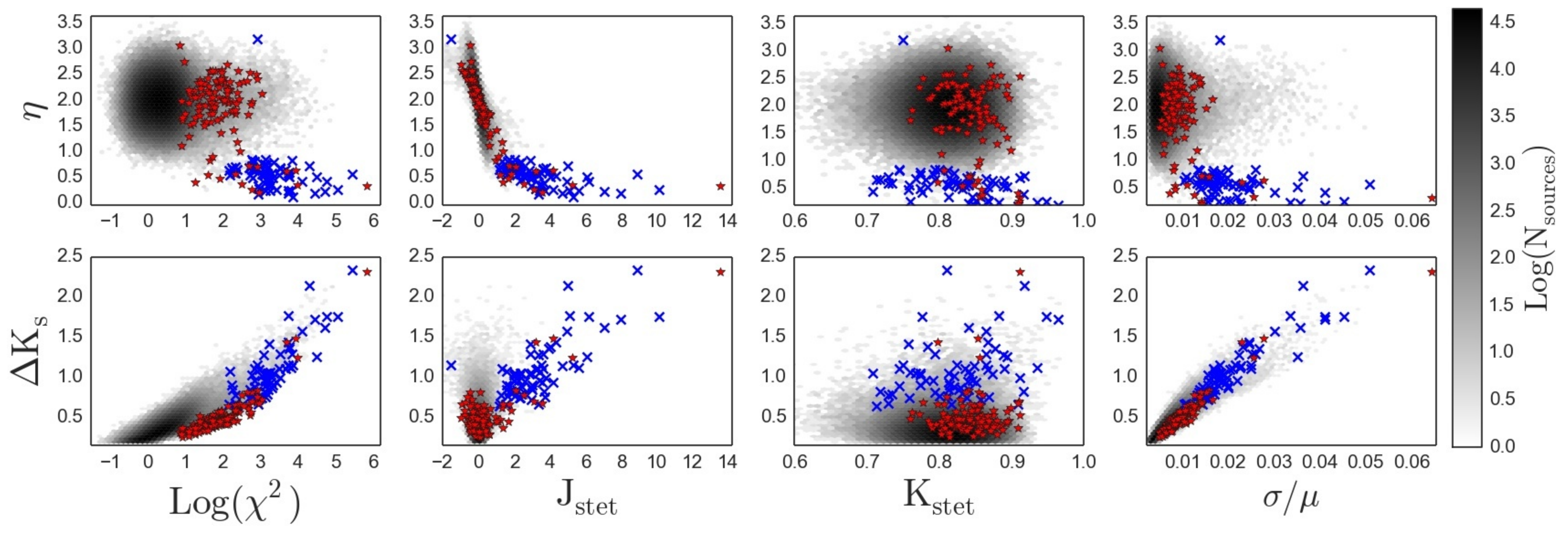}}
  \caption{Correlation plots using the main features in this analysis. The blue crosses are the irregular variables, and the red stars are the periodic sources.}
    \label{Variability_indices}
\end{figure}

Figure~\ref{Variability_indices} presents the main features $\eta$ (top row) and $\rm \Delta K_{s}$ (bottom row) as a function of different variability indices used in this article (see Table~\ref{Table_indices}). The distribution of all detected sources in tile d001 is shown in background. The red stars represent the periodic sources, and the blue crosses are the irregular ones. As can be seen from the figure, the irregular variable sources are well separated from the distribution and follow different trends in the plots. The source d001-75 is an exceptional case, having an amplitude of $\rm \Delta K_{s}=1.151$ mag,  $\eta = 3.193$, and a low number of observations (26 epochs). Excluding d001-75, the irregular sources fulfill $\rm J_{stet}>0.95$, which agrees with the limit used in~\cite{Rebull2014}. In the case of the periodic sources with short periods, in general they are located in the place where the main distribution is contained given the apparently uncorrelated shape of its time series, produced by the lack of observations.

\subsection{Variable stars around open clusters in the d001 and d002 tiles}

\begin{figure}[htbp]
\begin{center}
      \noindent\makebox[\textwidth]{\includegraphics[height=9cm,width=8cm]{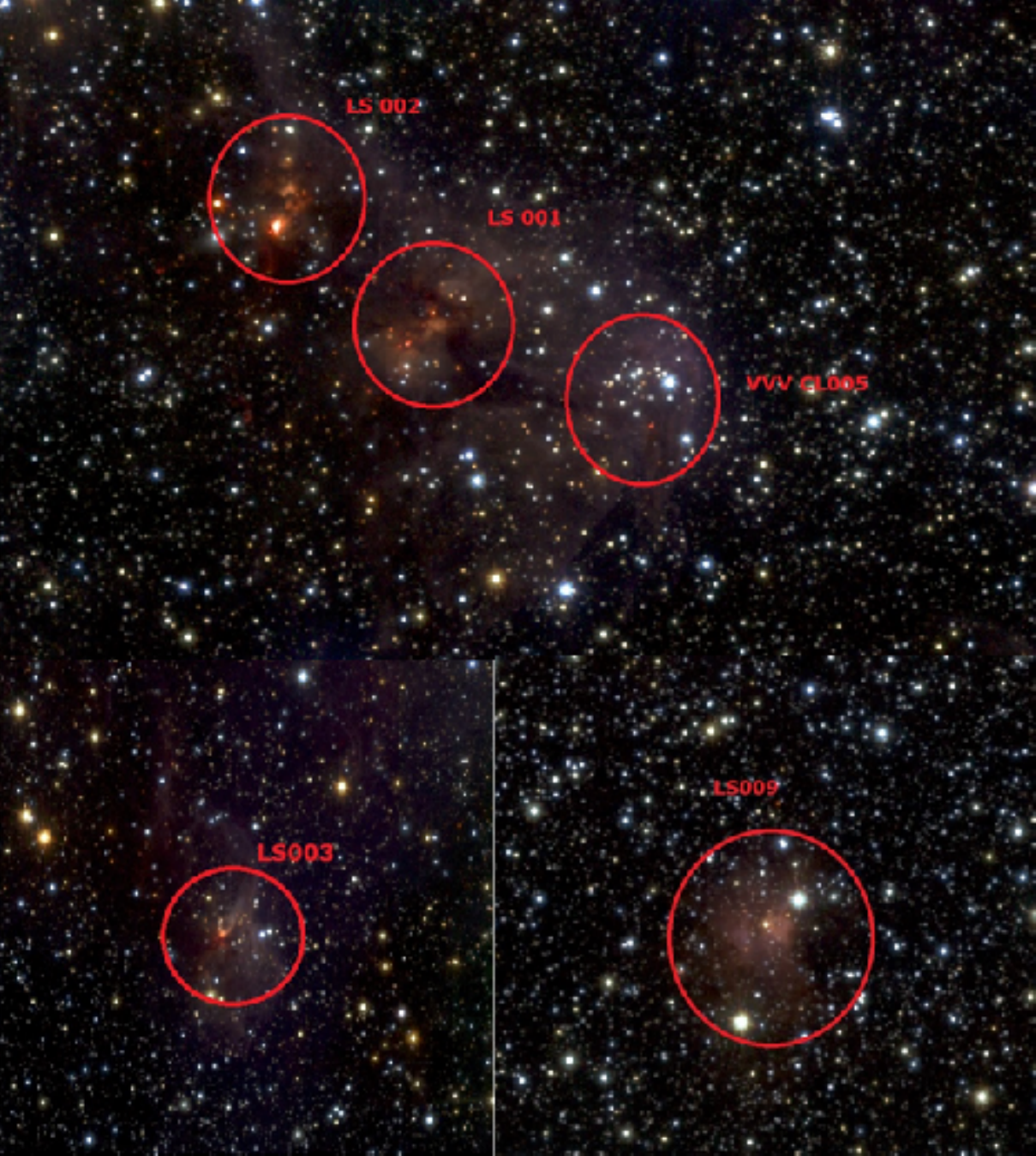}}
      \caption{The false color VVV images of star clusters projected in d001. North is up, East is to the left. The red circles represent visual diameter of the clusters as defined in the text (see below).}
	  \label{clusters}
\end{center}
\end{figure}

Nine open cluster candidates are projected in the field of view of d001 and d002, namely VVV CL005, 007 and 009~\citep{Borissova2011} and La Serena 001, 002, 003, 009 and 015~\citep{Barba2015}. The coordinates of the clusters are listed in Table~\ref{par_clusters} and the VVV color images of clusters in d001 are shown in Figure~\ref{clusters} for illustration. Little is known about their properties, except for VVV CL009, which has been investigated in~\cite{Chene2013} and ~\cite{Herve2016}. According to these papers, the VVV CL009 is young (4-6 Myr), moderately massive stellar cluster (total mass greater than 1000 $\rm M_{\odot}$), containing at least two O8-9V and an OIf/WN7 stars. The cluster distance of 5 kpc is estimated by ~\cite{Chene2013} using spectroscopic parallaxes. In the context of this paper, we search for variable YSOs around these clusters, that could be cluster members. To do this, we use the shape of the light curves, their position on the color-magnitude diagrams, the projection within the visual diameter of the clusters and proper motion diagrams taken from~\cite{Smith2018}.

\begin{figure}[htbp]
\begin{center}
      \noindent\makebox[\textwidth]{\includegraphics[height=6cm,width=15cm]{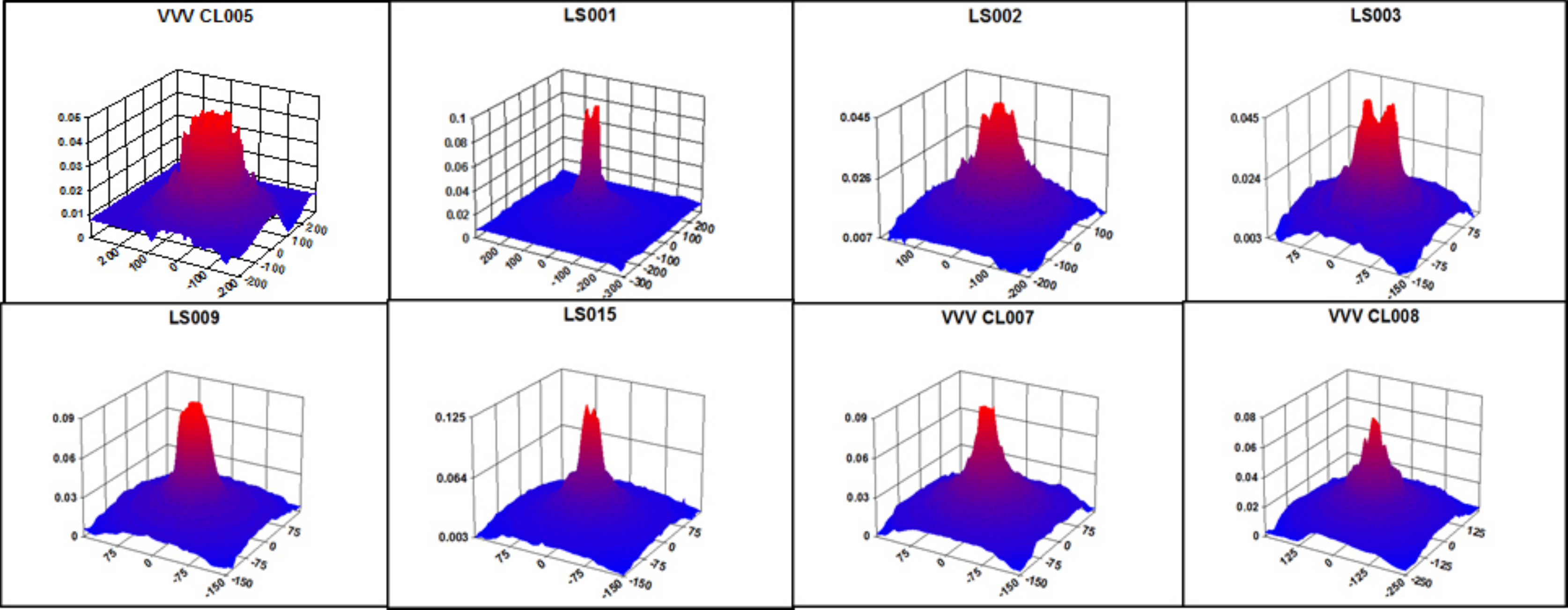}}
      \caption{Stellar surface-density maps $\rm \sigma$ (stars/$\rm arcmin^{2}$) of the clusters.}
	  \label{3D}
\end{center}
\end{figure} 

To estimate the projected cluster radius, we combined the existing VVV $\rm K_{s}$-band images (55 in case of d001 and 41 for d002 tiles) using the standard $\mathtt{IRAF}$ procedures and then, the PSF photometry with $\rm \mathtt{Daophot}$ in $\rm \mathtt{IRAF}$ was performed. The obtained magnitudes were transformed to 2MASS system using common stars. The details of these procedures can be seen in~\cite{Borissova2011,Borissova2014}. These photometric catalogs were used to construct the stellar surface-density maps, by performing direct star counting in the $\rm K_{s}$-band with a 5$''$ bin radius, assuming spherical symmetry. The maps are normalized by the area of the rings to determine the stellar density. The resulting spatial distribution maps of the stellar surface density are shown in Figure~\ref{3D}.

\begin{table}[h]
\begin{center}
\centering
\caption{Basic information of the star cluster candidates in the region.}
\label{par_clusters}
\begin{tabular}{clcccccc}
\hline
Tile ID & Name          & Ra$_{2000}$ & Dec$_{2000}$&  $\ell$  &    $b$  &  Radius       & $\rm E(J-K)$\\
        &               &  (h:m:s)      &  (d:m:s)      &  (deg)        &    (deg)    &   (arcmin)   & (mag)\\
\hline \hline  
d001    & VVVCL005      & 11:38:59    & -63:28:42   & 294.9481 & -1.7353 & 0.50$\pm$0.15 & 1.4$\pm$0.3\\ 
		& La Serena 001 & 11:39:13    & -63:29:04   & 294.9726 & -1.7292 & 0.52$\pm$0.10 & 2.8$\pm$0.6\\ 
		& La Serena 002 & 11:39:22    & -63:28:11   & 294.9896 & -1.7188 & 0.52$\pm$0.17 & 3.1$\pm$0.4\\ 
		& La Serena 003 & 11:40:28    & -63:27:58   & 295.1026 & -1.6779 & 0.48$\pm$0.09 & 3.2$\pm$0.5\\ 
		& La Serena 009 & 11:45:04    & -63:17:44   & 295.5571 & -1.3788 & 0.72$\pm$0.10 & 2.2$\pm$0.3\\ \hline

d002 & VVVCL007  & 11:53:50     & -64:20:28   & 296.7463 & -2.1629 & 0.33$\pm$0.08 & 2.2$\pm$0.2\\
           & VVVCL008   & 11:55:29    & -63:56:24   & 296.7611 & -1.7328 & 0.42$\pm$0.10 & 1.4$\pm$0.2\\ 
		   & VVVCL009   & 11:56:03    & -63:19:00   & 296.8331 & -1.1105 & 0.60$\pm$0.20 & 1.0$\pm$0.1$^{a}$ \\ 
& La Serena 015 & 11:55:23    & -63:25:30   & 296.7131 & -1.2336 & 0.33$\pm$0.07 & 1.4$\pm$0.2 \\ \hline          

\footnotesize{a: \cite{Chene2013}}
\end{tabular}
\end{center}
\end{table}

As can be seen from Figure~\ref{3D}, the over-density of the stars is clearly visible, the density peaks are at least 3 times higher that the surface density of surrounding fields, thus confirming the cluster/group nature of the candidates. The cluster boundary was determined by fitting the theoretical profile presented in~\cite{Elson1987}. The obtained visual radii of the clusters are listed in Table~\ref{par_clusters}, where the errors correspond to uncertainties from the model fit. 

The procedure employed for determining the fundamental cluster parameters such as age, reddening, and distance is described in~\cite{Borissova2011,Borissova2014} and~\cite{Chene2012,Chene2013}. Briefly, to construct the color-magnitude diagram we perform PSF photometry of $5\times5$ arcmin $\rm J$, $\rm H$, and $\rm K_s$ fields surrounding the selected candidate, using the $\rm \mathtt{Dophot}$ pipeline. Data for saturated stars (usually $\rm K_s \leq 11.5$ mag, depending from the crowding) were replaced by data from the 2\,MASS Point Source Catalog~\citep{Skrutskie2006}. Since 2MASS has a much lower angular resolution than VVV, when replacing stars we carefully examined each cluster to avoid contamination effects of crowding, using the Point Source Catalog Quality Flags available in 2MASS catalog. 

\begin{figure}[htbp]
\begin{center}
      \noindent\makebox[\textwidth]{\includegraphics[height=11.5cm,width=12cm]{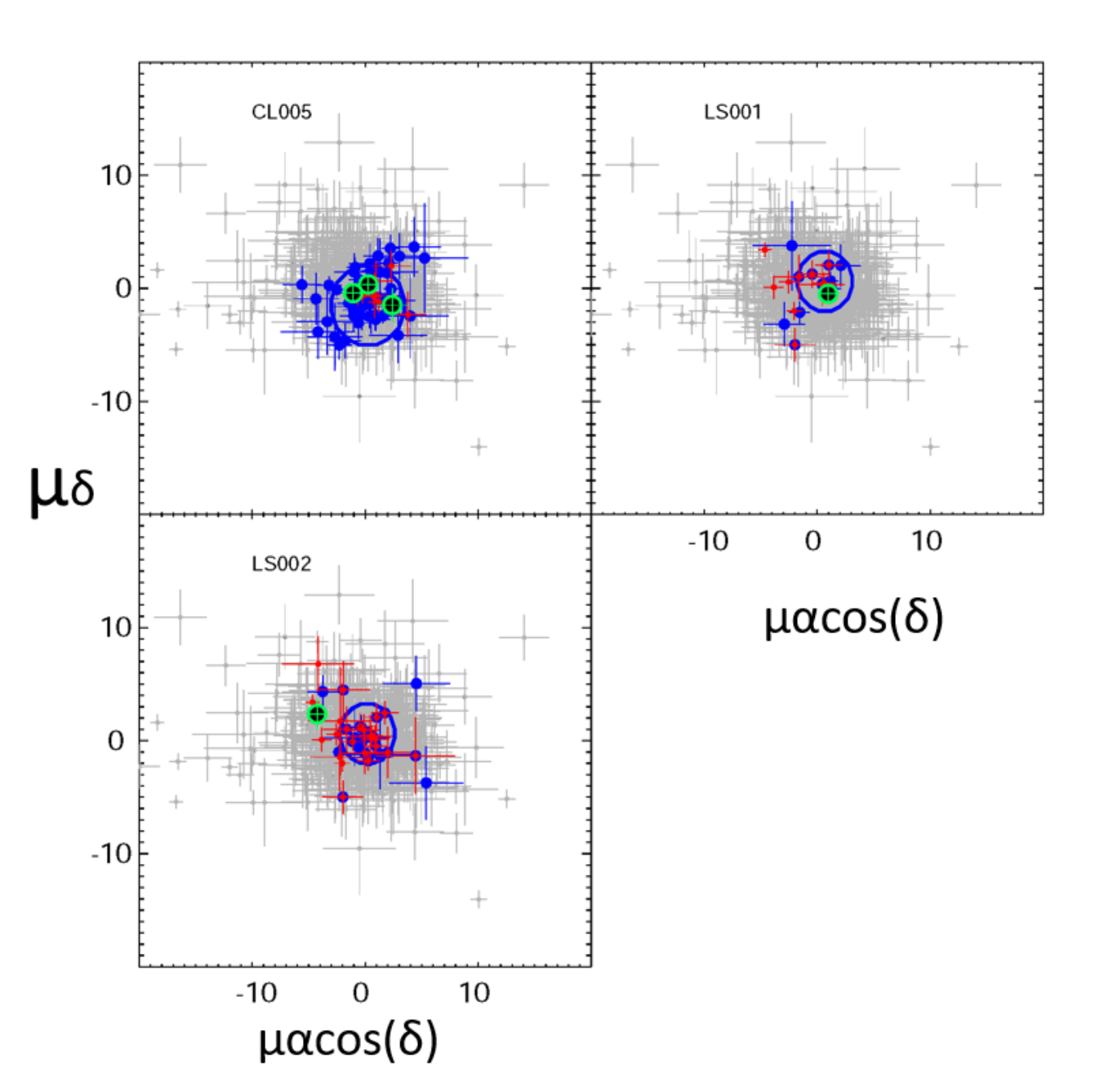}}
      \caption{The relative proper motion of the CL005, LS001 and LS002. The gray points are all stars in 5x5 arcmin area, blue ones stand for most probable cluster members, obtained after statistical photometric decontamination procedure, red point are emission candidates (see text), and green ones are the variable stars. The big blue circle marks the suggested area of cluster members.}
	  \label{pm}
\end{center}
\end{figure}

To separate the field stars from probable cluster members we used the field-star decontamination algorithm of~\cite{Bonatto2010}. The algorithm divides the $\rm K_s$, ($\rm H-K_s$) and ($\rm J-K_s$) quantities into a grid of cells. For each cell, the algorithm estimated the expected number density of member stars by subtracting the respective field-star number density. Thus, each grid setup produced a total number of member stars $\rm N_{mem}$. Repeating the above
procedure for different setups, we obtained the average
number of member stars. Each star was ranked according to the number of times it survived after all runs (survival
frequency), and only the $\rm N_{mem}$  highest ranked stars were taken as cluster members. For the present cases we obtained survival frequencies higher than 85$\%$.
To additionally clean-up the diagrams, we used the relative proper motion catalog recently constructed by \cite{Smith2018}. In general, the cluster members should form clearly visible overdensity with respect to field stars in the proper motion diagram. In our case (see Figure~\ref{pm}) it is impossible to separate the cluster members from the field stars, because the cluster members closely follow the motion of the Galactic disk. Nevertheless, they mark compact groups, slightly shifted from the disk population. To calculate the radius of the group, we started from the photometrically decontaminated candidates, calculated the mean proper motion and its error (quadratically adding to this error the mean of the individual proper motion errors), and drew the circle with 3$\sigma$ radius (blue circle in Figure~\ref{pm}). Thus, the stars with motion projected farther than 3$\sigma$ from the circle are rejected.

Taking into account that our candidates are classified as young clusters in the discovery papers of~\cite{Borissova2011} and~\cite{Barba2015}, the photometry/astrometry alone can not give accurate distance and age determinations. Usually, spectroscopic parallaxes from follow-up observations of selected members are needed. Here, only VVV CL005 has one follow-up object observed, thus it was impossible to obtain accurate basic parameters of the clusters. Instead, we use the PARSEC isochrones compilation for solar metallicity and ages of 5 and 400 Myr~\citep{Bressan2012,Marigo2017} to illustrate the position of the most probable cluster members and to estimate their mean reddening (last column in Table~\ref{par_clusters}). 

\begin{figure}[htbp]
\begin{center}
      \noindent\makebox[\textwidth]{\includegraphics[height=16cm,width=14cm]{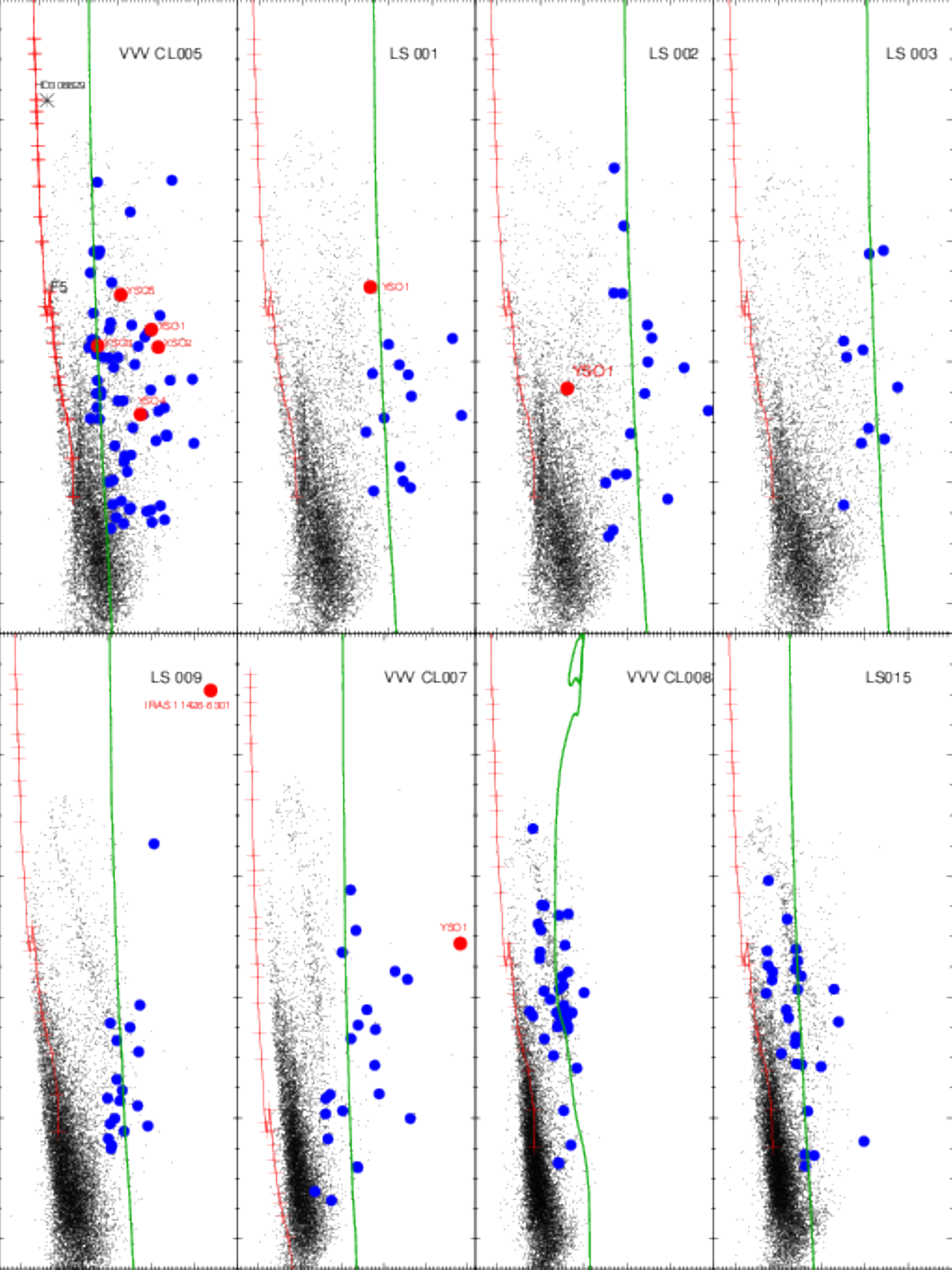}}
      \caption{The $\rm K_{\rm S}$ vs. $\rm (J-K_{\rm S})$  color-magnitude diagrams for the clusters in the field of view of d001 and d002. Black points are all stars in $5\times 5$ arcmin field around cluster centers, blue circles are the most probable cluster members, after statistical photometric decontamination. Red circles are variable stars (see text). The Geneva isochrones of 5 Myr (400 Myr for VVV CL008) and Z=0.020 are plotted with green lines. Red line represents the sequence of the zero-reddening stars of luminosity classes V~\citep{Schmidt-Kaler1982} for illustration.} \label{cmds}
\end{center}
\end{figure} 

\subsubsection{Notes of individual clusters: VVV CL005}
The cluster VVV CL005 is a young star cluster candidate, defined in~\cite{Borissova2011} as a small group of 24 stars; it is projected close to the IC 2944 H\,{\small II} region and to the IC\,2948 cluster (3.8 arcmin), on the part of the cloud [SMN83] Lam Cen 1. The brightest star within the cluster radius, namely HD308829, is classified as a Be star of B8 spectral type and is suggested to be a member of IC\,2944 cluster (Cl* IC 2944 THA 51). The star is found to be a periodic variable star with P=0.8709 days~\citep{Pojmanski1998}, but we can not follow its variability with VVV, because the star is saturated in our images ($\rm K=9.68$ mag). 
The published distance to the IC 2944 H\,{\small II} region varies from 1.8 kpc~\citep{McSwain2005} to 2 kpc~\citep{Sana2011} in the literature. The color-magnitude diagram of VVV CL005 (Figure~\ref{cmds}) shows a poorly populated main-sequence and some reddened stars, suggesting indeed a very young stellar group. With regards to the distance, it is impossible to estimate it from comparison with the theoretical isochrones, because the isochrones for ages younger than 5 Myr in this mass interval in the near infrared bands are vertical and practically identical. 
Then, the spectroscopic distance is calculated using the spectral classification of Obj1 (Ra=11:38:57.73 and  Dec=$-$63:28:22.4). The object is observed with the ARCoIRIS (Astronomy Research Cornell Infra Red Imaging Spectrograph). This is a cross-dispersed, single-object, longslit, infrared imaging spectrograph, mounted on Blanco 4-m Telescope, CTIO. The wavelength range is from  0.80 to 2.47 $\mu m$, with spectral resolving power about 3500. The comparison with different spectral templates taken from VOSA~\citep{Bayo2008} gives the most probable spectral type F4-F6V, used to estimate the spectroscopic parallax. We calculated reddening and distance modulus of $\rm E(J-K_{s})=1.4\pm0.3$ and $\rm (M-m)_{0}=10.45\pm0.43$ mag (1.23$\pm0.24$ kpc), respectively. The distance is comparable with distance estimates of 1 kpc as measured from the interstellar silicon monoxide (SiO) sources \citep{Harju1998}.
Five variables (d001-25, 27, 28, 29, 30, see Table~\ref{v4_table}) are probable cluster members, taking into account their projected position radius from the cluster center and the position in the color-magnitude and proper motion diagrams (Figure~\ref{cmds}, Figure~\ref{pm}). All of them show irregular variability, with relatively large amplitudes ($\rm 0.63 < \Delta K_{s}<0.88$ mag) and thus can be classified as YSOs. 

\subsubsection{La Serena 001}
The star cluster candidate La Serena 001 (LS001) is projected very close to VVV CL005, on the same H\,{\small II} region and contains a few very reddened stars (see Figure~\ref{cmds}). These objects are deeply embedded in dust and gas. We adopted the same distance as for CL005, and using 5 Myr isochrone determine $\rm E(J-K)=2.8\pm0.6$. Note the large uncertainty of the reddening determination, which can be result of a strong differential reddening inside the region. One variable star (d001-32) is found in the vicinity of this group in formation. It shows an irregular time series and $\rm \Delta K_{s}=0.96$ mag. Taking into account the time series and its position of the CMD (in the 'P' region), most probably it is an YSO candidate.

\subsubsection{La Serena 002}
The star cluster candidate La Serena 002 (LS002) is projected close to LS001 in the North-East direction. Several embedded and very red sources are visible, indicating stars in formation. The color-magnitude diagram is poorly populated, thus as in the case of LS\,001 we determined the mean reddening of $\rm E(J-K)=3.1\pm0.4$ adopting the same distance modulus as above. One periodic variable star (d001-38) is found, with $\rm \Delta K_{s}=0.453$ mag and period $P=42.706$ days, which we classified  as a Cepheid. Additionally, this source is far from the main locus of most probable cluster members in the color-magnitude and proper motion diagrams, thus we conclude that is a projected field star. Two more irregular variables (d001-35, d001-36, see Table~\ref{v4_table}) with amplitudes between $\rm  0.84 < \Delta K_{s}<0.91$ mag are most probably cluster members and YSO candidates.

\begin{figure}[htbp]
\begin{center}
  \hspace{1cm}
      {\includegraphics[height=3cm,width=14cm]{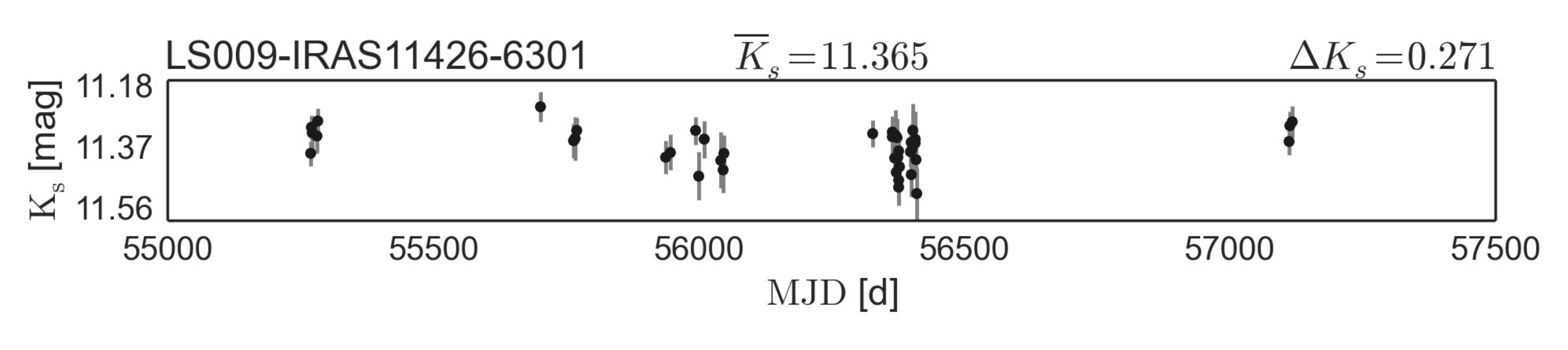}}
     ;   \hspace{-2cm}
       {\includegraphics[height=12cm,width=12cm]{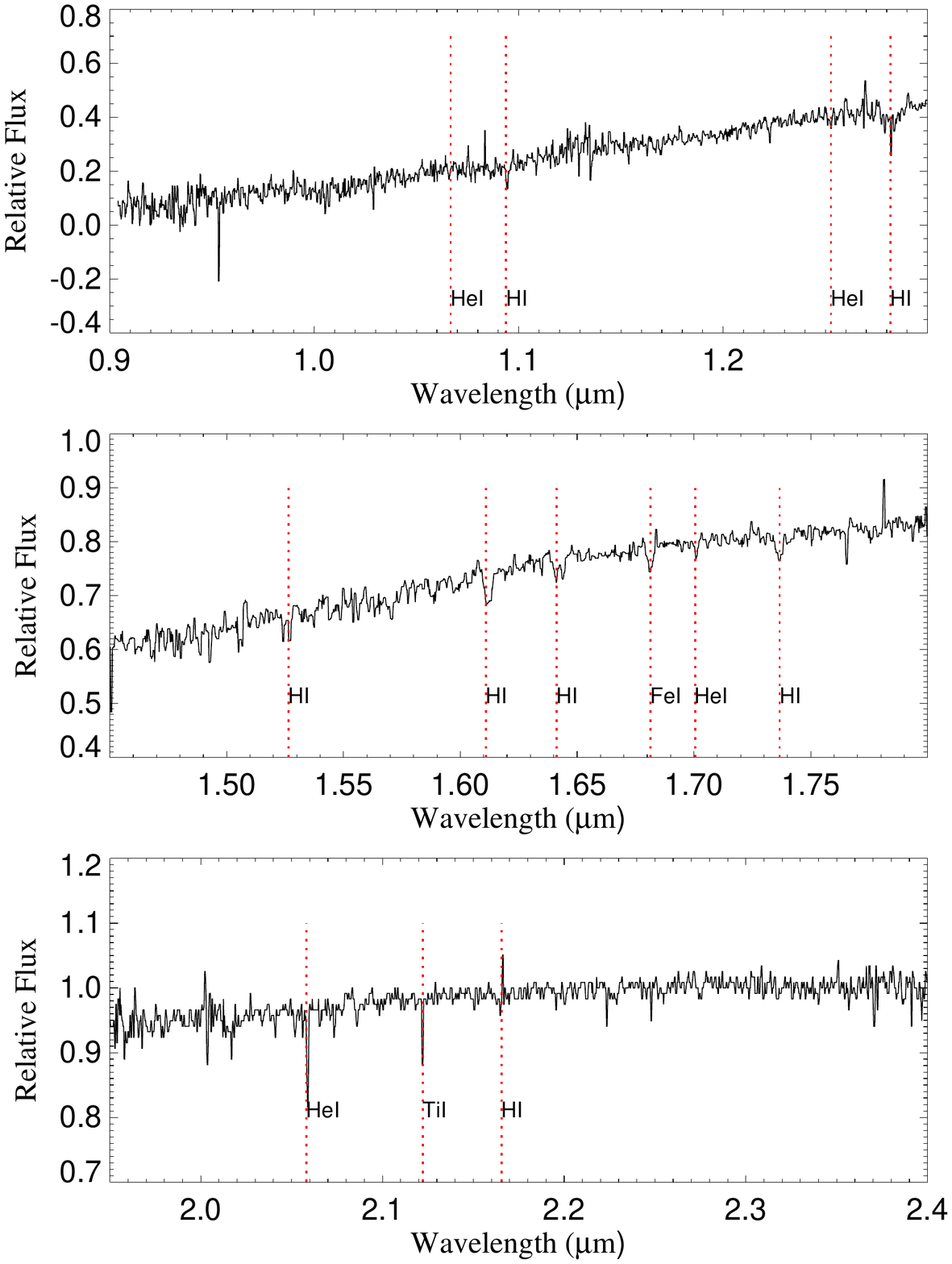}}
      \caption{The VVV time series and the ARCoIRIS spectrum of IRAS\,11426-6301 YSO candidate. }
	  \label{iras_sp}
\end{center}
\end{figure}

\subsubsection{La Serena 003 and La Serena 009}
The star cluster candidates La Serena 003 and 009 are groups of stars in formation, deeply embedded in dust and gas. No stars with variation in the magnitude above our sensitivity limit of $\rm \Delta K_s > 0.2$ mag are found in the vicinities of clusters. One YSO candidate IRAS\,11426-6301 in the field of LS\,009 is reported by ~\cite{Kwok1997}. Later on, ~\cite{Mottram2011} and~\cite{Lumsden2013} resolved the source to YSO (G295.5570-01.3787A) and  H\,{\small II} region (G295.5570-01.3787B, separated by 5 arcsec) on the base of far-infrared MSX measurements. They determined the radial velocity of 37.2 km/s and kinematic distance to both sources of 10.4 kpc. The bolometric luminosities of the YSO is calculated as 5980 $\rm L_{\odot}$, while the H\,{\small II} region has 63570 $\rm L_{\odot}$, the log Mass of the whole clump is estimated to be 3.106 $\rm M_{\odot}$ according to the same authors. ~\cite{Navarete2015} reported an extended H\,{\small II} emission associated with the region. Our $\rm K_s$-band light-curve (Figure~\ref{iras_sp}) shows low amplitude variability of 0.27 mag during the 2010-2015 time interval, with $\rm \overline{K}_s=11.365$ mag. Note, however that the 2MASS magnitude (taken from ``The 2MASS Extended sources catalog''~\citep{Skrutskie2006}) is $\rm K=8.94$ magnitude, thus the star most probably shows a long-term variability. 

IRAS 11426-6301 was observed on May 2017 with ARCoIRIS spectrograph, with 480 sec. integration time, at 1.28 average airmass. We reduced the spectrum using the $\mathtt{Spextool}$ IDL package (version 4.1,~\cite{Cushing2004}), which is a data reduction algorithm specifically designed for the data format and characteristics of ARCoIRIS by Dr. Katelyn Allers. Telluric correction and flux calibration of the post-extraction spectra are achieved through the $\mathtt{xtellcorr}$ IDL package~\citep{Vacca2003}. Figure~\ref{iras_sp} shows the spectrum, normalized at 2.293 microns. As can be seen from the figure, the continuum level is flat to slightly rising.
The overall spectra energy distribution  is peaking at around 2.5 microns. The H\,{\small I} and He\,{\small I} lines are clearly visible in absorption, some Ti lines and Na\,{\small I} doublet (2.21 microns) in absorption also can be identified. The Ca\,{\small I} triplet (2.26 microns) and $^{12}$CO bands (2.29 microns) are missing. While the H\,{\small I} and He\,{\small I} reprecent in general early type stars,  the Ti and Na features are typical of low mass YSOs. The absence of CO band means a absence of circumstellar disk. A possible explanation for this contradiction could be related to the photodisociation of CO molecule. This phenomenon (where one even detects CI but no CO, or a much lower abundance than expected) has been observed around young A-type stars in deep searches for molecular gas in debris disks~\citep{Higuchi2017}. Thus, the spectrum shows mixed features arising from both high and low mass young stars. Taking into account the kinematic distance of 10 kpc this object could be an unresolved compact cluster or group of young stars.

\subsubsection{VVV CL007}
One variable (d002-35) is detected with our algorithm very close to the center of VVV CL007 cluster. Taking into account its high amplitude $\rm \Delta K_{s}=1.76$ mag, its position on the CMD (in the 'P' region) and its classification as Eruptive variable ~\citep{Contreras_Pena2017} most probably is it a YSO.

\subsubsection{VVV CL008, VVV CL009 and La Serena 015}
No variable stars are detected around  VVV CL008, CL009 and La Serena 015. In~\cite{Chene2013} we classify the Obj 4 (Ra=11:56:03.03 and Dec=$-$63:19:00.72) of VVV CL009 as a Be star.   

\begin{figure}[htbp]
\begin{center}
      \noindent\makebox[\textwidth]{\includegraphics[height=16cm,width=13cm]{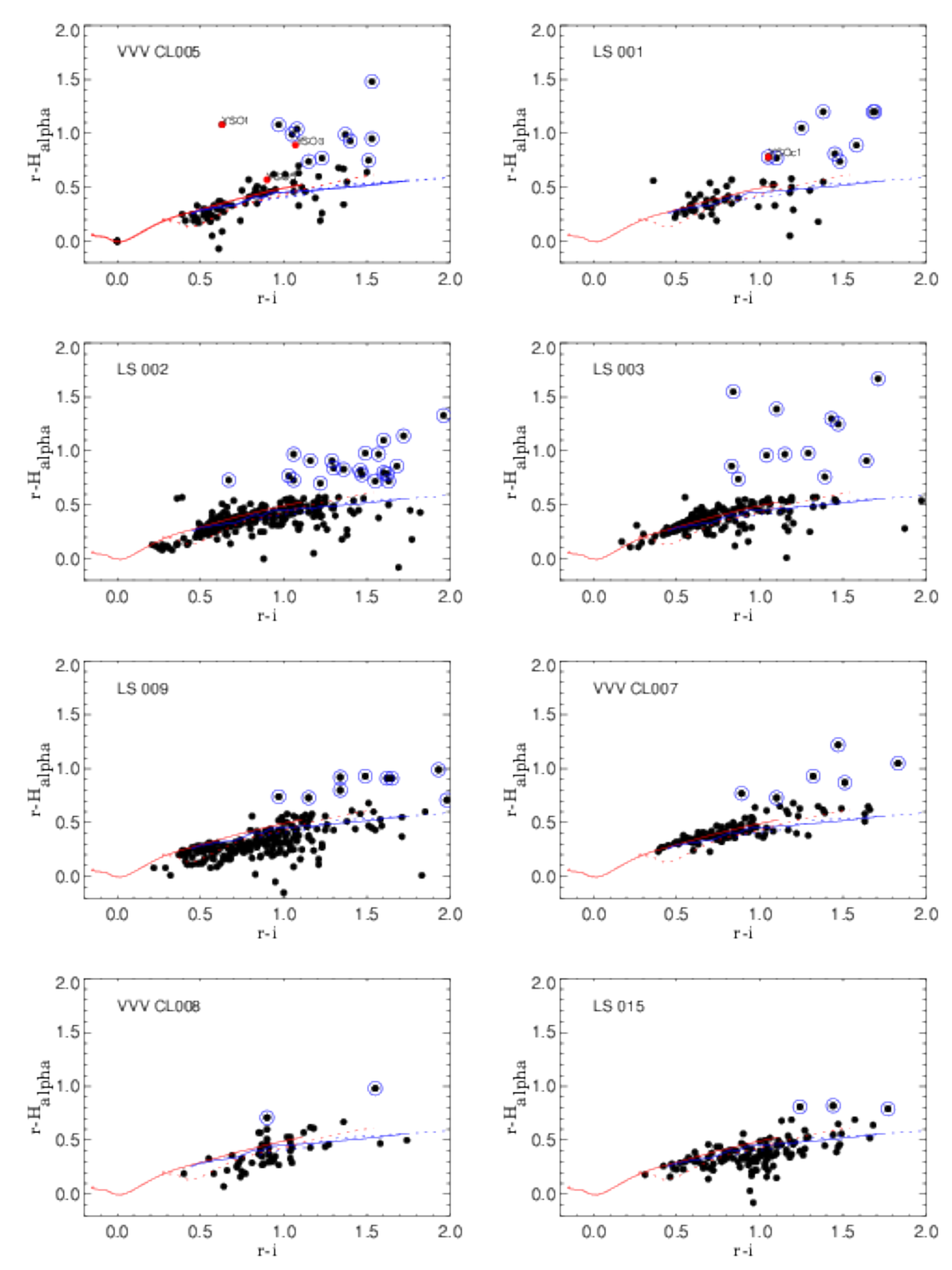}}
      \caption{The $\rm (r-H_{\alpha},r-i,)$ color-color diagrams. Blue circles show the selected emission stars, the red and blue solid lines are the synthetic unreddened main sequence and giant sequences, while the dashed ones corresponds to the corrected for the reddening sequences ~\citep{Drew2014}. Red circles in the top plots are infrared variable YSOs found in this work.}
	  \label{halphaa}
\end{center}
\end{figure} 

\begin{table}[!tbp]
\centering
\caption{VPHAS+ $r$, $(r-i)$ and $(r-\rm H_{\alpha})$ magnitudes and colors of the YSO candidates. The full version is available online.}
\begin{tabular}{cccccc}
\hline
ID        & RA$_{2000}$  & DEC$_{2000}$  &        $r$       &       $r-i$        & $r-\rm H_{\alpha}$  \\  
       & (deg)          & (deg)           &      (mag)       &       (mag)        &    (mag)  \\ \hline \hline
CL005 C1	  & 174.71433	&	-63.48757	&	17.98$\pm$0.02	&	1.08	$\pm$0.011	&	1.04	$\pm$0.01	\\
CL005 C2	  & 174.71582	&	-63.46827	&	19.39$\pm$0.04	&	0.97	$\pm$0.025	&	1.08	$\pm$0.03	\\
CL005 C3	  & 174.71664	&	-63.48768	&	20.29$\pm$0.10	&	1.53	$\pm$0.054	&	0.95	$\pm$0.09	\\
CL005 C4	  & 174.71783	&	-63.48682	&	18.37$\pm$0.02	&	1.05	$\pm$0.011	&	0.99	$\pm$0.02	\\
CL005 C5	  & 174.71942	&	-63.4668	&	19.67$\pm$0.05	&	1.37	$\pm$0.032	&	0.99	$\pm$0.04	\\
 \hline
\label{halpha}
\end{tabular}
\end{table}

\begin{figure}[htbp]
\begin{center}
  \noindent\makebox[\textwidth]{
  \includegraphics[height=7cm,width=14cm]{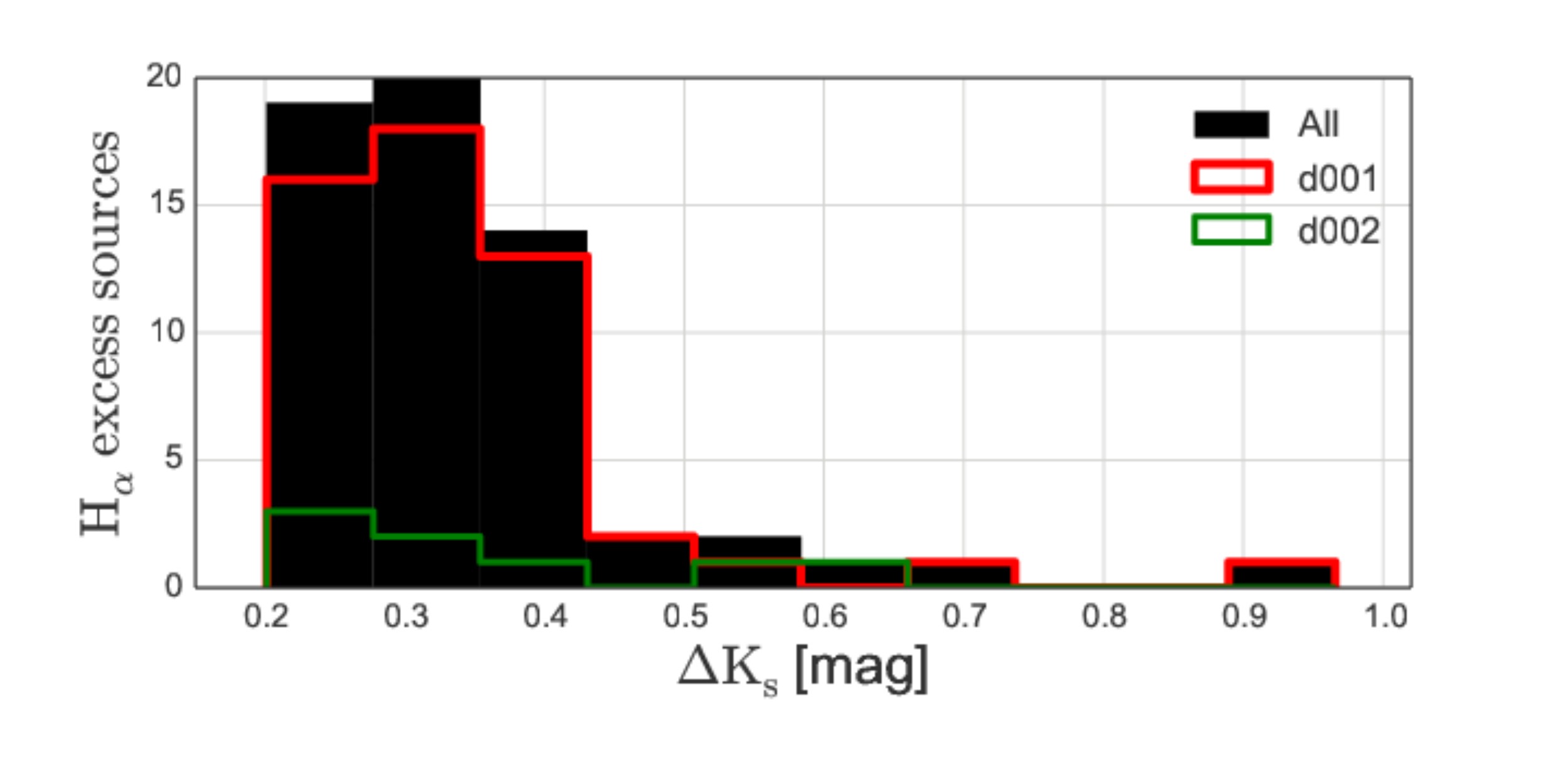}}
  \caption{The distribution of amplitudes in the $\rm K_s$-band of the emission line candidates.}
  \label{halpha_his}
\end{center}
\end{figure}

In summary, only a few variable YSOs around the clusters are detected in the near infrared. 

\subsection{Stars with $\rm H\alpha$ photometric emission}
The clusters in the tile d001 are very young, still in formation and are surrounded by dust and gas, thus we can use the photometric catalogs from VPHAS+ survey ~\citep{Drew2014} in order to search for additional YSOs and $\rm H\alpha$ emission candidates. The VPHAS+ catalog contains magnitudes in five filters: $u\ (\rm \lambda_{eff}=354nm)$, $g\ (\rm \lambda_{eff}=475nm)$, $r\ (\rm \lambda_{eff}=622nm)$, $i\ (\rm \lambda_{eff}=763nm)$ and $H_{\alpha}\ (\rm \lambda_{eff}=659nm)$.  Following ~\cite{Kalari2015}, we construct the $(r-i,r-\rm H\alpha)$ color-color diagrams using DR2 of VPHAS+ catalog. The main-sequence stars do not show any $\rm H\alpha$ emission with respect to the $r$-band photospheric continuum. We use the VST/OmegaCAM synthetic colors for main-sequence and giant stars in the $(r-H\alpha,r-i)$ plane~\citep{Drew2014} and selected the stars with more than 5$\rm \sigma$ deviation from these synthetic sequences, corrected for the corresponding reddening (dashed lines in Figure~\ref{halphaa}). The 73 selected sources are shown in Figure~\ref{halphaa}  and summarized in Table~\ref{halpha}.

The cross-identification of these sources with the $\rm K_s$ variability catalogs shows that the peak of amplitudes is around 0.3 mag (Figure~\ref{halpha_his}). As we stated in section~\ref{Ks_lightcurve}, we set our detection limit of spread of the magnitude measurements with time to be greater than 0.2 mag. For sources below this threshold, we treat these objects as constant at our sensitivity level. Thus, only a few of the infrared variable YSOs (d001-25, d001-32) are emission line object candidates.  
 
\section{Summary}\label{Summary}
We have developed an automated process for identification, classification and analysis of variable sources using the VVV $\rm K_s$-band time series, which are extracted directly from the $1.5\times1.2$ degrees image tiles. This process was created to automatically analyze the VVV tiles, given the huge amount of available data. The sources that present variability in the NIR are cataloged, in order to understand the physical process behind its variability, its spatial distribution, evolutionary state and relation with its environment. The gathered information from these sources will be collected in "VVV Variables ($\rm V^{4}$)" catalog, which will be publicaly avaliable in VISTA Science Archive (VSA, http://vsa.roe.ac.uk/index.html) database and constantly updated adding newly processed VVV tiles. 

This process is based on $\mathtt{Dophot}$ PSF photometry to create a multi-epoch $\rm K_{s}$-band catalog of detected sources in the FoV. We also obtained the $\rm J$ and $\rm H$ band photometry, using the data taken in 2010. All the PSF photometry was calibrated using the aperture catalogs made by CASU. To test this method, we select d001 and d002 tile regions, covering an area of $\rm \sim3.6$ deg$^{2}$. A total of 1,308,626 point sources between $\rm 10.8 \leq \overline{K}_{s} \leq 17.2$ mag were detected. The time series of sources with more than 25 epochs and amplitude $\rm \Delta K_{s}>0.2$ mag were selected and analyzed through different methods in order to detect real variables. To avoid outliers, suspicious photometric measurements were removed from the light-curves using the modified Thompson $\tau$ technique.

Our automated tool identified 200 sources with prominent NIR variability. Using two main variability indices ($\rm \Delta K_s$, $\eta$), we identified 70 variable sources without periodic or semi-periodic behavior, which are bona-fide irregular variables. On the other hand, we used periodograms (GLS, IP) to identify periodic sources, finding 130 of them. 
All identified sources have an average $\rm K_{s}$ magnitude distributed between $ 11.2 < \rm \overline{K}_{s} < 16.4$ mag, and a total amplitude contained between $0.2 < \rm \Delta K_{s} < 3.3$ mag.  About 90\% of these sources are previously unknown as variable stars.

For each source with available $\rm J$ and $\rm H$ photometric measurements, its position in the ($\rm H-K_{s},J-H $) color-color diagram was reported, dividing this diagram into three parts (the 'F', 'T' and 'P' region), following the procedures of~\cite{Ojha2004}, in order to extract information about the evolutionary state of those sources. 

In the case of the irregular variable sources, they were classified using the framework of~\cite{Contreras_Pena2017}, who use the time series morphology to classify the sources into 5 categories (Dippers, Eruptives, LPV-YSOs, STVs, Faders). 
The LPV-YSO sources, together with the periodic variable sample, will be analyzed in a subsequent paper. Sources that could not be classified unambiguously into the aforementioned categories, are marked as "unclassified". In total, we classified 20 STVs, 12 Eruptives, 5 Dippers, 7 Faders, 8 Low Amplitude Eruptive Variables, and 18 sources remained unclassified. These variable stars have amplitudes $\rm \Delta K_{s} > 0.63$ mag, $\eta<$ 0.82 and $\rm J_{stet}>0.95$, without counting the source d001-75, which has $\eta = 3.193$. and $\rm J_{stet}=-1.55$.

We examined some parameter distributions in the specific parameter space ~\citep{Graczyk2010} in order to separate the variability types. For example a $\chi^{2}$ parameter (in logarithmic scale) can separate the periodic sources from the main distribution when $\rm \log(\chi^{2})>0.8$. In the future, we will use this to optimize the process of generation and analysis of time series and light curves in the VVV in the supervised algorithms of machine learning.

We classified 25 RRab, 42 RRc, 13 Classic Cepheids, 33 Binaries, 7 LPV, 7 LPV-YSOs and three periodic variables remain unclassified. The periodic sources have a distribution of periods between $\rm 0.2 < P < 1430$ days and an average $\rm K_{s}$ magnitude distributed between $ 11.3 < \rm \overline{K}_{s} < 16.32$ mag.

We also analyzed nine open cluster candidates using surface-density maps, color-based decontamination  and proper-motion decontamination algorithms to determine the radii of the clusters. We have estimated the mean reddening $\rm E(J-Ks)$ by comparison with the PARSEC isochrones for solar metallicity. We were only able to determine a distance modulus for VVV CL005 cluster, using the spectroscopic parallax. This cluster also has the larges number of irregular variables as probable cluster members (d001-25, 27, 28, 29 and 30).  The cluster LS002 has two irregular variables, while clusters LS001 and VVV CL007 have one irregular variable, each other, as probable cluster members. All irregular sources projected close to open clusters in the region are young stellar object candidates.

Given the low number of irregular variable sources that we found around open clusters, we used the VPHAS+ survey to identify excess in $\rm H_{\alpha}$-band. We have selected 73 stars with more than 5$\rm \sigma$ difference from the VST/OmegaCAM synthetic colors for main-sequence and giant stars in the $(r-i,r-\rm H_\alpha)$ plane~\citep{Drew2014}. We noted that 64\% of this sample has low amplitude $\rm \Delta K_{s}<0.4$ mag, so this could be a good complementary method to find YSOs. 

\section{Acknowledgments}
We gratefully acknowledge data from the ESO Public Survey program ID 179.B-2002 taken with the VISTA telescope, and products from the Cambridge Astronomical Survey Unit (CASU), also on data products from observations made with ESO Telescopes at the La Silla, Paranal Observatory under programme ID 177.D-3023, as part of the 
VST Photometric H$_\alpha$ Survey of the Southern Galactic Plane and Bulge (VPHAS+). Support is provided by the Ministry for the Economy, Development and Tourism, Programa Iniciativa Cient\'ifica Milenio grant IC120009, awarded to the Millennium Institute of Astrophysics (MAS). We also thank the referee for its comments which greatly improved the present manuscript. A.B. thanks for support from the Millennium Science Initiative (Chilean Ministry of Economy). D.M. is supported by the BASAL Center for Astrophysics and Associated Technologies (CATA) through grant PFB-06, by the Ministry for the Economy, Development and Tourism, Programa Iniciativa Cientifica Milenio grant IC120009, awarded to the Millennium Institute of Astrophysics (MAS), and by FONDECYT Regular grant No. 1170121.

%\end{document}


\software{ $\mathtt{Matplotlib}$~\citep{Hunter2007}, $\mathtt{NumPy}$~\citep{Oliphant}, $\mathtt{AstroPy}$~\citep{Astropy2018}, $\mathtt{Sklearn}$~\citep{scikit-learn}, $\mathtt{AstroML}$~\citep{astroML}, $\mathtt{Dophot}$~\citep{Schechter1993, Alonso2012}, 
$\mathtt{VOSA}$~\citep{Bayo2008}, 
$\mathtt{STILTS}$~\citep{Taylor2006},
$\mathtt{Spextool}$~\cite{Cushing2004}, $\mathtt{xtellcorr}$~\citep{Vacca2003}.
}



\newpage
\appendix
\section{Basic information of $\rm V^{4}$ catalog.}

\begin{table}
\centering
\caption{Information of content of  V$^{4}$ Catalog}
\label{v4_catalog}
\begin{tabular}{ccc}
\hline
Column         & Units       & Description                                                  												 \\ \hline
ID                     & -                & Identification given by the catalog.                          							\\
Ra$_{2000}$            & Degrees & Right Ascension of VVV source.                                							\\
Dec$_{2000}$          & Degrees & Declination of VVV source.                                    									\\
GLon               & Degrees & Galactic longitude of VVV source.                         							   \\
GLat                & Degrees & Galactic latitude of VVV source.                             							   \\
$\rm \overline{K}_{s}$                    & mag        & Photometric mean value of $\rm K_{s}$-band.                     						 \\
$\rm \overline{K}_{s}^{err}$                 & mag       & Estimation of $\rm \overline{K}_{s}$ error using bootstrap technique.         	           \\
Epochs           & -                & Observation number of the source in their $\rm K_{s}$ time serie.    \\
$\rm \Delta K_{s}$     		   & mag     & Photometric total amplitude of $\rm K_{s}$-band.                 						 \\
$\rm J$                       & mag     & Photometric J-band.                                          										 \\
$\rm J_{err}$        		  & mag     & Photometric J-band error.                                    									   \\
$\rm H$                      & mag     & Photometric H-band.                                          											\\
$\rm H_{err}$          	     & mag     & Photometric H-band error.                                    									  \\
$u$                       & mag     & Photometric $u$-band.                                        										    \\
$u \rm_{err}$                  & mag     & Photometric $u$-band error.                                  									  \\
$g$                       & mag     & Photometric $g$-band.                                        											  \\
$g \rm_{err}$                  & mag     & Photometric $g$-band error.                                  								        \\
$r$                       & mag     & Photometric $r$-band.                                        											  \\
$r \rm_{err}$                  & mag     & Photometric $r$-band error.                                  									    \\
$r2$                     & mag     & Photometric $r2$-band.                                        											  \\
$r2 \rm_{err}$                & mag     & Photometric $r2$-band error.																		\\       
$i$                        & mag     & Photometric $i$-band.                                        												\\
$i \rm_{err}$                   & mag     & Photometric $i$-band error.                                											  \\
$\rm H_\alpha$                   & mag     & Photometric $\rm H_\alpha$-band.                           													  \\
$\rm H_\alpha$ err.           & mag     & Photometric $\rm H_\alpha$-band error.                     												  \\
$\eta$                       & -             & Value of the variability index $\eta$.                        										  \\
Class               & -            & Classification via VVV templates light curve or shape of the time serie. 		 \\
Tile ID             & -            & VVV tile where the source is located.                                                    \\    
Period            & days    & Identified period for a VVV source.                          								  \\
$\rm A_{K_{s}}$                & mag     & Extinction measured using~\cite{Nishiyama2009}.       		              \\ 
Distance       & Kpc        & Distance measured from PL relations of RRab sources.   			\\ 
CCD                & -             & Position in ($\rm H-K_{s}$,$\rm J-H$) color-color diagram.             		   \\
Reference     & -            &Reference to catalog of a documented source. 									\\ 

\hline
\end{tabular}
\end{table}

\begin{longrotatetable}
\begin{table}
\centering
\caption{Parameters of the VVV variables in the $\rm V^{4}$ catalog, description of each column can be found in~\ref{v4_catalog}. The complete catalog will be available online. }
\label{v4_table}
\resizebox*{1.3\textheight}{!}{
\tabcolsep=2pt
\begin{tabular}{ccccccccccccccccccccccccccccccccc}
\hline
ID      & $\rm Ra_{2000}$     & $\rm Dec_{2000}$    & $\rm G_{Lon}$          & $\rm G_{Lat}$           & $\rm \overline{K}_{s}$ & $\rm \overline{K}_{s}^{err}$ & $\rm N_{obs}$ & $\rm \Delta K_{s}$   & $\rm J$      & $\rm J_{err}$ & $\rm H $     & $\rm H_{err}$ & $u$       & $u_{err}$   & $g$       & $g_{err}$   & $r2$      & $r2_{err}$  & $r$       & $r_{err} $  & $i$       & $i_{err}$   & $\rm H_{\alpha}$  & $\rm H_{\alpha}^{err}$ & $\eta$   & Class    & TileID & Period   & $\rm A_{K_{s}}$ & Dist & CCD & Reference \\ \hline \hline
d001-1  & 174.13270153 & -63.58779258 & 294.716587832 & -1.91720472164 & 14.212                  & 0.043 & 40     & 0.951 &    -   &    -   &    -   &    -   &    -    &    -     &    -    &    -     &    -    &    -     &    -    &    -     &    -    &    -     &    -    &      -      & 0.568 & Eruptive & d001   &    -     &           -          &           -          &         -             &       -              \\
d001-2  & 174.14585121 & -63.70149127 & 294.754700509 & -2.02449224282 & 14.84                   & 0.033 & 42     & 0.852 &    -   &    -   &    -   &    -   &    -    &    -     &    -    &    -     &    -    &    -     &    -    &    -     &    -    &    -     &    -    &      -      & 0.548 & Eruptive & d001   &    -     &           -          &           -          &         -             &       -              \\
d001-3  & 174.28369391 & -63.36618123 & 294.718116325 & -1.68556805445 & 14.486                  & 0.03  & 43     & 0.776 & 15.598 & 0.048  & 14.717 & 0.055  &    -    &    -     &    -    &    -     & 20.5685 & 0.088321 & 20.5945 & 0.072345 & 19.535  & 0.046155 & 19.6652 & 0.116184    & 0.812 & STV      & d001   &    -     &           -          &           -          &         -             &       -              \\
d001-4  & 174.34896745 & -63.51237462 & 294.787537864 & -1.81751321427 & 13.105                  & 0.06  & 33     & 1.295 &    -   &    -   &    -   &    -   &    -    &    -     &    -    &    -     &    -    &    -     &    -    &    -     &    -    &    -     &    -    &      -      & 0.829 & Unclass  & d001   &    -     &           -          &           -          &         -             &       -              \\
d001-5  & 174.37960641 & -63.40070793 & 294.769122051 & -1.70652630648 & 15.247                  & 0.058 & 41     & 1.439 & 17.631 & 0.049  & 16.346 & 0.055  &    -    &    -     &    -    &    -     &    -    &    -     &    -    &    -     & 21.2187 & 0.208125 &    -    &      -      & 0.61  & LPV-YSO  & d001   & 90.2765  &           -          &           -          & T                     &       -              \\
d001-6  & 174.38660225 & -63.30768225 & 294.745873368 & -1.61639543528 & 15.796                  & 0.044 & 43     & 1.049 &    -   &    -   &    -   &    -   &    -    &    -     &    -    &    -     &    -    &    -     &    -    &    -     &    -    &    -     &    -    &      -      & 0.678 & Unclass  & d001   &    -     &           -          &           -          &         -             &       -              \\
d001-7  & 174.41869    & -63.49125342 & 294.811433652 & -1.78847231725 & 16.224                  & 0.057 & 41     & 1.279 & 17.947 & 0.05   & 16.868 & 0.055  &    -    &    -     &    -    &    -     &    -    &    -     &    -    &    -     & 20.8014 & 0.195378 &    -    &      -      & 0.544 & Unclass  & d001   &    -     &           -          &           -          & T                     &       -              \\
d001-8  & 174.44546193 & -63.78777894 & 294.906362304 & -2.06966169414 & 14.517                  & 0.04  & 38     & 1.033 & 17.677 & 0.05   & 15.829 & 0.055  &    -    &    -     &    -    &    -     &    -    &    -     &    -    &    -     &    -    &    -     &    -    &      -      & 0.743 & STV      & d001   &    -     &           -          &           -          & T                     &       -              \\
d001-9  & 174.47649923 & -63.05980135 & 294.715017585 & -1.36712007129 & 15.372                  & 0.043 & 43     & 1.052 &    -   &    -   &    -   &    -   &    -    &    -     &    -    &    -     &    -    &    -     &    -    &    -     &    -    &    -     &    -    &      -      & 0.793 & STV      & d001   &    -     &           -          &           -          &         -             &       -              \\
d001-10 & 174.49856436 & -63.21059924 & 294.766918008 & -1.50906184322 & 13.841                  & 0.018 & 43     & 0.476 & 14.774 & 0.048  & 14.06  & 0.055  & 21.0621 & 0.130539 & 19.124  & 0.018167 & 17.7393 & 0.008073 & 17.719  & 0.006832 & 17.0737 & 0.006334 & 17.1584 & 0.011451    & 0.544 & ClCeph   & d001   & 129.935  &           -          &           -          & F                     &       -              \\
d001-11 & 174.49901457 & -63.39755563 & 294.819565274 & -1.68845849266 & 16.332                  & 0.065 & 42     & 1.416 & 18.11  & 0.065  & 17.018 & 0.059  &    -    &    -     &    -    &    -     &    -    &    -     &    -    &    -     &    -    &    -     &    -    &      -      & 0.47  & Eruptive & d001   &    -     &           -          &           -          & T                     &       -              \\
d001-12 & 174.50256753 & -63.33327797 & 294.803061717 & -1.62631338008 & 16.037                  & 0.085 & 41     & 1.773 &    -   &    -   &    -   &    -   &    -    &    -     &    -    &    -     &    -    &    -     &    -    &    -     & 21.2009 & 0.223168 &    -    &      -      & 0.239 & Eruptive & d001   &    -     &           -          &           -          &         -             &       -              \\
d001-13 & 174.51180389 & -63.40093732 & 294.826013197 & -1.69009896834 & 14.93                   & 0.075 & 37     & 1.575 & 17.067 & 0.049  & 16.48  & 0.056  &    -    &    -     &    -    &    -     & 19.8391 & 0.047478 & 19.9979 & 0.045909 & 18.6628 & 0.022566 & 19.2114 & 0.088951    & 0.416 & Unclass  & d001   &    -     &           -          &           -          & P                     &       -              \\
d001-14 & 174.52235901 & -63.34128111 & 294.81383427  & -1.63150679146 & 15.188                  & 0.018 & 43     & 0.504 & 17.147 & 0.049  & 16.004 & 0.055  &    -    &    -     &    -    &    -     & 20.6715 & 0.102151 & 20.7031 & 0.087819 & 19.1605 & 0.035454 & 19.759  & 0.149883    & 0.826 & LPV-YSO  & d001   & 107.339  &           -          &           -          & T                     &       -              \\
d001-15 & 174.52851341 & -63.18521381 & 294.772769717 & -1.48090926299 & 14.69                   & 0.104 & 43     & 1.754 & 14.879 & 0.049  & 14.306 & 0.055  &    -    &    -     & 20.2052 & 0.052012 & 18.3015 & 0.013344 & 19.0201 & 0.02034  & 18.0251 & 0.013782 & 18.502  & 0.048707    & 0.258 & Dipper   & d001   &    -     &           -          &           -          &         -             &       -              \\
d001-16 & 174.55346357 & -63.20233576 & 294.788368391 & -1.49419878729 & 14.484                  & 0.017 & 39     & 0.404 &    -   &    -   &    -   &    -   &    -    &    -     &    -    &    -     &    -    &    -     &    -    &    -     &    -    &    -     &    -    &      -      & 2.336 & ClCeph   & d001   & 1.370011 &           -          &           -          &         -             &       -              \\
d001-17 & 174.55353029 & -63.50330822 & 294.87259933  & -1.78315447314 & 14.654                  & 0.013 & 43     & 0.295 & 15.938 & 0.049  & 15.144 & 0.055  &    -    &    -     & 20.8151 & 0.075503 & 19.0761 & 0.023867 & 19.2534 & 0.032048 & 18.3236 & 0.021309 & 18.8916 & 0.082716    & 1.298 & RRab     & d001   & 0.89999  & 0.231                & 10.7548              & T                     &       -              \\
d001-18 & 174.59579417 & -63.67813221 & 294.939525621 & -1.945765696   & 15.633                  & 0.042 & 43     & 1.021 & 16.838 & 0.049  & 15.753 & 0.055  &    -    &    -     &    -    &    -     &    -    &    -     &    -    &    -     &    -    &    -     &    -    &      -      & 0.778 & STV      & d001   &    -     &           -          &           -          &         -             &       -              \\
d001-19 & 174.59656925 & -63.41454926 & 294.866267627 & -1.69255674231 & 13.445                  & 0.033 & 44     & 0.747 & 15.3   & 0.049  & 14.406 & 0.055  & 21.1605 & 0.158495 & 18.6954 & 0.011662 & 17.2673 & 0.005715 & 18.0402 & 0.011509 & 17.3119 & 0.00965  & 17.6514 & 0.027395    & 0.556 & STV      & d001   &    -     &           -          &           -          & P                     &       -              \\
d001-20 & 174.60442126 & -63.45118855 & 294.879867262 & -1.72676178207 & 14.609                  & 0.038 & 43     & 0.938 & 15.921 & 0.049  & 15.079 & 0.055  & 20.8038 & 0.093883 & 20.278  & 0.045344 & 19.2298 & 0.026469 & 19.2354 & 0.030699 & 18.4935 & 0.024037 & 18.247  & 0.045281    & 0.345 & Unclass  & d001   &    -     &           -          &           -          & F                     &       -              \\
d001-21 & 174.6383814  & -63.49830505 & 294.907575466 & -1.76778557191 & 11.899                  & 0.041 & 41     & 0.968 &    -   &    -   &    -   &    -   & 20.4134 & 0.067918 & 18.1969 & 0.008596 & 16.5044 & 0.003441 & 16.4875 & 0.003746 & 15.752  & 0.003137 & 16.034  & 0.006969    & 0.341 & Unclass  & d001   &    -     &           -          &           -          &         -             &       -              \\
d001-22 & 174.65561041 & -63.33730748 & 294.87016198  & -1.61100194885 & 15.097                  & 0.032 & 43     & 0.773 & 17.537 & 0.054  & 16.223 & 0.056  &    -    &    -     &    -    &    -     &    -    &    -     & 21.1751 & 0.13503  & 20.8092 & 0.155988 &    -    &      -      & 0.845 & Unclass  & d001   &    -     &           -          &           -          & T                     &       -              \\
d001-23 & 174.6585263  & -63.19363264 & 294.831450436 & -1.47263041196 & 13.425                  & 0.048 & 25     & 0.794 &    -   &    -   &    -   &    -   &    -    &    -     &    -    &    -     &    -    &    -     &    -    &    -     &    -    &    -     &    -    &      -      & 0.586 & Eruptive & d001   &    -     &           -          &           -          &         -             &       -              \\
d001-24 & 174.71432507 & -63.38800177 & 294.909545064 & -1.65239163568 & 14.16                   & 0.025 & 41     & 0.683 & 16.474 & 0.049  & 15.179 & 0.055  &    -    &    -     &    -    &    -     & 21.2082 & 0.155063 & 20.8412 & 0.092301 & 19.5928 & 0.048271 & 18.8067 & 0.067593    & 0.697 & STV      & d001   &    -     &           -          &           -          & T                     &       -              \\
d001-25 & 174.71781737 & -63.48681575 & 294.938455506 & -1.74689858382 & 13.675                  & 0.021 & 42     & 0.674 & 15.217 & 0.005  & 14.473 & 0.004  & 20.3832 & 0.080564 & 19.6707 & 0.026482 & 18.4603 & 0.014426 & 18.5193 & 0.017226 & 17.7066 & 0.013369 & 17.6257 & 0.027176    & 0.724 & STV      & d001   &    -     &           -          &           -          & P                     &       -              \\
d001-26 & 174.73091611 & -63.69810021 & 295.002646171 & -1.9482910051  & 15.77                   & 0.039 & 40     & 1.027 & 19.081 & 0.055  & 17.299 & 0.058  &    -    &    -     &    -    &    -     &    -    &    -     &    -    &    -     &    -    &    -     &    -    &      -      & 0.715 & STV      & d001   &    -     &           -          &           -          & P                     &       -              \\
d001-27 & 174.74286153 & -63.48269181 & 294.948060319 & -1.73983841729 & 13.83                   & 0.033 & 42     & 0.752 &    -   &    -   &    -   &    -   &    -    &    -     &    -    &    -     &    -    &    -     &    -    &    -     &    -    &    -     &    -    &      -      & 0.682 & Unclass  & d001   &    -     &           -          &           -          &         -             &       -              \\
d001-28 & 174.74528875 & -63.46014502 & 294.942857561 & -1.71787278569 & 14.867                  & 0.021 & 43     & 0.636 & 17.429 & 0.03   & 16.03  & 0.016  &    -    &    -     &    -    &    -     & 20.9011 & 0.115453 & 21.0445 & 0.156534 &    -    &    -     & 19.6055 & 0.147164    & 0.592 & STV      & d001   &    -     &           -          &           -          & T                     &       -              \\
d001-29 & 174.74557089 & -63.45944988 & 294.942786234 & -1.71716991329 & 12.864                  & 0.022 & 41     & 0.641 & 14.955 & 0.004  & 13.99  & 0.003  &    -    &    -     & 19.9152 & 0.032117 & 18.5332 & 0.014602 & 18.3544 & 0.014671 & 17.7229 & 0.013218 & 17.8446 & 0.030147    & 0.675 & Unclass  & d001   &    -     &           -          &           -          & P                     &       -              \\
d001-30 & 174.7575568  & -63.49116246 & 294.956712439 & -1.74616262187 & 13.508                  & 0.028 & 40     & 0.879 & 16.3   & 0.012  & 14.857 & 0.006  &    -    &    -     & 21.6045 & 0.145009 & 20.1021 & 0.0567   & 20.3227 & 0.081997 & 19.7599 & 0.078043 & 19.9596 & 0.214782    & 0.96  & STV      & d001   &    -     &           -          &           -          & P                     &       -              \\
d001-31 & 174.78495    & -63.72963    & 295.034383471 & -1.97196912303 & 13.889                  & 0.033 & 43     & 0.833 & 15.666 & 0.049  & 14.689 & 0.055  &    -    &    -     &    -    &    -     &    -    &    -     &    -    &    -     &    -    &    -     & 17.6626 & 0.022099    & 0.698 & LPV-YSO  & d001   & 55.248   &           -          &           -          & T                     &       -              \\
d001-32 & 174.79571361 & -63.47490768 & 294.968597603 & -1.72583279486 & 12.934                  & 0.038 & 42     & 0.966 & 15.404 & 0.049  & 14.12  & 0.055  &    -    &    -     & 20.5187 & 0.054421 & 18.8488 & 0.018933 & 18.7404 & 0.020188 & 17.9998 & 0.016647 & 18.112  & 0.038448    & 0.758 & STV      & d001   &    -     &           -          &           -          & P                     &       -              \\
d001-33 & 174.81003286 & -63.44894739 & 294.967581872 & -1.69911506978 & 14.127                  & 0.029 & 43     & 0.811 & 16.799 & 0.049  & 15.169 & 0.055  &    -    &    -     &    -    &    -     &    -    &    -     &    -    &    -     &    -    &    -     &    -    &      -      & 0.65  & STV      & d001   &    -     &           -          &           -          & T                     &       -              \\
d001-34 & 174.81706237 & -63.44317121 & 294.969009881 & -1.69269619385 & 13.876                  & 0.032 & 44     & 0.824 & 15.43  & 0.049  & 14.642 & 0.055  &    -    &    -     &    -    &    -     &    -    &    -     &    -    &    -     &    -    &    -     & 16.8896 & 0.013655    & 0.65  & Eruptive & d001   &    -     &           -          &           -          & T                     &       -              \\
d001-35 & 174.8439099  & -63.46865193 & 294.987573089 & -1.71388406535 & 15.152                  & 0.044 & 32     & 0.903 &    -   &    -   &    -   &    -   &    -    &    -     &    -    &    -     &    -    &    -     &    -    &    -     &    -    &    -     &    -    &      -      & 0.682 & Eruptive & d001   &    -     &           -          &           -          &         -             &       -              \\
d001-36 & 174.84504516 & -63.46939365 & 294.988265138 & -1.71445747185 & 15.348                  & 0.048 & 31     & 0.849 &    -   &    -   &    -   &    -   &    -    &    -     &    -    &    -     &    -    &    -     &    -    &    -     &    -    &    -     &    -    &      -      & 0.65  & Unclass  & d001   &    -     &           -          &           -          &         -             &       -              \\
d001-37 & 174.85457043 & -63.79202108 & 295.081204793 & -2.02345227097 & 16.226                  & 0.044 & 41     & 1.112 &    -   &    -   &    -   &    -   &    -    &    -     &    -    &    -     &    -    &    -     &    -    &    -     &    -    &    -     &    -    &      -      & 0.814 & Eruptive & d001   &    -     &           -          &           -          &         -             &       -              \\
d001-38 & 174.8623778  & -63.46894123 & 294.995587011 & -1.71189166068 & 14.421                  & 0.018 & 43     & 0.453 & 16.126 & 0.049  & 15.11  & 0.055  &    -    &    -     &    -    &    -     &    -    &    -     &    -    &    -     & 19.7414 & 0.075651 & 19.5196 & 0.135836    & 0.906 & ClCeph   & d001   & 42.706   &           -          &           -          & T                     &       -              \\
d001-39 & 174.87844157 & -62.86599294 & 294.836730884 & -1.13018754516 & 14.177                  & 0.024 & 41     & 0.668 & 15.157 & 0.049  & 14.619 & 0.055  & 20.5803 & 0.096708 & 18.5343 & 0.011593 & 17.3772 & 0.006149 & 17.6678 & 0.006628 & 17.1084 & 0.006615 & 17.3543 & 0.012501    & 2.564 & RRab     & d001   & 0.395346 & 0.308                & 5.8502               & T                     &       -              \\
d001-40 & 174.887108   & -62.988212   & 294.874109473 & -1.24661992286 & 13.363                  & 0.031 & 40     & 0.667 & 14.748 & 0.048  & 14.118 & 0.055  &    -    &    -     &    -    &    -     &    -    &    -     &    -    &    -     &    -    &    -     & 18.9151 & 0.047509    & 0.716 & LPV-YSO  & d001   & 107.505  &  &  & P & - \\
d001-41  & 174.90503438 & -63.74297262 & 295.089173301 & -1.97016174451 & 15.01                   & 0.018 & 44     & 0.395 & 16.275 & 0.049  & 15.453 & 0.055  &    -    &    -     &    -    &    -     & 21.2008 & 0.13217  & 21.0537 & 0.136432 & 19.5579 & 0.06387  &    -    &      -      & 2.561 & RRab     & d001   & 0.348842  & 0.358 & 7.955      & F   &          -                      \\
d001-42  & 174.91787563 & -63.05083632 & 294.90472739  & -1.30300841895 & 15.2                    & 0.059 & 44     & 1.441 & 17.221 & 0.052  & 16.048 & 0.057  &    -    &    -     &    -    &    -     &    -    &    -     &    -    &    -     & 20.6985 & 0.134086 &    -    &      -      & 0.52  & Eruptive & d001   &    -      &   -   &     -      & T   &          -                      \\
d001-43  & 174.93715784 & -63.46692966 & 295.027169272 & -1.70078635325 & 14.149                  & 0.043 & 43     & 0.936 & 15.276 & 0.049  & 14.352 & 0.055  &    -    &    -     & 19.5502 & 0.023194 & 18.0523 & 0.010214 & 18.0581 & 0.011675 & 17.3699 & 0.010238 & 17.6976 & 0.02775     & 0.268 & Eruptive & d001   &    -      &   -   &     -      &  -  &          -                      \\
d001-44  & 175.00491951 & -63.47621668 & 295.058834462 & -1.70144196523 & 15.48                   & 0.085 & 43     & 1.616 & 15.912 & 0.049  & 15.116 & 0.055  &    -    &    -     &    -    &    -     & 20.8883 & 0.118532 & 21.0967 & 0.167048 & 19.7478 & 0.078924 &    -    &      -      & 0.223 & Dipper   & d001   &    -      &   -   &     -      &  -  &          -                      \\
d001-45  & 175.21297019 & -63.41614807 & 295.132074225 & -1.61838242178 & 15.826                  & 0.074 & 34     & 1.361 &    -   &    -   &    -   &    -   &    -    &    -     &    -    &    -     &    -    &    -     &    -    &    -     &    -    &    -     &    -    &      -      & 0.592 & Eruptive & d001   &    -      &   -   &     -      &  -  &          -                      \\
d001-46  & 175.28487967 & -62.95803377 & 295.039895249 & -1.16846143672 & 15.921                  & 0.052 & 41     & 1.097 &    -   &    -   &    -   &    -   &    -    &    -     &    -    &    -     &    -    &    -     &    -    &    -     &    -    &    -     &    -    &      -      & 0.279 & Dipper   & d001   &    -      &   -   &     -      &  -  &          -                      \\
d001-47  & 175.29780354 & -63.00170576 & 295.057288737 & -1.20894985896 & 14.655                  & 0.041 & 39     & 1.05  & 18.641 & 0.066  & 17.187 & 0.061  &    -    &    -     &    -    &    -     &    -    &    -     &    -    &    -     &    -    &    -     & 20.7969 & 0.22459     & 0.432 & Eruptive & d001   &    -      &   -   &     -      & P   &          -                      \\
d001-48  & 175.31075221 & -63.03529086 & 295.071968921 & -1.23972425333 & 14.793                  & 0.019 & 41     & 0.469 & 15.699 & 0.049  & 15.142 & 0.055  &    -    &    -     &    -    &    -     & 19.2997 & 0.027752 & 19.3908 & 0.023753 & 18.4209 & 0.017972 & 19.2435 & 0.059381    & 1.991 & RRab     & d001   & 0.64087   & 0.277 & 9.7024     & T   &          -                      \\
d001-49  & 175.32699139 & -63.8683344  & 295.302594471 & -2.04026812799 & 14.541                  & 0.021 & 41     & 0.5   & 15.674 & 0.049  & 15.074 & 0.055  &    -    &    -     &    -    &    -     &    -    &    -     &    -    &    -     &    -    &    -     & 18.7247 & 0.033165    & 1.958 & RRab     & d001   & 0.387783  & 0.326 & 6.8066     & T   &          -                      \\
d001-50  & 175.36822776 & -63.15863488 & 295.130094026 & -1.35158707292 & 15.552                  & 0.061 & 40     & 1.148 &    -   &    -   &    -   &    -   &    -    &    -     &    -    &    -     &    -    &    -     &    -    &    -     &    -    &    -     &    -    &      -      & 0.193 & Eruptive & d001   &    -      &   -   &     -      &  -  &          -                      \\
d001-51  & 175.4259293  & -63.7143567  & 295.303503771 & -1.88021329349 & 13.428                  & 0.021 & 43     & 0.462 & 14.808 & 0.049  & 13.962 & 0.055  &    -    &    -     &    -    &    -     &    -    &    -     &    -    &    -     &    -    &    -     & 18.8363 & 0.063551    & 1.648 & RRab     & d001   & 0.5259608 & 0.242 & 4.8253     & T   &          -                      \\
d001-52  & 175.46935098 & -63.48317698 & 295.260496299 & -1.65224628207 & 11.99                   & 0.013 & 42     & 0.339 &    -   &    -   &    -   &    -   &    -    &    -     & 20.3607 & 0.038877 & 18.1937 & 0.010413 & 18.2312 & 0.012316 & 17.0607 & 0.007652 & 17.8651 & 0.024501    & 1.526 & ClCeph   & d001   & 22.212    &   -   &     -      &  -  &          -                      \\
d001-53  & 175.52256545 & -63.32281578 & 295.240848997 & -1.49131895783 & 15.839                  & 0.052 & 44     & 1.114 &    -   &    -   &    -   &    -   &    -    &    -     &    -    &    -     &    -    &    -     &    -    &    -     &    -    &    -     &    -    &      -      & 0.538 & Eruptive & d001   &    -      &   -   &     -      &  -  &          -                      \\
d001-54  & 175.58388882 & -63.02467179 & 295.188564598 & -1.19649097221 & 14.24                   & 0.042 & 40     & 0.902 & 16.743 & 0.055  & 15.054 & 0.056  &    -    &    -     &    -    &    -     &    -    &    -     &    -    &    -     &    -    &    -     &    -    &      -      & 0.49  & Eruptive & d001   &    -      &   -   &     -      & F   &          -                      \\
d001-55  & 175.711813   & -63.350653   & 295.330139057 & -1.49575844137 & 14.137                  & 0.061 & 43     & 1.489 & 16.976 & 0.049  & 15.267 & 0.055  &    -    &    -     &    -    &    -     & 21.5195 & 0.196079 &    -    &    -     &    -    &    -     &    -    &      -      & 0.64  & LPV-YSO  & d001   & 467.629   &   -   &     -      & T   &          -                      \\
d001-56  & 175.73341657 & -63.06706191 & 295.265127558 & -1.21954365139 & 14.532                  & 0.02  & 41     & 0.549 & 15.143 & 0.049  & 14.632 & 0.055  & 20.3815 & 0.076347 & 18.5672 & 0.010873 & 17.3396 & 0.005601 & 17.7351 & 0.006404 & 17.2211 & 0.006573 & 17.4129 & 0.011349    & 2.337 & RRab     & d001   & 0.293773  & 0.358 & 5.927      & F   &          -                      \\
d001-57  & 175.76631501 & -63.36340507 & 295.357076394 & -1.50165931731 & 15.945                  & 0.059 & 43     & 1.38  & 18.986 & 0.06   & 17.272 & 0.057  &    -    &    -     &    -    &    -     &    -    &    -     &    -    &    -     &    -    &    -     &    -    &      -      & 0.516 & Unclass  & d001   &    -      &   -   &     -      & T   &          -                      \\
d001-58  & 175.80425    & -63.7939947  & 295.485993944 & -1.91286173744 & 14.985                  & 0.017 & 40     & 0.462 & 15.734 & 0.049  & 15.196 & 0.055  &    -    &    -     & 20.2293 & 0.039338 & 18.7087 & 0.015028 & 18.6598 & 0.016156 & 17.8476 & 0.013988 & 18.3986 & 0.026967    & 2.112 & RRab     & d001   & 0.221178  & 0.411 & 6.3103     & F   &          -                      \\
d001-59  & 175.81602886 & -63.76981629 & 295.484703931 & -1.88816330724 & 13.497                  & 0.03  & 42     & 0.621 & 14.257 & 0.049  & 13.738 & 0.055  &    -    &    -     & 20.1591 & 0.037196 & 18.172  & 0.009816 & 18.2049 & 0.011309 & 17.1849 & 0.008223 & 17.9447 & 0.01732     & 1.184 & Binary   & d001   & 2.03445   &   -   &     -      & F   &          -                      \\

\end{tabular}}
\end{table}
\end{longrotatetable}




\bibliography{references} 

\begin{thebibliography}{}
\expandafter\ifx\csname natexlab\endcsname\relax\def\natexlab#1{#1}\fi
\providecommand{\url}[1]{\href{#1}{#1}}

\bibitem[{{Alonso-Garc{\'{\i}}a} {et~al.}(2015){Alonso-Garc{\'{\i}}a},
  {D{\'e}k{\'a}ny}, {Catelan}, {Contreras Ramos}, {Gran}, {Amigo}, {Leyton}, \&
  {Minniti}}]{Alonso2015}
{Alonso-Garc{\'{\i}}a}, J., {D{\'e}k{\'a}ny}, I., {Catelan}, M., {et~al.} 2015,
  \aj, 149, 99

\bibitem[{{Alonso-Garc{\'{\i}}a} {et~al.}(2012){Alonso-Garc{\'{\i}}a}, {Mateo},
  {Sen}, {Banerjee}, {Catelan}, {Minniti}, \& {von Braun}}]{Alonso2012}
{Alonso-Garc{\'{\i}}a}, J., {Mateo}, M., {Sen}, B., {et~al.} 2012, \aj, 143, 70

\bibitem[{{Angeloni} {et~al.}(2014){Angeloni}, {Contreras Ramos}, {Catelan},
  {D{\'e}k{\'a}ny}, {Gran}, {Alonso-Garc{\'{\i}}a}, {Hempel}, {Navarrete},
  {Andrews}, {Aparicio}, {Beam{\'{\i}}n}, {Berger}, {Borissova}, {Contreras
  Pe{\~n}a}, {Cunial}, {de Grijs}, {Espinoza}, {Eyheramendy}, {Ferreira Lopes},
  {Fiaschi}, {Hajdu}, {Han}, {He{\l}miniak}, {Hempel}, {Hidalgo}, {Ita},
  {Jeon}, {Jord{\'a}n}, {Kwon}, {Lee}, {Mart{\'{\i}}n}, {Masetti}, {Matsunaga},
  {Milone}, {Minniti}, {Morelli}, {Murgas}, {Nagayama}, {Navarro}, {Ochner},
  {P{\'e}rez}, {Pichara}, {Rojas-Arriagada}, {Roquette}, {Saito}, {Siviero},
  {Sohn}, {Sung}, {Tamura}, {Tata}, {Tomasella}, {Townsend}, \&
  {Whitelock}}]{Angeloni2014}
{Angeloni}, R., {Contreras Ramos}, R., {Catelan}, M., {et~al.} 2014, \aap, 567,
  A100

\bibitem[{{Arnaboldi} {et~al.}(2007){Arnaboldi}, {Neeser}, {Parker}, {Rosati},
  {Lombardi}, {Dietrich}, \& {Hummel}}]{Arnaboldi2007}
{Arnaboldi}, M., {Neeser}, M.~J., {Parker}, L.~C., {et~al.} 2007, The
  Messenger, 127

\bibitem[{{Arnaboldi} {et~al.}(2012){Arnaboldi}, {Rejkuba}, {Retzlaff},
  {Delmotte}, {Hanuschik}, {Hilker}, {H{\"u}mmel}, {Hussain}, {Ivanov},
  {Micol}, {Neeser}, {Petr-Gotzens}, {Szeifert}, {Comeron}, {Primas}, \&
  {Romaniello}}]{Arnaboldi2012}
{Arnaboldi}, M., {Rejkuba}, M., {Retzlaff}, J., {et~al.} 2012, The Messenger,
  149, 7

\bibitem[{{Bailey}(1902)}]{Bailey1902}
{Bailey}, S.~I. 1902, Annals of Harvard College Observatory, 38

\bibitem[{{Barb{\'a}} {et~al.}(2015){Barb{\'a}}, {Roman-Lopes}, {Nilo
  Castell{\'o}n}, {Firpo}, {Minniti}, {Lucas}, {Emerson}, {Hempel}, {Soto}, \&
  {Saito}}]{Barba2015}
{Barb{\'a}}, R.~H., {Roman-Lopes}, A., {Nilo Castell{\'o}n}, J.~L., {et~al.}
  2015, \aap, 581, A120

\bibitem[{{Bayo} {et~al.}(2008){Bayo}, {Rodrigo}, {Barrado Y Navascu{\'e}s},
  {Solano}, {Guti{\'e}rrez}, {Morales-Calder{\'o}n}, \& {Allard}}]{Bayo2008}
{Bayo}, A., {Rodrigo}, C., {Barrado Y Navascu{\'e}s}, D., {et~al.} 2008, \aap,
  492, 277

\bibitem[{{Bessell} \& {Brett}(1988)}]{Bessell1988}
{Bessell}, M.~S., \& {Brett}, J.~M. 1988, \pasp, 100, 1134

\bibitem[{{Bonatto} \& {Bica}(2010)}]{Bonatto2010}
{Bonatto}, C., \& {Bica}, E. 2010, \aap, 516, A81

\bibitem[{{Borissova} {et~al.}(2011){Borissova}, {Bonatto}, {Kurtev}, {Clarke},
  {Pe{\~n}aloza}, {Sale}, {Minniti}, {Alonso-Garc{\'{\i}}a}, {Artigau},
  {Barb{\'a}}, {Bica}, {Baume}, {Catelan}, {Chen{\`e}}, {Dias}, {Folkes},
  {Froebrich}, {Geisler}, {de Grijs}, {Hanson}, {Hempel}, {Ivanov}, {Kumar},
  {Lucas}, {Mauro}, {Moni Bidin}, {Rejkuba}, {Saito}, {Tamura}, \&
  {Toledo}}]{Borissova2011}
{Borissova}, J., {Bonatto}, C., {Kurtev}, R., {et~al.} 2011, \aap, 532, A131

\bibitem[{{Borissova} {et~al.}(2014){Borissova}, {Chen{\'e}}, {Ram{\'{\i}}rez
  Alegr{\'{\i}}a}, {Sharma}, {Clarke}, {Kurtev}, {Negueruela}, {Marco},
  {Amigo}, {Minniti}, {Bica}, {Bonatto}, {Catelan}, {Fierro}, {Geisler},
  {Gromadzki}, {Hempel}, {Hanson}, {Ivanov}, {Lucas}, {Majaess}, {Moni Bidin},
  {Popescu}, \& {Saito}}]{Borissova2014}
{Borissova}, J., {Chen{\'e}}, A.-N., {Ram{\'{\i}}rez Alegr{\'{\i}}a}, S.,
  {et~al.} 2014, \aap, 569, A24

\bibitem[{{Borissova} {et~al.}(2016){Borissova}, {Ram{\'{\i}}rez
  Alegr{\'{\i}}a}, {Alonso}, {Lucas}, {Kurtev}, {Medina}, {Navarro}, {Kuhn},
  {Gromadzki}, {Retamales}, {Fernandez}, {Agurto-Gangas}, {Chen{\'e}},
  {Minniti}, {Contreras Pena}, {Catelan}, {Decany}, {Thompson}, {Morales}, \&
  {Amigo}}]{Borissova2016}
{Borissova}, J., {Ram{\'{\i}}rez Alegr{\'{\i}}a}, S., {Alonso}, J., {et~al.}
  2016, \aj, 152, 74

\bibitem[{{Bressan} {et~al.}(2012){Bressan}, {Marigo}, {Girardi}, {Salasnich},
  {Dal Cero}, {Rubele}, \& {Nanni}}]{Bressan2012}
{Bressan}, A., {Marigo}, P., {Girardi}, L., {et~al.} 2012, \mnras, 427, 127

\bibitem[{{Carpenter} {et~al.}(2001){Carpenter}, {Hillenbrand}, \&
  {Skrutskie}}]{Carpenter2001}
{Carpenter}, J.~M., {Hillenbrand}, L.~A., \& {Skrutskie}, M.~F. 2001, \aj, 121,
  3160

\bibitem[{{Chen{\'e}} {et~al.}(2012){Chen{\'e}}, {Borissova}, {Clarke},
  {Bonatto}, {Majaess}, {Moni Bidin}, {Sale}, {Mauro}, {Kurtev}, {Baume},
  {Feinstein}, {Ivanov}, {Geisler}, {Catelan}, {Minniti}, {Lucas}, {de Grijs},
  \& {Kumar}}]{Chene2012}
{Chen{\'e}}, A.-N., {Borissova}, J., {Clarke}, J.~R.~A., {et~al.} 2012, \aap,
  545, A54

\bibitem[{{Chen{\'e}} {et~al.}(2013){Chen{\'e}}, {Borissova}, {Bonatto},
  {Majaess}, {Baume}, {Clarke}, {Kurtev}, {Schnurr}, {Bouret}, {Catelan},
  {Emerson}, {Feinstein}, {Geisler}, {de Grijs}, {Herv{\'e}}, {Ivanov},
  {Kumar}, {Lucas}, {Mahy}, {Martins}, {Mauro}, {Minniti}, \& {Moni
  Bidin}}]{Chene2013}
{Chen{\'e}}, A.-N., {Borissova}, J., {Bonatto}, C., {et~al.} 2013, \aap, 549,
  A98

\bibitem[{{Cody} {et~al.}(2014){Cody}, {Stauffer}, {Baglin}, {Micela},
  {Rebull}, {Flaccomio}, {Morales-Calder{\'o}n}, {Aigrain}, {Bouvier},
  {Hillenbrand}, {Gutermuth}, {Song}, {Turner}, {Alencar}, {Zwintz},
  {Plavchan}, {Carpenter}, {Findeisen}, {Carey}, {Terebey}, {Hartmann},
  {Calvet}, {Teixeira}, {Vrba}, {Wolk}, {Covey}, {Poppenhaeger}, {G{\"u}nther},
  {Forbrich}, {Whitney}, {Affer}, {Herbst}, {Hora}, {Barrado}, {Holtzman},
  {Marchis}, {Wood}, {Medeiros Guimar{\~a}es}, {Lillo Box}, {Gillen},
  {McQuillan}, {Espaillat}, {Allen}, {D'Alessio}, \& {Favata}}]{Cody2014}
{Cody}, A.~M., {Stauffer}, J., {Baglin}, A., {et~al.} 2014, \aj, 147, 82

\bibitem[{{Contreras Pe{\~n}a} {et~al.}(2017{\natexlab{a}}){Contreras
  Pe{\~n}a}, {Lucas}, {Minniti}, {Kurtev}, {Stimson}, {Navarro Molina},
  {Borissova}, {Kumar}, {Thompson}, {Gledhill}, {Terzi}, {Froebrich}, \&
  {Caratti o Garatti}}]{Contreras_Pena2017}
{Contreras Pe{\~n}a}, C., {Lucas}, P.~W., {Minniti}, D., {et~al.}
  2017{\natexlab{a}}, \mnras, 465, 3011

\bibitem[{{Contreras Pe{\~n}a} {et~al.}(2017{\natexlab{b}}){Contreras
  Pe{\~n}a}, {Lucas}, {Kurtev}, {Minniti}, {Caratti o Garatti}, {Marocco},
  {Thompson}, {Froebrich}, {Kumar}, {Stimson}, {Navarro Molina}, {Borissova},
  {Gledhill}, \& {Terzi}}]{Contreras_Pena_b2017}
{Contreras Pe{\~n}a}, C., {Lucas}, P.~W., {Kurtev}, R., {et~al.}
  2017{\natexlab{b}}, \mnras, 465, 3039

\bibitem[{{Cross} {et~al.}(2012){Cross}, {Collins}, {Mann}, {Read}, {Sutorius},
  {Blake}, {Holliman}, {Hambly}, {Emerson}, {Lawrence}, \&
  {Noddle}}]{Cross2012}
{Cross}, N.~J.~G., {Collins}, R.~S., {Mann}, R.~G., {et~al.} 2012, \aap, 548,
  A119

\bibitem[{{Cushing} {et~al.}(2004){Cushing}, {Vacca}, \&
  {Rayner}}]{Cushing2004}
{Cushing}, M.~C., {Vacca}, W.~D., \& {Rayner}, J.~T. 2004, \pasp, 116, 362

\bibitem[{{Dalton} {et~al.}(2006){Dalton}, {Caldwell}, {Ward}, {Whalley},
  {Woodhouse}, {Edeson}, {Clark}, {Beard}, {Gallie}, {Todd}, {Strachan},
  {Bezawada}, {Sutherland}, \& {Emerson}}]{Dalton2006}
{Dalton}, G.~B., {Caldwell}, M., {Ward}, A.~K., {et~al.} 2006, in \procspie,
  Vol. 6269, Society of Photo-Optical Instrumentation Engineers (SPIE)
  Conference Series, 62690X

\bibitem[{{D{\'e}k{\'a}ny} {et~al.}(2015){D{\'e}k{\'a}ny}, {Minniti},
  {Majaess}, {Zoccali}, {Hajdu}, {Alonso-Garc{\'{\i}}a}, {Catelan}, {Gieren},
  \& {Borissova}}]{Dekany2015b}
{D{\'e}k{\'a}ny}, I., {Minniti}, D., {Majaess}, D., {et~al.} 2015, \apjl, 812,
  L29

\bibitem[{{Dong} {et~al.}(2017){Dong}, {Sch{\"o}del}, {Williams},
  {Nogueras-Lara}, {Gallego-Cano}, {Gallego-Calvente}, {Wang}, {Rich},
  {Morris}, {Do}, \& {Ghez}}]{Dong2017}
{Dong}, H., {Sch{\"o}del}, R., {Williams}, B.~F., {et~al.} 2017, \mnras, 471,
  3617

\bibitem[{{Drake} {et~al.}(2009){Drake}, {Djorgovski}, {Mahabal}, {Beshore},
  {Larson}, {Graham}, {Williams}, {Christensen}, {Catelan}, {Boattini},
  {Gibbs}, {Hill}, \& {Kowalski}}]{Drake2009}
{Drake}, A.~J., {Djorgovski}, S.~G., {Mahabal}, A., {et~al.} 2009, \apj, 696,
  870

\bibitem[{{Drew} {et~al.}(2014){Drew}, {Gonzalez-Solares}, {Greimel}, {Irwin},
  {K{\"u}pc{\"u} Yoldas}, {Lewis}, {Barentsen}, {Eisl{\"o}ffel}, {Farnhill},
  {Martin}, {Walsh}, {Walton}, {Mohr-Smith}, {Raddi}, {Sale}, {Wright},
  {Groot}, {Barlow}, {Corradi}, {Drake}, {Fabregat}, {Frew}, {G{\"a}nsicke},
  {Knigge}, {Mampaso}, {Morris}, {Naylor}, {Parker}, {Phillipps}, {Ruhland},
  {Steeghs}, {Unruh}, {Vink}, {Wesson}, \& {Zijlstra}}]{Drew2014}
{Drew}, J.~E., {Gonzalez-Solares}, E., {Greimel}, R., {et~al.} 2014, \mnras,
  440, 2036

\bibitem[{{Elorrieta} {et~al.}(2016){Elorrieta}, {Eyheramendy}, {Jord{\'a}n},
  {D{\'e}k{\'a}ny}, {Catelan}, {Angeloni}, {Alonso-Garc{\'{\i}}a},
  {Contreras-Ramos}, {Gran}, {Hajdu}, {Espinoza}, {Saito}, \&
  {Minniti}}]{Elorrieta2016}
{Elorrieta}, F., {Eyheramendy}, S., {Jord{\'a}n}, A., {et~al.} 2016, \aap, 595,
  A82

\bibitem[{{Elson} {et~al.}(1987){Elson}, {Fall}, \& {Freeman}}]{Elson1987}
{Elson}, R.~A.~W., {Fall}, S.~M., \& {Freeman}, K.~C. 1987, \apj, 323, 54

\bibitem[{{Findeisen} {et~al.}(2013){Findeisen}, {Hillenbrand}, {Ofek},
  {Levitan}, {Sesar}, {Laher}, \& {Surace}}]{Findeisen2013}
{Findeisen}, K., {Hillenbrand}, L., {Ofek}, E., {et~al.} 2013, \apj, 768, 93

\bibitem[{{Gavrilchenko} {et~al.}(2014){Gavrilchenko}, {Klein}, {Bloom}, \&
  {Richards}}]{Gavrilchenko2014}
{Gavrilchenko}, T., {Klein}, C.~R., {Bloom}, J.~S., \& {Richards}, J.~W. 2014,
  \mnras, 441, 715

\bibitem[{{Graczyk} \& {Eyer}(2010)}]{Graczyk2010}
{Graczyk}, D., \& {Eyer}, L. 2010, \actaa, 60, 109

\bibitem[{{Gran} {et~al.}(2016){Gran}, {Minniti}, {Saito}, {Zoccali},
  {Gonzalez}, {Navarrete}, {Catelan}, {Contreras Ramos}, {Elorrieta},
  {Eyheramendy}, \& {Jord{\'a}n}}]{Gran2016}
{Gran}, F., {Minniti}, D., {Saito}, R.~K., {et~al.} 2016, \aap, 591, A145

\bibitem[{{Harju} {et~al.}(1998){Harju}, {Lehtinen}, {Booth}, \&
  {Zinchenko}}]{Harju1998}
{Harju}, J., {Lehtinen}, K., {Booth}, R.~S., \& {Zinchenko}, I. 1998, \aaps,
  132, 211

\bibitem[{{Herbig}(1966)}]{Herbig1966}
{Herbig}, G.~H. 1966, Vistas in Astronomy, 8, 109

\bibitem[{{Herbig}(1989)}]{Herbig89}
{Herbig}, G.~H. 1989, in European Southern Observatory Conference and Workshop
  Proceedings, Vol.~33, European Southern Observatory Conference and Workshop
  Proceedings, ed. B.~{Reipurth}, 233--246

\bibitem[{{Herv{\'e}} {et~al.}(2016){Herv{\'e}}, {Martins}, {Chen{\'e}},
  {Bouret}, \& {Borissova}}]{Herve2016}
{Herv{\'e}}, A., {Martins}, F., {Chen{\'e}}, A.-N., {Bouret}, J.-C., \&
  {Borissova}, J. 2016, \na, 45, 84

\bibitem[{{Higuchi} {et~al.}(2017){Higuchi}, {Sato}, {Tsukagoshi}, {Sakai},
  {Iwasaki}, {Momose}, {Kobayashi}, {Ishihara}, {Watanabe}, {Kaneda}, \&
  {Yamamoto}}]{Higuchi2017}
{Higuchi}, A.~E., {Sato}, A., {Tsukagoshi}, T., {et~al.} 2017, \apjl, 839, L14

\bibitem[{{Hillenbrand} {et~al.}(1992){Hillenbrand}, {Strom}, {Vrba}, \&
  {Keene}}]{Hillenbrand1992}
{Hillenbrand}, L.~A., {Strom}, S.~E., {Vrba}, F.~J., \& {Keene}, J. 1992, \apj,
  397, 613

\bibitem[{{Huijse} {et~al.}(2011){Huijse}, {Estevez}, {Zegers}, {Principe}, \&
  {Protopapas}}]{Huijse2011}
{Huijse}, P., {Estevez}, P.~A., {Zegers}, P., {Principe}, J.~C., \&
  {Protopapas}, P. 2011, IEEE Signal Processing Letters, 18, 371

\bibitem[{Hunter(2007)}]{Hunter2007}
Hunter, J.~D. 2007, Computing In Science \& Engineering, 9, 90

\bibitem[{{Kaiser} {et~al.}(2002){Kaiser}, {Aussel}, {Burke}, {Boesgaard},
  {Chambers}, {Chun}, {Heasley}, {Hodapp}, {Hunt}, {Jedicke}, {Jewitt},
  {Kudritzki}, {Luppino}, {Maberry}, {Magnier}, {Monet}, {Onaka}, {Pickles},
  {Rhoads}, {Simon}, {Szalay}, {Szapudi}, {Tholen}, {Tonry}, {Waterson}, \&
  {Wick}}]{Kaiser2002}
{Kaiser}, N., {Aussel}, H., {Burke}, B.~E., {et~al.} 2002, in \procspie, Vol.
  4836, Survey and Other Telescope Technologies and Discoveries, ed. J.~A.
  {Tyson} \& S.~{Wolff}, 154--164

\bibitem[{{Kalari} {et~al.}(2015){Kalari}, {Vink}, {Drew}, {Barentsen},
  {Drake}, {Eisl{\"o}ffel}, {Mart{\'{\i}}n}, {Parker}, {Unruh}, {Walton}, \&
  {Wright}}]{Kalari2015}
{Kalari}, V.~M., {Vink}, J.~S., {Drew}, J.~E., {et~al.} 2015, \mnras, 453, 1026

\bibitem[{{Krabbendam} \& {Sweeney}(2010)}]{Krabbendam2010}
{Krabbendam}, V.~L., \& {Sweeney}, D. 2010, in \procspie, Vol. 7733,
  Ground-based and Airborne Telescopes III, 77330D

\bibitem[{{Kwok} {et~al.}(1997){Kwok}, {Volk}, \& {Bidelman}}]{Kwok1997}
{Kwok}, S., {Volk}, K., \& {Bidelman}, W.~P. 1997, \apjs, 112, 557

\bibitem[{{Lomb}(1976)}]{Lomb1976}
{Lomb}, N.~R. 1976, \apss, 39, 447

\bibitem[{{Lumsden} {et~al.}(2013){Lumsden}, {Hoare}, {Urquhart}, {Oudmaijer},
  {Davies}, {Mottram}, {Cooper}, \& {Moore}}]{Lumsden2013}
{Lumsden}, S.~L., {Hoare}, M.~G., {Urquhart}, J.~S., {et~al.} 2013, \apjs, 208,
  11

\bibitem[{{Marigo} {et~al.}(2017){Marigo}, {Girardi}, {Bressan}, {Rosenfield},
  {Aringer}, {Chen}, {Dussin}, {Nanni}, {Pastorelli}, {Rodrigues}, {Trabucchi},
  {Bladh}, {Dalcanton}, {Groenewegen}, {Montalb{\'a}n}, \& {Wood}}]{Marigo2017}
{Marigo}, P., {Girardi}, L., {Bressan}, A., {et~al.} 2017, \apj, 835, 77

\bibitem[{{McSwain} \& {Gies}(2005)}]{McSwain2005}
{McSwain}, M.~V., \& {Gies}, D.~R. 2005, \apjs, 161, 118

\bibitem[{{Meyer} {et~al.}(1997){Meyer}, {Calvet}, \&
  {Hillenbrand}}]{Meyer1997}
{Meyer}, M.~R., {Calvet}, N., \& {Hillenbrand}, L.~A. 1997, \aj, 114, 288

\bibitem[{{Minniti} {et~al.}(2010){Minniti}, {Lucas}, {Emerson}, {Saito},
  {Hempel}, {Pietrukowicz}, {Ahumada}, {Alonso}, {Alonso-Garcia}, {Arias},
  {Bandyopadhyay}, {Barb{\'a}}, {Barbuy}, {Bedin}, {Bica}, {Borissova},
  {Bronfman}, {Carraro}, {Catelan}, {Clari{\'a}}, {Cross}, {de Grijs},
  {D{\'e}k{\'a}ny}, {Drew}, {Fari{\~n}a}, {Feinstein}, {Fern{\'a}ndez
  Laj{\'u}s}, {Gamen}, {Geisler}, {Gieren}, {Goldman}, {Gonzalez}, {Gunthardt},
  {Gurovich}, {Hambly}, {Irwin}, {Ivanov}, {Jord{\'a}n}, {Kerins}, {Kinemuchi},
  {Kurtev}, {L{\'o}pez-Corredoira}, {Maccarone}, {Masetti}, {Merlo},
  {Messineo}, {Mirabel}, {Monaco}, {Morelli}, {Padilla}, {Palma}, {Parisi},
  {Pignata}, {Rejkuba}, {Roman-Lopes}, {Sale}, {Schreiber}, {Schr{\"o}der},
  {Smith}, {}, {Soto}, {Tamura}, {Tappert}, {Thompson}, {Toledo}, {Zoccali}, \&
  {Pietrzynski}}]{Minniti2010}
{Minniti}, D., {Lucas}, P.~W., {Emerson}, J.~P., {et~al.} 2010, \na, 15, 433

\bibitem[{{Minniti} {et~al.}(2015){Minniti}, {Contreras Ramos},
  {Alonso-Garc{\'{\i}}a}, {Anguita}, {Catelan}, {Gran}, {Motta}, {Muro},
  {Rojas}, \& {Saito}}]{Minniti2015}
{Minniti}, D., {Contreras Ramos}, R., {Alonso-Garc{\'{\i}}a}, J., {et~al.}
  2015, \apjl, 810, L20

\bibitem[{{Minniti} {et~al.}(2017){Minniti}, {D{\'e}k{\'a}ny}, {Majaess},
  {Palma}, {Pullen}, {Rejkuba}, {Alonso-Garc{\'{\i}}a}, {Catelan}, {Contreras
  Ramos}, {Gonzalez}, {Hempel}, {Irwin}, {Lucas}, {Saito}, {Tissera},
  {Valenti}, \& {Zoccali}}]{Minniti2017}
{Minniti}, D., {D{\'e}k{\'a}ny}, I., {Majaess}, D., {et~al.} 2017, \aj, 153,
  179

\bibitem[{{Mottram} {et~al.}(2011){Mottram}, {Hoare}, {Urquhart}, {Lumsden},
  {Oudmaijer}, {Robitaille}, {Moore}, {Davies}, \& {Stead}}]{Mottram2011}
{Mottram}, J.~C., {Hoare}, M.~G., {Urquhart}, J.~S., {et~al.} 2011, \aap, 525,
  A149

\bibitem[{{Navarete} {et~al.}(2015){Navarete}, {Damineli}, {Barbosa}, \&
  {Blum}}]{Navarete2015}
{Navarete}, F., {Damineli}, A., {Barbosa}, C.~L., \& {Blum}, R.~D. 2015,
  \mnras, 450, 4364

\bibitem[{{Navarro Molina} {et~al.}(2016){Navarro Molina}, {Borissova},
  {Catelan}, {Alonso-Garc{\'{\i}}a}, {Kerins}, {Kurtev}, {Lucas}, {Medina},
  {Minniti}, \& {D{\'e}k{\'a}ny}}]{Navarro2016}
{Navarro Molina}, C., {Borissova}, J., {Catelan}, M., {et~al.} 2016, \mnras,
  462, 1180

\bibitem[{{Nishiyama} {et~al.}(2009){Nishiyama}, {Tamura}, {Hatano}, {Kato},
  {Tanab{\'e}}, {Sugitani}, \& {Nagata}}]{Nishiyama2009}
{Nishiyama}, S., {Tamura}, M., {Hatano}, H., {et~al.} 2009, \apj, 696, 1407

\bibitem[{{Ojha} {et~al.}(2004){Ojha}, {Tamura}, {Nakajima}, {Fukagawa},
  {Sugitani}, {Nagashima}, {Nagayama}, {Nagata}, {Sato}, {Vig}, {Ghosh},
  {Pickles}, {Momose}, \& {Ogura}}]{Ojha2004}
{Ojha}, D.~K., {Tamura}, M., {Nakajima}, Y., {et~al.} 2004, \apj, 616, 1042

\bibitem[{Oliphant(2006)}]{Oliphant}
Oliphant, T. 2006, Guide to NumPy

\bibitem[{{Palma} {et~al.}(2016){Palma}, {Minniti}, {D{\'e}k{\'a}ny},
  {Clari{\'a}}, {Alonso-Garc{\'{\i}}a}, {Gramajo}, {Ram{\'{\i}}rez
  Alegr{\'{\i}}a}, \& {Bonatto}}]{Palma2016}
{Palma}, T., {Minniti}, D., {D{\'e}k{\'a}ny}, I., {et~al.} 2016, \na, 49, 50

\bibitem[{Pedregosa {et~al.}(2011)Pedregosa, Varoquaux, Gramfort, Michel,
  Thirion, Grisel, Blondel, Prettenhofer, Weiss, Dubourg, Vanderplas, Passos,
  Cournapeau, Brucher, Perrot, \& Duchesnay}]{scikit-learn}
Pedregosa, F., Varoquaux, G., Gramfort, A., {et~al.} 2011, Journal of Machine
  Learning Research, 12, 2825

\bibitem[{{Perryman}(2005)}]{Perryman2005}
{Perryman}, M.~A.~C. 2005, in Astronomical Society of the Pacific Conference
  Series, Vol. 338, Astrometry in the Age of the Next Generation of Large
  Telescopes, ed. P.~K. {Seidelmann} \& A.~K.~B. {Monet}, 3

\bibitem[{{Pojmanski}(1998)}]{Pojmanski1998}
{Pojmanski}, G. 1998, \actaa, 48, 35

\bibitem[{{Rebull} {et~al.}(2014){Rebull}, {Cody}, {Covey}, {G{\"u}nther},
  {Hillenbrand}, {Plavchan}, {Poppenhaeger}, {Stauffer}, {Wolk}, {Gutermuth},
  {Morales-Calder{\'o}n}, {Song}, {Barrado}, {Bayo}, {James}, {Hora}, {Vrba},
  {Alves de Oliveira}, {Bouvier}, {Carey}, {Carpenter}, {Favata}, {Flaherty},
  {Forbrich}, {Hernandez}, {McCaughrean}, {Megeath}, {Micela}, {Smith},
  {Terebey}, {Turner}, {Allen}, {Ardila}, {Bouy}, \& {Guieu}}]{Rebull2014}
{Rebull}, L.~M., {Cody}, A.~M., {Covey}, K.~R., {et~al.} 2014, \aj, 148, 92

\bibitem[{{Saito} {et~al.}(2012{\natexlab{a}}){Saito}, {Minniti}, {Dias},
  {Hempel}, {Rejkuba}, {Alonso-Garc{\'{\i}}a}, {Barbuy}, {Catelan}, {Emerson},
  {Gonzalez}, {Lucas}, \& {Zoccali}}]{Saito2012}
{Saito}, R.~K., {Minniti}, D., {Dias}, B., {et~al.} 2012{\natexlab{a}}, \aap,
  544, A147

\bibitem[{{Saito} {et~al.}(2012{\natexlab{b}}){Saito}, {Hempel}, {Minniti},
  {Lucas}, {Rejkuba}, {Toledo}, {Gonzalez}, {Alonso-Garc{\'{\i}}a}, {Irwin},
  {Gonzalez-Solares}, {Hodgkin}, {Lewis}, {Cross}, {Ivanov}, {Kerins},
  {Emerson}, {Soto}, {Am{\^o}res}, {Gurovich}, {D{\'e}k{\'a}ny}, {Angeloni},
  {Beamin}, {Catelan}, {Padilla}, {Zoccali}, {Pietrukowicz}, {Moni Bidin},
  {Mauro}, {Geisler}, {Folkes}, {Sale}, {Borissova}, {Kurtev}, {Ahumada},
  {Alonso}, {Adamson}, {Arias}, {Bandyopadhyay}, {Barb{\'a}}, {Barbuy},
  {Baume}, {Bedin}, {Bellini}, {Benjamin}, {Bica}, {Bonatto}, {Bronfman},
  {Carraro}, {Chen{\`e}}, {Clari{\'a}}, {Clarke}, {Contreras}, {Corvill{\'o}n},
  {de Grijs}, {Dias}, {Drew}, {Fari{\~n}a}, {Feinstein},
  {Fern{\'a}ndez-Laj{\'u}s}, {Gamen}, {Gieren}, {Goldman},
  {Gonz{\'a}lez-Fern{\'a}ndez}, {Grand}, {Gunthardt}, {Hambly}, {Hanson},
  {He{\l}miniak}, {Hoare}, {Huckvale}, {Jord{\'a}n}, {Kinemuchi}, {Longmore},
  {L{\'o}pez-Corredoira}, {Maccarone}, {Majaess}, {Mart{\'{\i}}n}, {Masetti},
  {Mennickent}, {Mirabel}, {Monaco}, {Morelli}, {Motta}, {Palma}, {Parisi},
  {Parker}, {Pe{\~n}aloza}, {Pietrzy{\'n}ski}, {Pignata}, {Popescu}, {Read},
  {Rojas}, {Roman-Lopes}, {Ruiz}, {Saviane}, {Schreiber}, {Schr{\"o}der},
  {Sharma}, {Smith}, {Sodr{\'e}}, {Stead}, {Stephens}, {Tamura}, {Tappert},
  {Thompson}, {Valenti}, {Vanzi}, {Walton}, {Weidmann}, \&
  {Zijlstra}}]{Saito2012DR1}
{Saito}, R.~K., {Hempel}, M., {Minniti}, D., {et~al.} 2012{\natexlab{b}}, \aap,
  537, A107

\bibitem[{{Sana} {et~al.}(2011){Sana}, {James}, \& {Gosset}}]{Sana2011}
{Sana}, H., {James}, G., \& {Gosset}, E. 2011, \mnras, 416, 817

\bibitem[{{Scargle}(1982)}]{Scargle1982}
{Scargle}, J.~D. 1982, \apj, 263, 835

\bibitem[{{Schechter} {et~al.}(1993){Schechter}, {Mateo}, \&
  {Saha}}]{Schechter1993}
{Schechter}, P.~L., {Mateo}, M., \& {Saha}, A. 1993, \pasp, 105, 1342

\bibitem[{{Schmidt-Kaler}(1982)}]{Schmidt-Kaler1982}
{Schmidt-Kaler}, T. 1982, Bulletin d'Information du Centre de Donnees
  Stellaires, 23, 2

\bibitem[{{Shin} {et~al.}(2009){Shin}, {Sekora}, \& {Byun}}]{Shin2009}
{Shin}, M.-S., {Sekora}, M., \& {Byun}, Y.-I. 2009, \mnras, 400, 1897

\bibitem[{{Skrutskie} {et~al.}(2006){Skrutskie}, {Cutri}, {Stiening},
  {Weinberg}, {Schneider}, {Carpenter}, {Beichman}, {Capps}, {Chester},
  {Elias}, {Huchra}, {Liebert}, {Lonsdale}, {Monet}, {Price}, {Seitzer},
  {Jarrett}, {Kirkpatrick}, {Gizis}, {Howard}, {Evans}, {Fowler}, {Fullmer},
  {Hurt}, {Light}, {Kopan}, {Marsh}, {McCallon}, {Tam}, {Van Dyk}, \&
  {Wheelock}}]{Skrutskie2006}
{Skrutskie}, M.~F., {Cutri}, R.~M., {Stiening}, R., {et~al.} 2006, \aj, 131,
  1163

\bibitem[{{Smith} {et~al.}(2018){Smith}, {Lucas}, {Kurtev}, {Smart}, {Minniti},
  {Borissova}, {Jones}, {Zhang}, {Marocco}, {Contreras Pe{\~n}a}, {Gromadzki},
  {Kuhn}, {Drew}, {Pinfield}, \& {Bedin}}]{Smith2018}
{Smith}, L.~C., {Lucas}, P.~W., {Kurtev}, R., {et~al.} 2018, \mnras, 474, 1826

\bibitem[{{Sokolovsky} {et~al.}(2017){Sokolovsky}, {Gavras}, {Karampelas},
  {Antipin}, {Bellas-Velidis}, {Benni}, {Bonanos}, {Burdanov}, {Derlopa},
  {Hatzidimitriou}, {Khokhryakova}, {Kolesnikova}, {Korotkiy}, {Lapukhin},
  {Moretti}, {Popov}, {Pouliasis}, {Samus}, {Spetsieri}, {Veselkov}, {Volkov},
  {Yang}, \& {Zubareva}}]{Sokolovsky2017}
{Sokolovsky}, K.~V., {Gavras}, P., {Karampelas}, A., {et~al.} 2017, \mnras,
  464, 274

\bibitem[{{Stetson}(1996)}]{Stetson1996}
{Stetson}, P.~B. 1996, \pasp, 108, 851

\bibitem[{{Taylor}(2006)}]{Taylor2006}
{Taylor}, M.~B. 2006, in Astronomical Society of the Pacific Conference Series,
  Vol. 351, Astronomical Data Analysis Software and Systems XV, ed.
  C.~{Gabriel}, C.~{Arviset}, D.~{Ponz}, \& S.~{Enrique}, 666

\bibitem[{{The Astropy Collaboration} {et~al.}(2018){The Astropy
  Collaboration}, {Price-Whelan}, {Sip{\H o}cz}, {G{\"u}nther}, {Lim},
  {Crawford}, {Conseil}, {Shupe}, {Craig}, {Dencheva}, {Ginsburg},
  {VanderPlas}, {Bradley}, {P{\'e}rez-Su{\'a}rez}, {de Val-Borro}, {Aldcroft},
  {Cruz}, {Robitaille}, {Tollerud}, {Ardelean}, {Babej}, {Bachetti}, {Bakanov},
  {Bamford}, {Barentsen}, {Barmby}, {Baumbach}, {Berry}, {Biscani}, {Boquien},
  {Bostroem}, {Bouma}, {Brammer}, {Bray}, {Breytenbach}, {Buddelmeijer},
  {Burke}, {Calderone}, {Cano Rodr{\'{\i}}guez}, {Cara}, {Cardoso},
  {Cheedella}, {Copin}, {Crichton}, {D{\'A}vella}, {Deil}, {Depagne},
  {Dietrich}, {Donath}, {Droettboom}, {Earl}, {Erben}, {Fabbro}, {Ferreira},
  {Finethy}, {Fox}, {Garrison}, {Gibbons}, {Goldstein}, {Gommers}, {Greco},
  {Greenfield}, {Groener}, {Grollier}, {Hagen}, {Hirst}, {Homeier}, {Horton},
  {Hosseinzadeh}, {Hu}, {Hunkeler}, {Ivezi{\'c}}, {Jain}, {Jenness}, {Kanarek},
  {Kendrew}, {Kern}, {Kerzendorf}, {Khvalko}, {King}, {Kirkby}, {Kulkarni},
  {Kumar}, {Lee}, {Lenz}, {Littlefair}, {Ma}, {Macleod}, {Mastropietro},
  {McCully}, {Montagnac}, {Morris}, {Mueller}, {Mumford}, {Muna}, {Murphy},
  {Nelson}, {Nguyen}, {Ninan}, {N{\"o}the}, {Ogaz}, {Oh}, {Parejko}, {Parley},
  {Pascual}, {Patil}, {Patil}, {Plunkett}, {Prochaska}, {Rastogi}, {Reddy
  Janga}, {Sabater}, {Sakurikar}, {Seifert}, {Sherbert}, {Sherwood-Taylor},
  {Shih}, {Sick}, {Silbiger}, {Singanamalla}, {Singer}, {Sladen}, {Sooley},
  {Sornarajah}, {Streicher}, {Teuben}, {Thomas}, {Tremblay}, {Turner},
  {Terr{\'o}n}, {van Kerkwijk}, {de la Vega}, {Watkins}, {Weaver}, {Whitmore},
  {Woillez}, \& {Zabalza}}]{Astropy2018}
{The Astropy Collaboration}, {Price-Whelan}, A.~M., {Sip{\H o}cz}, B.~M.,
  {et~al.} 2018, ArXiv e-prints, arXiv:1801.02634

\bibitem[{Thompson(1985)}]{Thompson1985}
Thompson, R. 1985, Journal of the Royal Statistical Society. Series B
  (Methodological), 47, 53.
\newblock \url{http://www.jstor.org/stable/2345543}

\bibitem[{{Vacca} {et~al.}(2003){Vacca}, {Cushing}, \& {Rayner}}]{Vacca2003}
{Vacca}, W.~D., {Cushing}, M.~C., \& {Rayner}, J.~T. 2003, \pasp, 115, 389

\bibitem[{{Vanderplas} {et~al.}(2012){Vanderplas}, {Connolly}, {Ivezi{\'c}}, \&
  {Gray}}]{astroML}
{Vanderplas}, J., {Connolly}, A., {Ivezi{\'c}}, {\v Z}., \& {Gray}, A. 2012, in
  Conference on Intelligent Data Understanding (CIDU), 47 --54

\bibitem[{von Neumann(1941)}]{Neumann1941}
von Neumann, J. 1941, Ann. Math. Statist., 12, 367.
\newblock \url{http://dx.doi.org/10.1214/aoms/1177731677}

\bibitem[{{Welch} \& {Stetson}(1993)}]{Welch1993}
{Welch}, D.~L., \& {Stetson}, P.~B. 1993, \aj, 105, 1813

\bibitem[{{Wolk} {et~al.}(2013){Wolk}, {Rice}, \& {Aspin}}]{Wolk2013}
{Wolk}, S.~J., {Rice}, T.~S., \& {Aspin}, C. 2013, \apj, 773, 145

\bibitem[{{Zechmeister} \& {K{\"u}rster}(2009)}]{Zechmeister2009}
{Zechmeister}, M., \& {K{\"u}rster}, M. 2009, \aap, 496, 577

\end{thebibliography}



\end{document}